\documentclass[%
 aip,
 amsmath,amssymb,
 reprint,%
]{revtex4-1}

\usepackage{graphicx}
\usepackage{dcolumn}
\usepackage{bm}

\usepackage[utf8]{inputenc}
\usepackage[T1]{fontenc}
\usepackage{mathptmx}
\usepackage{etoolbox}

\makeatletter
\def\@email#1#2{%
 \endgroup
 \patchcmd{\titleblock@produce}
  {\frontmatter@RRAPformat}
  {\frontmatter@RRAPformat{\produce@RRAP{*#1\href{mailto:#2}{#2}}}\frontmatter@RRAPformat}
  {}{}
}%
\makeatother

\begin{document}

\preprint{AIP/123-QED}

\title[Sample title]{Transition from chimera/solitary states to traveling waves}
\author{E. Rybalova}
\email{rybalovaev@gmail.com}
\affiliation{%
Institute of Physics, Saratov State University, 83 Astrakhanskaya Street, Saratov 410012, Russia
}%

\author{S. Muni}%
\email{sishushankarmuni@gmail.com}
\affiliation{ 
Department of Physical Sciences, Indian Institute of Science Education and Research Kolkata, Campus Road, Mohanpur, West Bengal 741246, India
}%

\author{G. Strelkova}
\email{strelkovagi@sgu.ru}
\affiliation{%
Institute of Physics, Saratov State University, 83 Astrakhanskaya Street, Saratov 410012, Russia
}%

\date{\today}

\begin{abstract}
We study numerically the spatiotemporal dynamics in a ring network of nonlocally coupled nonlinear oscillators, each represented by a two-dimensional discrete-time model of the classical van der Pol oscillator. It is shown that the discretized oscillator exhibits a richer behavior, combining the peculiarities of both the original system and its own dynamics. Moreover, a large variety of spatiotemporal structures is observed in the network of discrete van der Pol oscillators when the discretization parameter and the coupling strength are varied. Regimes such as the coexistence of  multichimera state/traveling wave and solitary state are revealed for the first time and are studied in detail. It is established that the majority of the observed chimera/solitary states, including the newly found ones, are transient towards the purely traveling wave mode. The peculiarities of the transition process and the lifetime (transient duration) of the chimera structures and the solitary state are analyzed depending on the system parameters, the observation time, initial conditions, and the influence of external noise. 
\end{abstract}

\maketitle

\begin{quotation}

Uncovering the peculiarities of various structure formation and analyzing phase transitions between them plays an essential role in studying complex  networks.  This knowledge can give a deeper insight into the processes observed in real-world systems and enables one to predict and control the effective functioning of vital systems. In the numerical simulation, the nature of  individual elements of networks and the topology of coupling often act as determining factors in the structure formation.  
In this case, different spatiotemporal patterns can be observed, such as synchronization, desynchronization, traveling waves, chimera states of various types, solitary states, and phase transitions  can be realized between some of them in time. Here, we investigate the formation of a variety of spatiotemporal patterns in a network of nonlocally coupled discrete-time van der Pol oscillators. 
We report the emergence of two dynamical regimes, such as the coexistence of chimera state/traveling wave and solitary states. Our numerical experiments show that some of the observed structures  are transient patterns which finally convert to the traveling wave mode. 
The lifetime of chimera and of solitary states is analyzed in detail depending on the network parameters, the observation time, initial conditions,  and noise influence. All the regimes are illustrated by snapshots and space-time plots of the network dynamics, mean phase velocity profiles, projections of multidimensional attractors of the system, and cross-correlation coefficient distributions. We also construct and analyze riddled basins of attraction for different spatiotemporal structures.

\end{quotation}

\section{\label{sec:intro}Introduction}

The formation of complex spatiotemporal structures and their evolution are important and is a challenging issue in studying the cooperative dynamics of interacting nonlinear dynamical systems. This analysis can give insight into diverse real-world systems \cite{Pikovsky:2001uj,Strogatz:2001aa,Albert:2002wu,Newman:2003wd,Balanov:2009tt,Boccaletti:2018up}
and, in particular, enables one  to deduce mechanisms and necessary conditions for  the proper functioning of real-world systems, e.g., power grids, communication, infrastructure, and transportation networks \cite{Cardillo:2013wg}, 
and to reveal factors and mechanisms of different phase transitions between various patterns \cite{Jaros:2015tv,Semenova:2017tt}.
Uncovering the peculiarities of spatiotemporal dynamics of complex networks can also be useful to develop efficient methods of controlling the structure formation and transitions to avoid any undesirable effects in the system dynamics.
 
 Different types of dynamics can be observed in complex networks, depending on the behavior and nature of individual elements and the character and topology of links between the nodes \cite{Amari:1977wi,Compte:2000ul,Panaggio:2015uu,Scholl:2016vm,Muni:2018vd,Belyaev:2020we}.
 Besides completely  synchronized periodic and chaotic oscillations, desynchronized chaotic dynamics, traveling waves, more complex spatiotemporal structures can be found in networks. One of them is a chimera state which was initially discovered in the network of identical phase oscillator with symmetric nonlocal coupling topology \cite{Kuramoto:2002uu,Abrams:2004vx}.  
 Chimera state represents a peculiar partial synchronization pattern which combines coexisting spatially separated domains of coherence and incoherence. Chimeras have recently been found and studied theoretically and numerically in networks with different types of individual elements \cite{Omelchenko:2011uc,Dudkowski:2014vm,Slepnev:2017tz,Zakharova:2014td,Ulonska:2016tx,Semenova:2016aa,Tsigkri-DeSmedt:2016wm} 
 and coupling topology \cite{Buscarino:2015vr,Omelchenko:2015uu,Banerjee:2016tw,Ghosh:2016vc,Majhi:2017to,Kasatkin:2017wl,Bukh:2017vp,Sawicki:2017um}
 and observed in experiments \cite{Hagerstrom:2012vd,Larger:2013ub,Martens:2013wq,Kapitaniak:2014vc,Gambuzza:2014vj,Rosin:2014wy,Schmidt:2014ui,Tinsley:2012tc,Wickramasinghe:2013tp,Shena:2017ul}. 
 
Another important type of complex partial synchronization patterns is exemplified by a solitary state which is recognized as a network state for which single or several elements behave differently compared with the majority of units which can exhibit
either coherent dynamics or be fully synchronized \cite{Jaros:2015tv,Maistrenko:2014tm,Berner:2020um}. Solitary states have recently been observed in networks of the Kuramoto–Sakaguchi models, Kuramoto oscillators with inertia \cite{Maistrenko:2014tm,Jaros:2015tv,Jaros:2018ve,Wu:2018ws,Berner:2020um}, discrete-time systems \cite{Semenova:2015tt,Rybalova:2017tl,Semenova:2017wn,Semenova:2018th}, the FitzHugh–Nagumo neurons \cite{Mikhaylenko:2019uq,Rybalova:2019vg,Schulen:2019td,Schulen:2021tn},  models of power grids \cite{Taher:2019wz,Hellmann:2020wr,Berner:2021wd}, and in experimental setups \cite{Kapitaniak:2014vc}.

It is  necessary to note that the properties and peculiarities of chimera states can essentially depend on the local dynamics of individual elements of networks. Besides the classical Kuramoto chimera \cite{Kuramoto:2002uu}, different types of chimera states have been revealed and then classified \cite{Kemeth:2016vc}. 
They include phase and amplitude chimeras \cite{Zakharova:2016vm,Bogomolov:2017wq}, chimera death \cite{Zakharova:2014td}, double-well  chimeras \cite{Shepelev:2017uy,MuZe22,Muni20,VaMu22}, multicluster and traveling chimeras \cite{Xie:2014uj},
multichimera states \cite{Omelchenko:2013uv}, 
intermittent chimeras \cite{Olmi:2015uv}, 
turbulent chimeras \cite{Bordyugov:2010wg,Shena:2017ul,Bolotov:2018uz},  coherence-resonance chimeras \cite{Semenova:2016aa}, 3D chimeras \cite{Maistrenko:2015vq}, spiral  and target wave chimeras \cite{Kuramoto:2003tj,Shima:2004aa,Bukh:2019wo}, solitary state chimeras \cite{Bukh:2017vp,Rybalova:2018we}, etc. 
There are cases when networks of nonlocally coupled nonlinear systems, e.g., the FitzHugh-Nagumo system network \cite{Mikhaylenko:2019uq,Rybalova:2019vg}, the Lorenz system ensemble \cite{Omelchenko:2012tv,Semenova:2015tt,Majhi:2019wh}, 
the Kuramoto–Sakaguchi models \cite{Wolfrum:2011wb,Maistrenko:2014tm} can demonstrate both chimeras and solitary states for appropriately chosen parameters of the individual nodes and the coupling strength. At the same time, only solitary states are typically observed in certain networks upon the transition from coherence to incoherence, for example, the Lozi map ring \cite{Semenova:2015tt,Rybalova:2017tl,Semenova:2017wn,Semenova:2018th}.

 Some types of chimera states, e.g., amplitude chimeras, are characterized by a finite lifetime and can be transient \cite{Loos:2016aa,Semenova:2017tt,Rybalova:2019un}. After that, these structures either no longer exist (chimera death), as shown for the network of nonlocally coupled Stuart-Landau oscillators \cite{Loos:2016aa}, or is replaced by phase chimeras in networks of nonlocally coupled logistic and Henon maps \cite{Semenova:2017tt,Rybalova:2019un}. Certain types of chimeras, e.g., phase ones, are stationary and stable structures against external perturbations \cite{Semenova:2017tt,Rybalova:2019un}. Amplitude chimeras are rather sensitive to the influence of noise which can either suppress them \cite{Loos:2016aa,Bukh:2018wa} or increase their lifetime \cite{Rybalova:2019un}. It has also been shown that in a network of Stuart-Landau oscillators, multichimera states can be transient to traveling waves  \cite{Loos:2016aa}.
 
  Thus, it is logical and justified to continue studying the influence of the local dynamics of individual elements of networks on the formation of various spatiotemporal structures. This can contribute to the expansion of knowledge about the features of existing complex structures and transitions between them, as well as to the possibility of observing new patterns and dynamical regimes. 

In connection to the classical continuous-time van der Pol oscillators, solitary states were found in a 2D lattice\cite{ShMu20a} which were responsible for the formation of a variety of spatiotemporal patterns. Spiral waves, spiral wave chimeras were observed and quantified in a lattice of van der Pol oscillators\cite{ShMu20b} and various types of coupling form such as 
repulsive coupling\cite{ShMu21a} were considered. Antiphase synchronization was observed in multiplex networks\cite{ShMu21b, ShMu21c}. Since the mechanism of various local and global bifurcations are different in continuous- and discrete-time systems, it is natural to understand a discrete version of a continuous-time dynamical oscillator and explore associated spatiotemporal patterns in a variety of network topologies. 

 In this paper, we provide a detailed systematic study of the spatiotemporal dynamics of a network of nonlocally coupled oscillators which are defined by a discrete analog of the classical  van der Pol oscillator. We apply an artificial discretization  procedure (the Euler-Cromer method\cite{Cromer:1981vm}) to the original continuous-time system to derive a discrete-time oscillator.   
 It was shown \cite{Popova-eng:2020tz}  that such discretized systems can demonstrate richer dynamics as compared with the system-prototype. Chaos can even arise from a low-dimensional discrete analog of continuous-time systems \cite{USHIO:1986vr, Sprott:2003vc}.
 Similar discretization schemes were applied to van der Pol oscillators, Thomson oscillators in Ref.\cite{Zaytcev:2017vj}. A new class of discrete systems was achieved through the method of integration gains\cite{NguyenVan:2013vc}.
 We explore various spatiotemporal structures in the network of discrete van der Pol oscillator and define regions of their existence in the parameter plane of the discretization parameter and the coupling strength. Two new spatiotemporal structures are detected, which represent the coexistence  of chimera state/traveling wave and solitary states. 

Multistability refers to the coexistence of attractors in a dynamical system. There has been many works which showcase multistability in many nonlinear systems \cite{Astakhov:2001wq,Muni:2020wy}.  Recently, researchers have found coexistence of a variety of spatiotemporal patterns in networks of oscillators\cite{VaMu22}. In this paper we will refer to the term of multistability as the coexistence of different spatiotemporal patterns for a fixed set of network parameters. 
In our work, we investigate in detail the transition from chimera states, the coexistence of traveling wave, solitary state to the traveling wave mode, which appears to be an interesting aspect in the considered network. It is shown that some structures observed in the network are characterized  by riddled basins of attraction for different states. We also analyze the influence of additive noise and variation of the coupling strength between elements on the lifetime of chimera and solitary states in the considered network. 
 
\section{\label{sec:model}Model under study}

\subsection{Network equations}

In our numerical simulation we deal with a network of coupled two-dimensional discrete-time nonlinear oscillators. The network has a ring topology, i.e., it is characterized by periodic boundary conditions, and is described by the following system of equations:
     \begin{eqnarray}\label{system}
x_{i}(n+1)&=&f(x_{i}(n),y_{i}(n))+\\\nonumber 
&+&\frac{\sigma}{2R}\sum\limits_{j=i-R}^{i+R}[f(x_{j}(n),y_{j}(n))-f(x_{i}(n),y_{i}(n))]\\\nonumber 
y_{i}(n+1)&=&g(x_{i}(n),y_{i}(n)),~~~~~~x_{i \pm N}(n) \equiv x_{i}(n),\\\nonumber
     \end{eqnarray}
where $x_{i}$ is the dynamical variable, $i=1,2,\dots,N$ denotes the element number, $N=1000$ is the total number of elements in the network, and $n$ is the discrete time. The functions $f(x,y)$ and $g(x,y)$ represent a two-dimensional discrete model of the  van der Pol oscillator, which is used as an individual element of the studied network, i.e. $f(x,y)$ and $g(x,y)$ are right-hand parts of the first and the second equation of a two-dimensional discrete model of the  van der Pol oscillator, respectively.  The coupling between the nodes is organized as nonlocal, i.e., each node is coupled with a certain (given) number of neighbors from each side. This number is defined by the coupling range $R$, and the strength between the elements is controlled by the parameter $\sigma$. Initial conditions are randomly and uniformly distributed within the interval $[-0.5,0.5]$. The choice of this range for the initial conditions enables us to exclude the system going to infinity and, at the same time, includes basins of attraction of the structures which are observed. All these facts allow us to construct properly a diagram of dynamical regimes of the system.  An exception is the study of basins of attraction of attractors corresponding to various structures, where the initial conditions are chosen from a different range, which will be noted in the text when describing these investigations.
     
\subsection{\label{sec:single_map}Dynamics of a single discrete oscillator}

In our work, a discrete-time model (map) of the classical van der Pol oscillator  is taken as an individual element of the network (\ref{system}). This map is derived by using an artificial discretization procedure applied to  the van der Pol oscillator, which is governed by the second-order ordinary differential equation: 

     \begin{eqnarray}\label{vdP}
\ddot{u}-(\lambda-u^2)\dot{u}+u=0.
     \end{eqnarray}
\noindent Here $\lambda$ is a parameter that characterizes the ``negative dissipation'', and the term $u^2$ is responsible for the nonlinear energy loss, which increases as the oscillation amplitude grows.

\begin{figure}[ht]
	\centering
\includegraphics[width=1.\columnwidth]{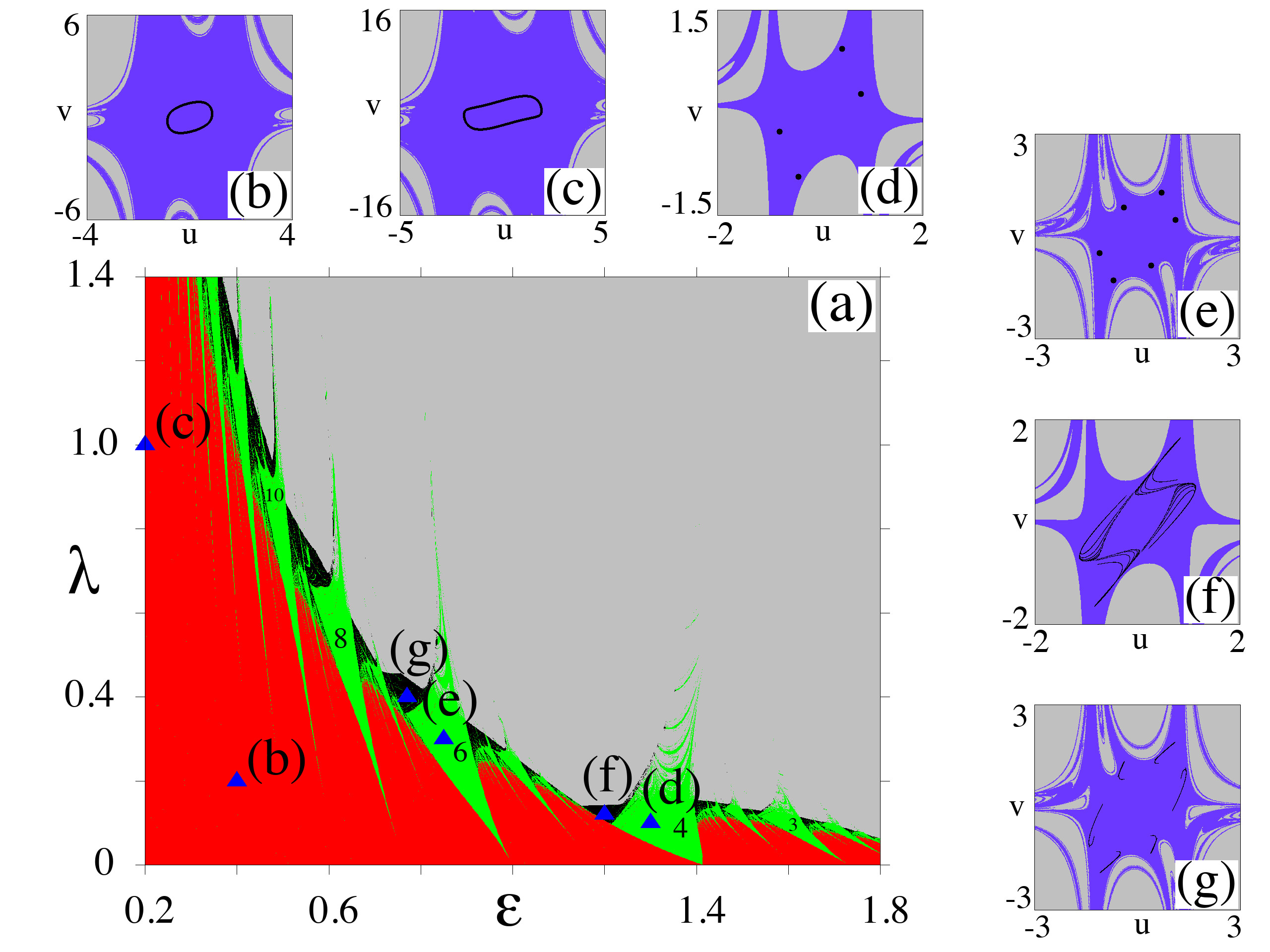} \\
\caption{Two-parameter diagram of dynamical regimes in the single discrete van der Pol oscillator \eqref{discrete_vdP} in the ($\epsilon, \lambda$) parameter plane (a), and attractors of certain typical regimes (black lines or points) with their basins of attraction (purple color), grey regions correspond to the divergence of trajectories to infinity (b)-(g). Quasiperiodic dynamics are marked by red colour, resonant cycles of different periods (marked by  numbers) are observed inside the green regions, chaotic dynamics are marked in black, and  trajectories diverging to infinity are marked in grey. Attractors for different values of the parameters $\epsilon$ and $\lambda$ (marked by blue triangles in the diagram) : (b) $\epsilon=0.4$, $\lambda=0.2$, (c) $\epsilon=0.2$, $\lambda=1.0$, (d) $\epsilon=1.3$, $\lambda=0.1$, (e) $\epsilon=0.85$, $\lambda=0.3$, (f) $\epsilon=1.2$, $\lambda=0.12$, (g) $\epsilon=0.77$, $\lambda=0.4$.}
	\label{regimes_single_map}
\end{figure}

We use the Euler-Cromer method\cite{Cromer:1981vm} to discretize the van der Pol oscillator \eqref{vdP}. This method is well suited for oscillatory problems since it conserves energy, unlike the forward Euler method which artificially increases the oscillator's energy with time. Applying such an artificial discretization technique for  continuous-time systems  to Eq.~\eqref{vdP}, we get the equations for the discrete van der Pol oscillator:

     \begin{eqnarray}\label{discrete_vdP}
u_{n+1}&=&u_{n}+\epsilon v_{n+1},\\\nonumber
v_{n+1}&=&v_{n}+\epsilon(\lambda v_{n}-u^2_n v_n -u_n),\\\nonumber
     \end{eqnarray}

\noindent where $u,v$ represent the state variables, $\epsilon$ is a parameter representing the integration (discretization) step. In such a system, $\epsilon$ can no longer be considered small and can take both negative and positive values. The latter discretization method enables one to obtain a map that can both possess the properties of the original oscillator and demonstrate a richer dynamics when $\epsilon$ is varied. To illustrate this, we calculate the Lyapunov exponent spectrum using Benettin's algorithm~\cite{Benettin:1980tq} for the single discrete van der Pol oscillator \eqref{discrete_vdP} and construct a dynamical regime diagram in the ($\epsilon, \lambda$) parameter plane, which is plotted in Fig.~\ref{regimes_single_map}(a).  Different colors in the diagram enables one to distinguish regions with different dynamics observed in \eqref{discrete_vdP}. They are exemplified by several attractors, also shown in Fig.~\ref{regimes_single_map}(b)-(g).
Moreover, Figs.~\ref{regimes_single_map}(b)-(g) demonstrate not only attractors but also their basins of attraction (blue color), and regions of initial conditions that correspond to divergence of trajectories to infinity.
It is seen that the discrete van der Pol oscillator can demonstrate  quasiperiodic dynamics (the red region) which corresponds to attractors in the form of closed invariant curves in \eqref{discrete_vdP} (Figs.~\ref{regimes_single_map}(b) and \ref{regimes_single_map}(c)) and to limit cycles in the original oscillator \eqref{vdP}. The diagram also reflects the presence of a countable set of resonance regions or Arnold tongues (the green areas in the diagram). Inside them, there are resonant cycles of different periods (marked by numbers in the diagram) as illustrated by exemplary attractors in Figs.~\ref{regimes_single_map}(d) and \ref{regimes_single_map}(e). As the parameter $\lambda$ increases, a transition to chaos in the discrete van der Pol oscillator is realized via a quasiperiodic dynamics destruction route (the Afraimovich-Shilnikov scenario \cite{Afraimovich:1991ul}), and chaotic oscillations (Figs.~\ref{regimes_single_map}(f) and \ref{regimes_single_map}(g)) can be observed in \eqref{discrete_vdP} (the black region in the diagram). Trajectories diverge to infinity in the grey region. 

\subsection{Quantitative measures}

In order to explore in detail a variety of spatiotemporal regimes which can be observed in the network of nonlocally coupled discrete van der Pol oscillators  \eqref{system}, we  construct snapshots of variable $x_i$, space-time plots of variable $x_i$, projections of multidimensional attractors of the system (where multidimensional attractors correspond to some established spatiotemporal regimes) and mean phase velocity profiles. The mean phase velocity for each element  in the ring is calculated as follows:
     \begin{eqnarray}\label{phase_velocity}
w_{i}=2\pi M_{i}/ \Delta T,
     \end{eqnarray}
\noindent where $i = 1,2,\ldots,N$, where $M_i$ is the number of complete rotations around the origin performed by the $i$th unit during the time interval $\Delta T$ (Ref.\cite{Omelchenko:2013uv}). 

We also evaluate the statistical relationship between the coupled elements in the network \eqref{system} by calculating the 
cross-correlation coefficient for the pair of elements $i$ and $k$:

     \begin{eqnarray}\label{CCC}
C_{ik}=\frac{\langle {\tilde{x}}_i(n) {\tilde{x}}_k(n) \rangle}{\sqrt{\langle ({\tilde{x}}_i(n))^2 \rangle  \langle ({\tilde{x}}_k(n))^2 \rangle}},
     \end{eqnarray}

\noindent where $\tilde{x}(n) = x(n) - \langle x \rangle$ is the fluctuation around the mean value $\langle x \rangle$. The angle brackets $\langle \ldots \rangle$ denote time-averaging. 

\section{\label{sec:regimes}Spatiotemporal regimes in the ring network of discrete van der Pol oscillators}

As can be seen from the diagram structure shown in Fig.~\ref{regimes_single_map}(a), 
various dynamical regimes, such as quasiperiodic, periodic and chaotic modes, can be observed in the single discrete van der Pol oscillator 
as one of the parameters $\epsilon$ or $\lambda$ is varied while the other one is fixed.  Thus, the dynamics of the network of coupled maps \eqref{system} will be sensitive to the changes in the control parameters of the individual elements. In order to uncover this dependence, we fix $\lambda=0.12$, and the coupling range $R=320$ and study the network dynamics when  $\epsilon$ and the coupling strength $\sigma$ are varied.

\begin{figure}[ht]
	\centering
\includegraphics[width=.95\columnwidth]{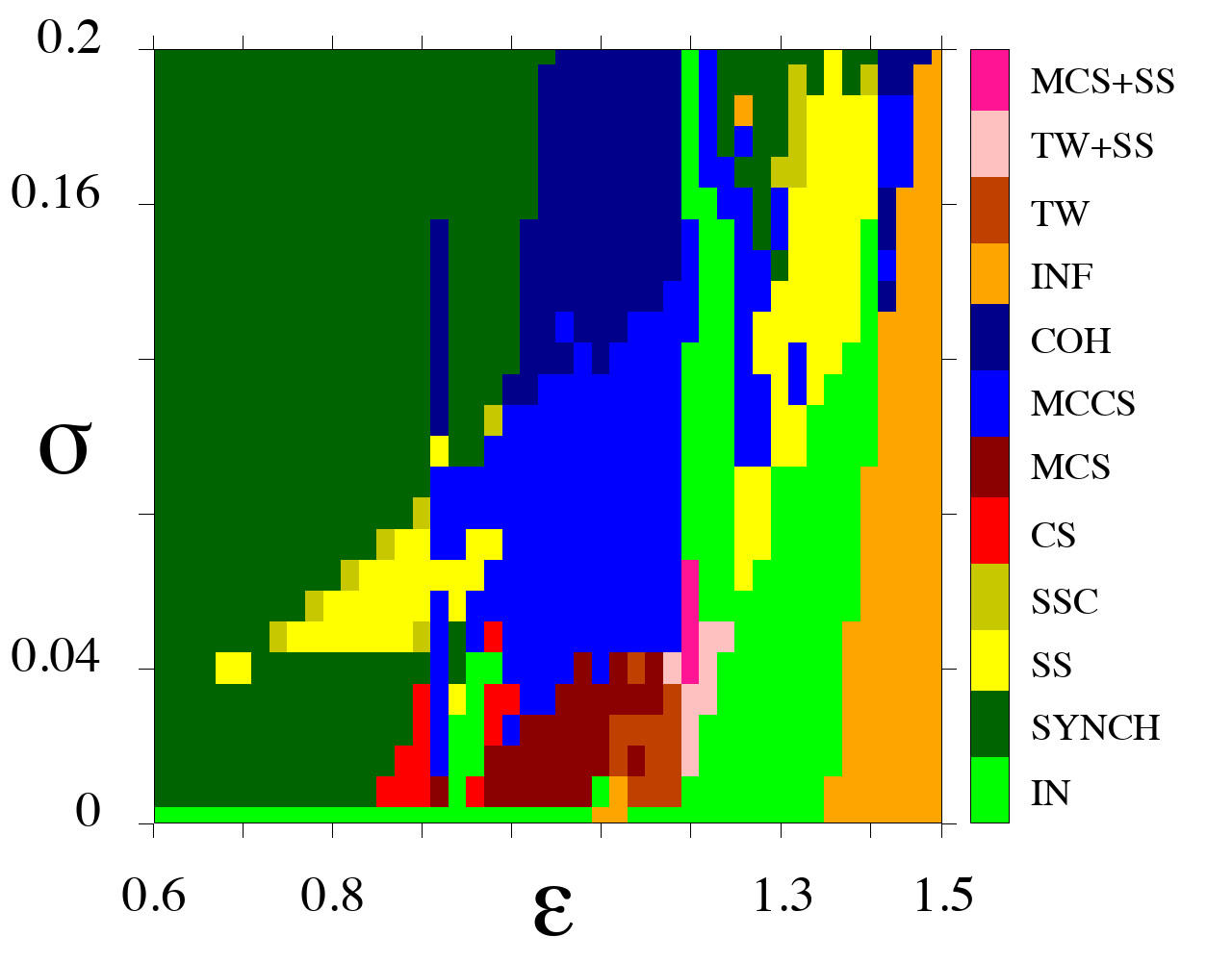} \\
\caption{Schematic diagram of spatiotemporal regimes in the network of nonlocally coupled discrete van der Pol oscillators \eqref{system} in  the ($\epsilon,\sigma$) parameter plane. Designations: IN -- incoherence, SYNCH -- complete synchronization, SS -- solitary state, SSC -- solitary state chimera, CS -- chimera state, MCS -- multichimera state, MCCS -- multicluster chimera state, COH -- coherence, INF -- going to infinity, TW -- traveling wave, TW+SS -- coexisting traveling wave and solitary state, MCS+SS -- coexisting multichimera state and solitary state. Other parameters: $\lambda=0.12$, $R=320$, $N=1000$, $T_{\rm obs}=100,000$ iterations. Sampling steps are fixed at $\Delta\epsilon=0.02$ and $\Delta\sigma=0.008$.}
	\label{map_regimes_ring}
\end{figure}

\begin{figure*}
	\centering
\includegraphics[width=1.98\columnwidth]{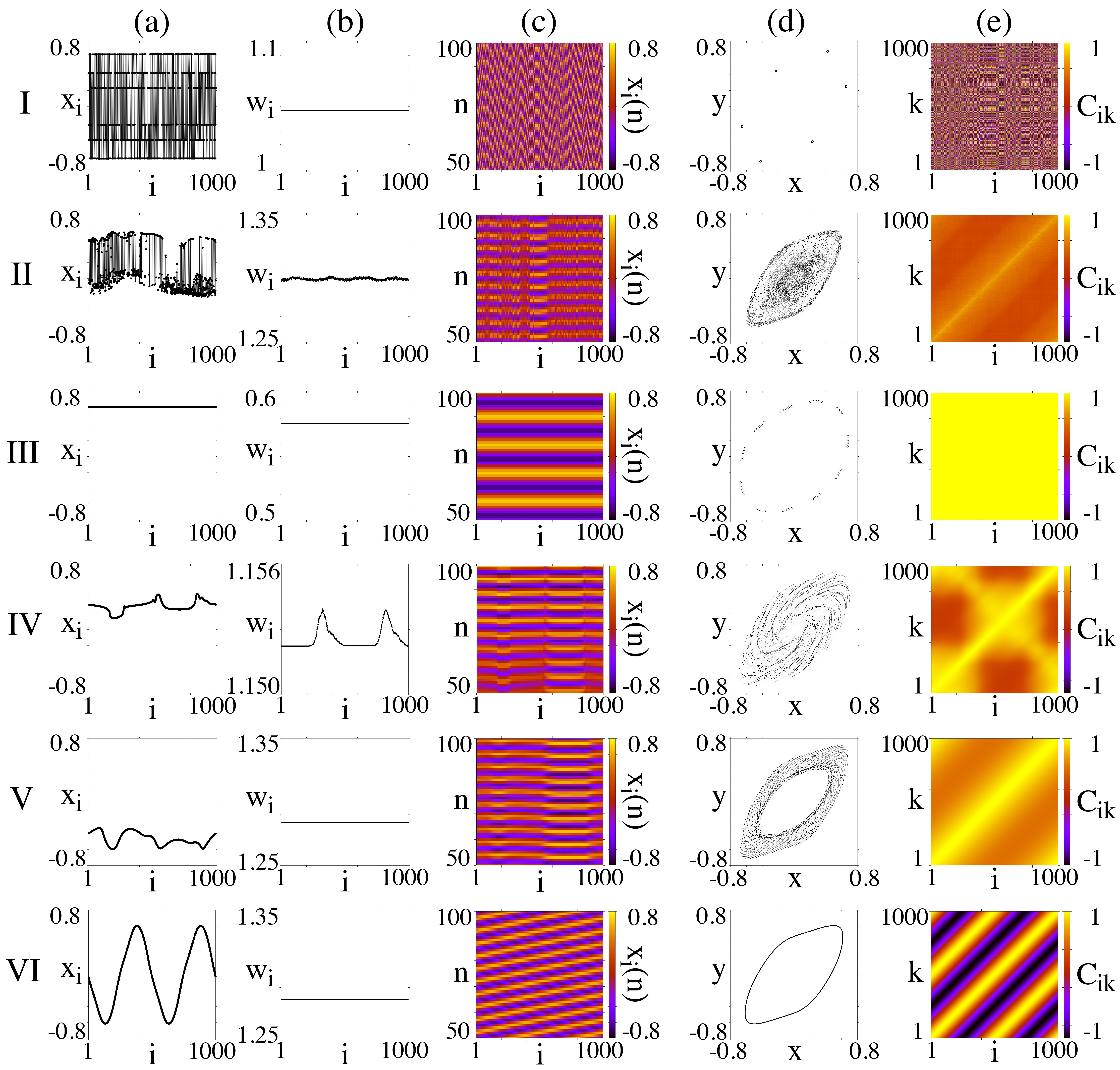} \\
\caption{Snapshots of variable $x_i$ (a), mean phase velocity profiles (b), space-time plots (c), projections of a multidimensional attractor of the system in the $(x,y)$ phase plane (d), and cross-correlation coefficient distributions (e) for different values of the parameters $\sigma$ and $\epsilon$: I panel: $\sigma=0.01$, $\epsilon=0.95$ (incoherence (IN)), II panel: $\sigma=0.11$, $\epsilon=1.2$ (incoherence (IN)), III panel: $\sigma=0.15$, $\epsilon=0.55$ (complete synchronization (SYNCH)), IV panel: $\sigma=0.12$, $\epsilon=1.06$, (coherence (COH)), V panel: $\sigma=0.18$, $\epsilon=1.15$ (coherence (COH)), VI panel: $\sigma=0.024$, $\epsilon=1.14$ ( traveling wave (TW)). The regime names and abbreviations correspond to Fig.~\ref{map_regimes_ring}. Other parameters: $\lambda=0.12$, $R=320$,  $N=1000$, $T_{\rm obs}=500,000$ and the averaging time for cross-correlation coefficient distributions is $T = 450,000$ iterations.}
	\label{regimes_ring_classical}
\end{figure*}

\begin{figure*}
	\centering
\includegraphics[width=1.98\columnwidth]{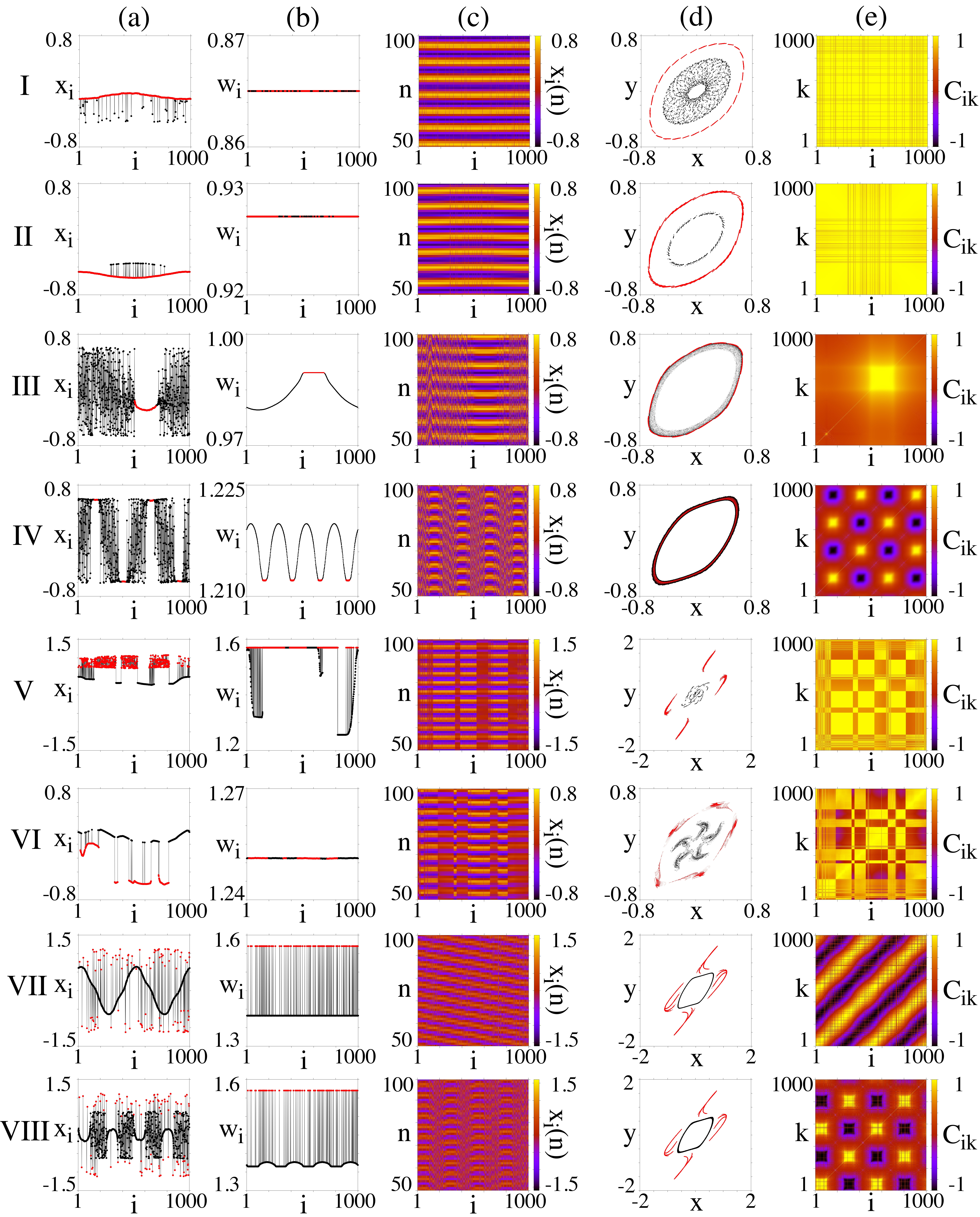} \\
\caption{Snapshots of variable $x_i$ (a), mean phase velocity profiles (b), space-time plots (c), projections of a multidimensional attractor of the system in the $(x,y)$ phase plane (d), and cross-correlation coefficient distributions (e) for different values of the parameters $\sigma$ and $\epsilon$: I panel: $\sigma=0.05$, $\epsilon=0.8$ (solitary state (SS)), II panel: $\sigma=0.07$, $\epsilon=0.85$ (solitary state chimera (SSC)), III panel: $\sigma=0.016$, $\epsilon=0.9$ (chimera state (CS)), IV panel: $\sigma=0.03$, $\epsilon=1.1$ (multichimera state (MCS)), V panel: $\sigma=0.15$, $\epsilon=1.2$ (multicluster chimera state (MCCS)), VI panel: $\sigma=0.12$, $\epsilon=1.15$  (multicluster chimera state (MCCS)), VII panel: $\sigma=0.024$, $\epsilon=1.2$ (coexisting traveling wave and solitary state (TW+SS)), VIII panel: $\sigma=0.03$, $\epsilon=1.2$ (coexisting multichimera state and solitary state (MCS+SS)). The regime names and abbreviations correspond to Fig.~\ref{map_regimes_ring}. Other parameters:  $\lambda=0.12$, $R=320$, $N=1000$, $T_{\rm obs}=500,000$ and the averaging time for cross-correlation coefficient distributions is $T = 450,000$ iterations.}
	\label{regimes_ring_other}
\end{figure*}

A schematically constructed diagram of spatiotemporal regimes in the network \eqref{system} is shown in Fig.~\ref{map_regimes_ring} in the ($\epsilon, \sigma$) parameter plane. The observation time is set to be $T_{\rm obs}=100,000$ iterations. As can be seen, the network dynamics appears to be rather diverse as  the discretization parameter $\epsilon$ changes. Our studies have enabled us to distinguish twelve different spatiotemporal regimes. The corresponding color scale with the regime abbreviations are given in the right side of the diagram (the full descriptions are provided in the figure caption). 
There exist so-called classical dynamical modes, such as incoherence, coherence, complete synchronization,  and traveling waves, which are typically observed in networks of coupled oscillators. Besides, the network of discrete van der Pol models can exhibit chimera states, solitary states, and the coexistence of different regimes, which are now intensively studied. Each of the found regimes is illustrated in Figs.~\ref{regimes_ring_classical} and~\ref{regimes_ring_other} by snapshots of the network dynamics (the $x_i$ variable values (column (a)),  mean phase velocity distributions (column (b)),  space-time plots (column (c)), projections of a multidimensional attractor of the system in the $(x,y)$ phase plane (column (d)), and cross-correlation coefficient distributions (column (e)).
Since the network under consideration is multidimensional (a 2000-dimensional system), the established spatiotemporal regime correspond to a multidimensional attractor. To visualize this attractor in the figures throughout the paper, the multidimensional attractor is represented as a projection in the $(x,y)$ phase plane, that is, in the plane of the values of the dynamical  variables of all network elements.

We can see that the incoherent dynamics of the network, which is exemplified by Figs.~\ref{regimes_ring_classical},I and \ref{regimes_ring_classical},II, is mainly observed for rather large values of the parameter $\epsilon$ (Fig.~\ref{map_regimes_ring},IN). 
In the first case (Fig.\ref{regimes_ring_classical},I), the control parameters of the elements lie in the area of quasiperiodic dynamics (Fig.~\ref{regimes_single_map}(a), red region), however, close to a resonant period-6 cycle within one of the Arnold tongues. Weak coupling between the elements causes the effective values of the control parameters to change, but it is still insufficient to synchronize the elements in the ring. 
Thus, the network is characterized by spatially incoherent dynamics (Fig.~\ref{regimes_ring_classical},I(a)) and periodic (or nearly periodic) oscillations of the elements in time  (Fig.~\ref{regimes_ring_classical},I(d)). It is obvious that all nodes have the same  mean phase velocity and thus, the mean phase velocity profile is flat (Fig.~\ref{regimes_ring_classical},I(b)).

The control parameters in the second case (Fig.~\ref{regimes_ring_classical},II) also belong to the chaotic dynamics domain (Fig.~\ref{regimes_single_map}(a), black region). Therefore, when the elements are coupled within a wide range of the coupling strength values (Fig.~\ref{map_regimes_ring} at $\epsilon=1.2$, green region), an incoherent regime with chaotic dynamics in time is observed (Fig.~\ref{regimes_ring_classical},II(d)).
However, as can be seen from Fig.~\ref{regimes_ring_classical},II(b), the mean phase velocity is rather identical for all nodes and  only slightly fluctuates around the value $1.29$. 
At larger values of $\epsilon$ (the orange region (INF) in Fig.~\ref{map_regimes_ring}), the trajectories of the nodes diverge to infinity for all values of the coupling strength.

On the contrary, complete synchronization (Fig.~\ref{regimes_ring_classical},III) occurs at small values of $\epsilon$, at which the individual elements demonstrate quasiperiodic dynamics (compare Fig.~\ref{map_regimes_ring},SYNCH and Fig.~\ref{regimes_single_map}(a), red region). It should be noted that for complete synchronization, the dynamics of the elements in time is periodic, and the phase space includes a countable set of points that form, as it were, a limit cycle (Fig.~\ref{regimes_ring_classical},III(d)). In this case, the mean phase velocity distribution is uniform (Fig.~\ref{regimes_ring_classical},III(b)). 

As the coupling strength increases, the coherent dynamics region also appears (Fig.~\ref{map_regimes_ring},COH). Exemplary figures are shown in  Figs.~\ref{regimes_ring_classical},IV and \ref{regimes_ring_classical},V). 
Depending on the values of $\epsilon$ and $\sigma$, the mean phase velocity profile can be either nonuniform (Fig.~\ref{regimes_ring_classical},IV(b)) or flat (Fig.~\ref{regimes_ring_classical},V(b)). However, chaotic attractors are  observed in the phase space (Figs.~\ref{regimes_ring_classical},IV(d) and \ref{regimes_ring_classical},V(d)), whose shape is also defined by the parameters of the system and can be quite diverse.

The last regime, which we refer to as the classical one, is related to traveling waves.  This network dynamics is observed within a small range of the parameters $\sigma$ and $\epsilon$ (Fig~.\ref{map_regimes_ring},TW) for the given parameters of the individual elements, the initial conditions, and the observation time.  A typical example of this regime is shown in Fig.~\ref{regimes_ring_classical},VI. Note that this mode is classical for ensembles of coupled elements which exhibit a self-sustained oscillation mode without coupling \cite{Endo:1978tf,Ermentrout:1985wo}. However, in our case of nonlocal coupling and individual discrete-time systems, this regime is observed only at  weak coupling  between the elements. Moreover, some of the found regimes disappear with time and only traveling waves exist  in the considered network \eqref{system}. These findings will be discussed in more detail below.

We now consider and describe dynamical regimes in the network of nonlocally coupled discrete van der Pol oscillators, which we cannot classify as classical ones. Firstly, approximately within the intervals $0.65<\epsilon<1.0$ and $0.04<\sigma<0.08$, the network exhibits a solitary state mode which occupies the yellow area in the ($\epsilon, \sigma$) parameter plane (Fig.~\ref{map_regimes_ring},SS) and is exemplified in  Fig.~\ref{regimes_ring_other},I. 
There is one more region with solitary state, which is ranged within the intervals $1.25<\epsilon<1.4$ and $0.06<\sigma<0.2$. 
On the boundary of the solitary state regions, the solitary state chimera (Fig.~\ref{regimes_ring_other},II) can be found in the network within a narrow region (Fig.~\ref{map_regimes_ring},SSC). It was shown earlier \cite{Bukh:2017vp,Rybalova:2018we}, this kind of chimeras is characterized by the coexistence of a coherent cluster and a cluster containing solitary nodes, as clearly seen in the snapshot in Fig.~\ref{regimes_ring_other},II(a).
Note that for these two cases, the mean phase velocity is identical for all elements (Figs.~\ref{regimes_ring_other},I(b) and \ref{regimes_ring_other},II(b)), and two non-intersecting sets exist in the phase space (Figs.~\ref{regimes_ring_other},I(d) and \ref{regimes_ring_other},II(d)). One of them corresponds to the oscillators in the coherent mode (also called as typical states), which form a smooth part of the snapshot, and the other one -- to the solitary nodes (Figs.~\ref{regimes_ring_other},I(d) and \ref{regimes_ring_other},II(d), red and black dots, respectively). Such structures for mean phase velocity profiles and the phase space organization are typical for the solitary state regime and the solitary state chimera \cite{Semenova:2018th,Rybalova:2018we,Rybalova:2019wd} and were also observed in a network of nonlocally coupled FitzHugh-Nagumo neurons\cite{Mikhaylenko:2019uq,Rybalova:2019vg}. Solitary states and solitary states chimeras from different regions of the parameter plane differ only in the projection of the multidimensional attractor of typical states and solitary nodes.

Using the cross-correlation coefficient distributions (Figs.~\ref{regimes_ring_other},I(e) and \ref{regimes_ring_other},II(e)), one can mention that the oscillators corresponding to the typical states are completely synchronized with each other ($C_{ik}\approx1$). On the other hand, the cross-correlation coefficients between the typical states and the solitary nodes are less than $1$ and their values lie in the interval $C_{ik}\in(0.4,0.44)$. At the same time, the cross-correlation coefficients between the solitary nodes are approximately $1$ and thus we can say  that the solitary nodes are also synchronized with each other.  

The next dynamical regime which is observed in the nonlocally coupled discrete van der Pol oscillators is a chimera state. In fact, several types of chimera states are distinguished in  the ($\epsilon,\sigma$) parameter plane in Fig.~\ref{map_regimes_ring}:
\par1. A classic chimera state  is observed within the red region in the diagram (Fig.~\ref{map_regimes_ring},CS and Fig.~\ref{regimes_ring_other},III). This state is  characterized by a domed mean phase velocity profile  (Fig.~\ref{regimes_ring_other},III(b)). However, in this case, the nodes from the  coherent cluster  have the largest mean phase velocity, while it decreases for the units from the incoherence domain and reaches a minimum in its center. There are two intersecting sets in the phase space (Fig.~\ref{regimes_ring_other},III(d)): the red almost closed curve corresponds to the coherent cluster, and the wider attractor (black one) corresponds to the incoherent cluster.
\par2. A multichimera state is found within the burgundy region (Fig.~\ref{map_regimes_ring},MCS) and is illustratively pictured in Fig.~\ref{regimes_ring_other},IV. This is a kind of the classical chimera but consists of four coherent and four incoherent clusters (Fig.~\ref{regimes_ring_other},IV(a)). The mean phase velocities of the nodes from the incoherent clusters  are slightly larger  as compared to those for the oscillators of the coherent domains (Fig.~\ref{regimes_ring_other},IV(b)).  
\par3. A multicluster chimera state exists within the blue region in the diagram (Fig.~\ref{map_regimes_ring},MCCS) and is exemplified in Figs.~\ref{regimes_ring_other},V  and \ref{regimes_ring_other},VI). This structure typically includes several coherent and incoherent clusters that can have different spatiotemporal dynamics (compare clusters in Figs.~\ref{regimes_ring_other},V(a) and \ref{regimes_ring_other},VI(a) and Figs.~\ref{regimes_ring_other},V(c) and \ref{regimes_ring_other},VI(c)). Moreover, there are two different attracting sets in the phase space projection, which the nodes from different clusters can belong to (Figs.~\ref{regimes_ring_other},V(d) and \ref{regimes_ring_other},VI(d)).

There are also two dynamical regimes in the network \eqref{system}, which represent the coexistence of different dynamical modes:
\par1. The coexistence of traveling waves and solitary nodes is observed in the light-pink region in the diagram (Fig.~\ref{map_regimes_ring},TW+SS). An exemplary snapshot and a projection of the multidimensional attractor of the system are shown in  Fig.~\ref{regimes_ring_other},VII.
\par2. The multichimera state and solitary nodes coexist within the hot pink region in the ($\epsilon, \sigma$) parameter plane (Fig~.\ref{map_regimes_ring},MCS+SS). This regime is illustrated in Fig~.\ref{regimes_ring_other},VIII.

The existence of two different attracting subsets in the phase space projections is typical for both dynamical regimes described above.  A closed invariant curve corresponds to the oscillators of the traveling wave and of the multichimera state (black dots in Figs.~\ref{regimes_ring_other},VII(d) and \ref{regimes_ring_other},VIII(d)), while a chaotic attractor is related to the solitary nodes (red dots in Figs.~\ref{regimes_ring_other},VII(d) and \ref{regimes_ring_other},VIII(d)).
Note that in this case the projection of the multidimensional attractor of the solitary nodes qualitatively corresponds to the attractor of the single oscillator (see the example in Fig.~\ref{regimes_single_map}(f)).

Note that the regimes of the multichimera state, of the coexistence of the multichimera state and solitary nodes, and of the coexistence of traveling waves and solitary nodes are generally transient and as the time goes on, the traveling wave  mode is observed in the network. The transient time depends on both the system parameters  and the initial conditions. These regimes are observed at the values of $\lambda$ and $\epsilon$ which correspond to  either the chaotic behavior of the isolated discrete van der Pol oscillator, or to the close vicinity of the region with such  dynamics.

\section{\label{sec:chimera}Transition from the multichimera to traveling waves}

\begin{figure}[ht]
	\centering
\includegraphics[width=.95\columnwidth]{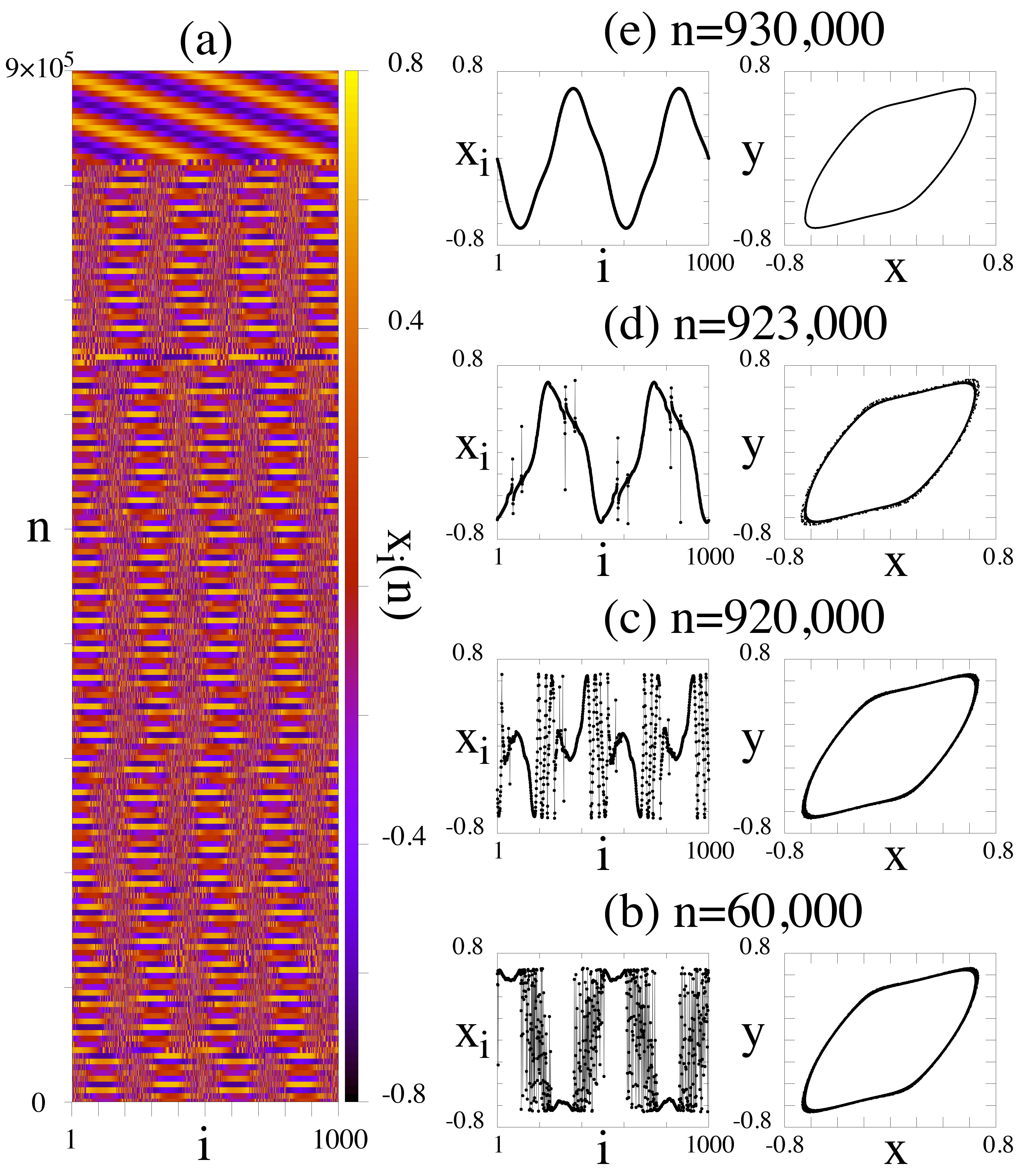} \\
\caption{Transition from multichimera state to  traveling waves. (a) Space-time plot of variable $x_i$, (b)-(e) snapshots of variable $x_i$ (left panel) and projections of  multidimensional attractors of the system in the $(x,y)$ phase plane (right panel) at different fixed times $n$: (b) $n=60,000$, (c) $n=920,000$, (d) $n=923,000$, and (e) $n=930,000$. Every 5000th iteration is displayed in the space-time plot (a). Other parameters: $\epsilon=1.17931$, $\lambda=0.12$,  $\sigma=0.02$, $R=320$, $N=1000$.}
	\label{chimera-to-traveling-wave}
\end{figure}

We now proceed to explore in more detail the dynamics of the network \eqref{system} for the parameter values which correspond to the appearance of the multichimera state. Our calculations demonstrate that this state can disappear over time, i.e., it has a finite lifetime,  and after this, only traveling waves are observed in the network under study. Figure~\ref{chimera-to-traveling-wave} shows the space-time plot of this regime (every 5000th iteration is displayed) and several snapshots and projections of  multidimensional attractors of the system at a different number of iterations $n$. As can be seen from the space-time plot (Fig.~\ref{chimera-to-traveling-wave}(a)), the incoherent clusters only slightly move around the ring with time, and for most of the observation time, there exists the regime illustrated in Fig.~\ref{chimera-to-traveling-wave}(b) for $n=60,000$. 
At these parameters and initial conditions, the incoherent clusters begin to destroy when $n>918,000$ (Fig.~\ref{chimera-to-traveling-wave}(c)). The wave number in these clusters gradually decreases, and the former coherent clusters transform into traveling waves.
Eventually, one can see traveling waves with solitary nodes (Fig.~\ref{chimera-to-traveling-wave}(d)), which rather quickly disappear, and only the traveling wave remains in the network (Fig.~\ref{chimera-to-traveling-wave}(e)). Note that the modes similar to that presented in Fig.~\ref{chimera-to-traveling-wave}(c)-(e)  can be observed until the complete disappearance (death) of the chimera state. However, small non-stationarities in the dynamics of the individual elements can return the network to the chimera mode.

\begin{figure}[ht]
	\centering
\includegraphics[width=.95\columnwidth]{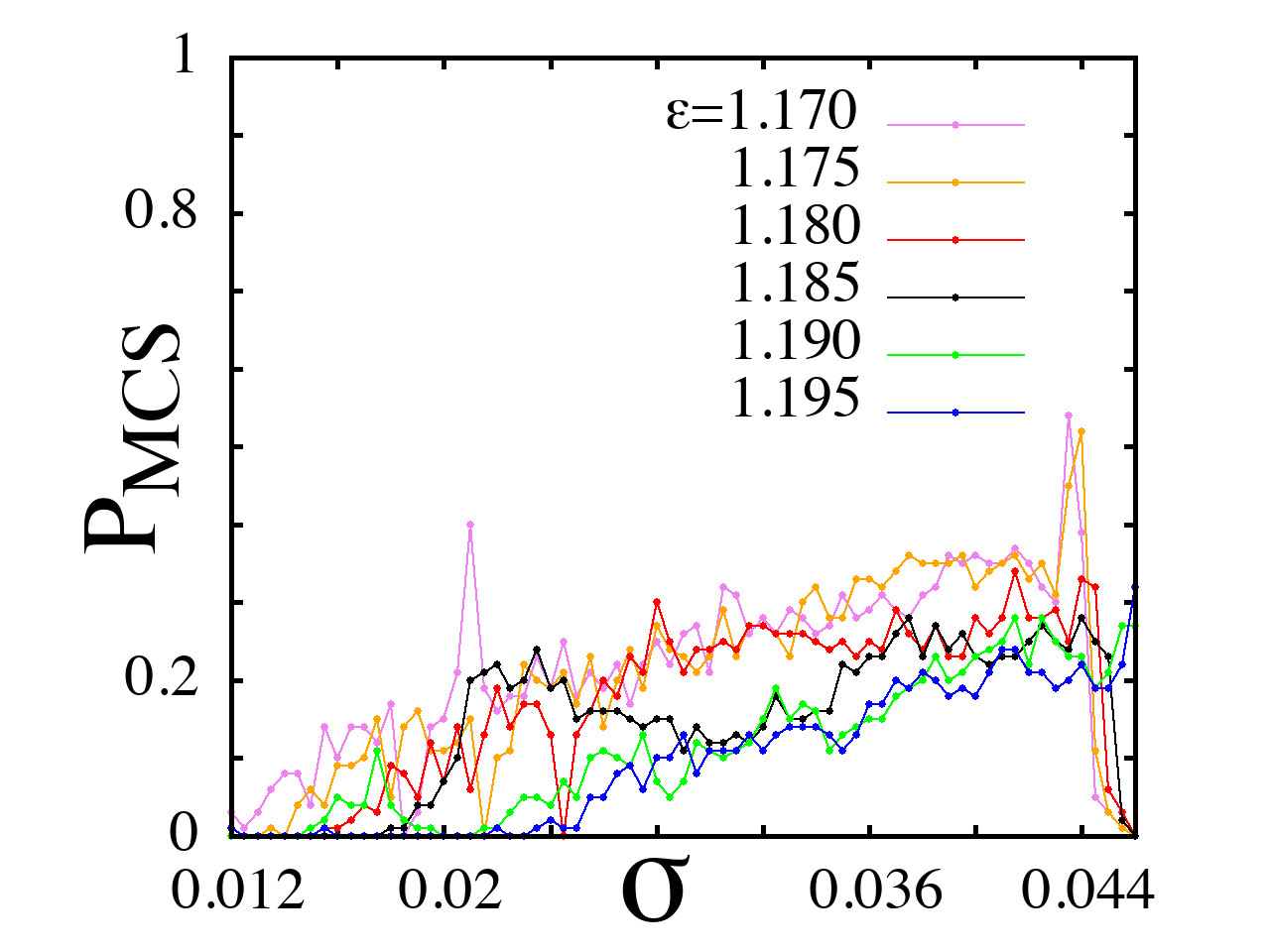} \\
\caption{Probability of observing the multichimera state $P_{\rm MCS}$ \eqref{P_MCS} versus the coupling strength $\sigma$ at the observation time $T_{\rm obs}=100,000$ iterations for different values of $\epsilon$ and at $R=320$ and $\lambda=0.12$. Hundred sets of initial conditions uniformly distributed in the interval $x_{i}(0),y_{i}(0)\in[-0.5,0.5]$, $i=1,2,\ldots,N=1000$ were used.}
	\label{probability-chimera}
\end{figure}

Figure~\ref{probability-chimera} shows the probability of observing the multichimera state at the observation time $T_{\rm obs}=100,000$ iterations as a function of the coupling strength $\sigma$ for different values of the parameter $\epsilon$. 
The probability is calculated by using the following relation:
      \begin{eqnarray}\label{P_MCS}
P_{MCS}=N_{MCS}/N_{IC},
     \end{eqnarray}
 \noindent where $N_{\rm MCS}$ is the number of the initial conditions which lead to the establishment of the stable multichimera state ($T_{\rm life}>T_{\rm obs}$), and $N_{\rm IC}=100$ is the total number of the used initial conditions.
It is seen from Fig.~\ref{probability-chimera} that for all the values of $\epsilon$, increasing the coupling strength leads to an increase in the probability of observing the multichimera, but the maximum probability is reached at $\epsilon\approx1.170$.

\subsection{\label{sec:chimera-basin}Riddled basins of attraction for multichimera states and traveling waves}

We explore numerically the phase space structure of the network \eqref{system} in the case of the transient multichimera state. Our calculations show that the basins of attraction of the existing limit sets are riddled, the such property have been established for chimera states\cite{Santos:2018vr,Dos-Santos:2020wv}, solitary nodes in solitary state chimera\cite{Rybalova:2018we}. The following technique is used to visualize the phase space structure. 
The whole space of the $(x,y)$ phase plane is divided into a grid with  constant steps $\Delta x$ and $\Delta y$, and the initial conditions for each element are chosen in the vicinity of the grid nodes with a size of $10^{-8}$.
The parameters are fixed to be  $\epsilon= 1.17931$, $\lambda=0.12$, $R=320$, and the calculation time is $T_{\rm obs} = 100,000$ iterations. 

\begin{figure}[ht]
	\centering
\begin{tabular}{ccc}
\includegraphics[width=.33\columnwidth]{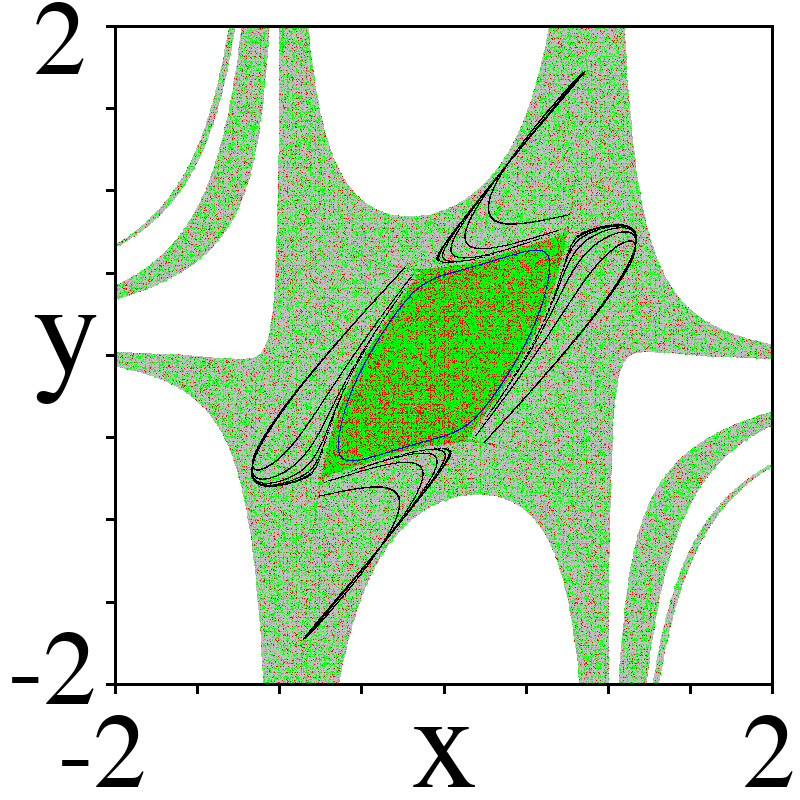} &
\includegraphics[width=.33\columnwidth]{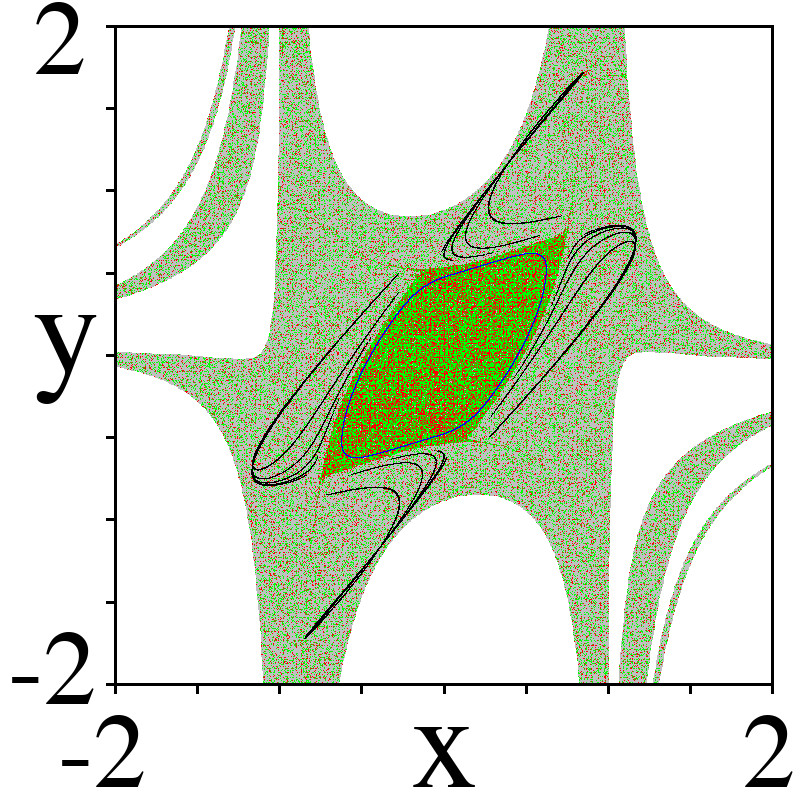} &
\includegraphics[width=.33\columnwidth]{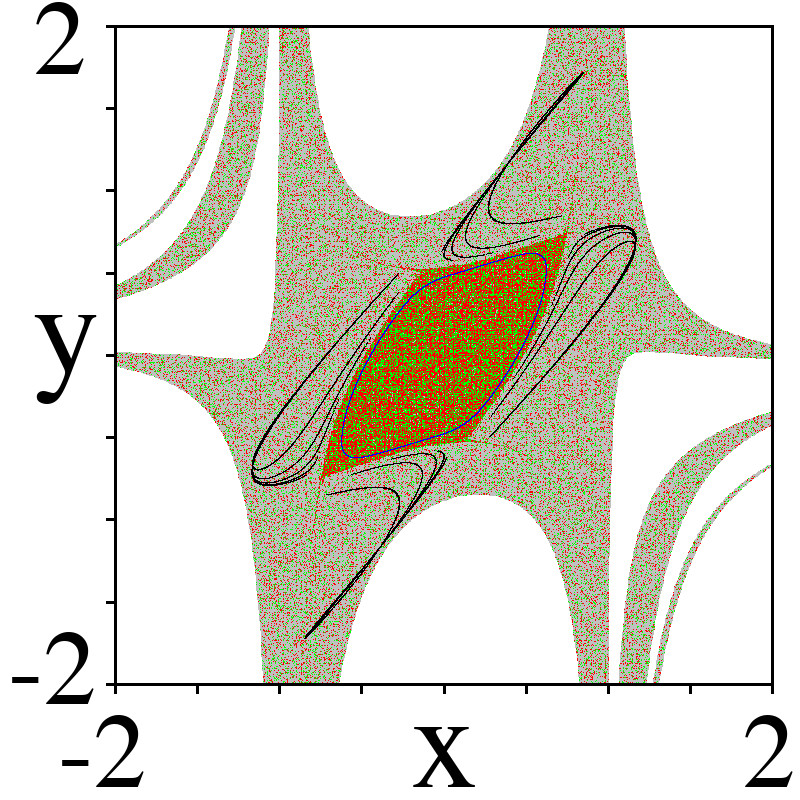}  \\
\hspace{8pt} (a) & \hspace{8pt} (b) & \hspace{8pt} (c)\\
\end{tabular}
\caption{Sections of basins of attraction and projections of multidimensional attractors for different values of the coupling strength $\sigma$: (a) $0.02$, (b) $0.025$, and (c) $0.03$. Grey dots show the basin of attraction for the attractor in the incoherent regime (black dots, Fig.~\ref{example-of-regimes-for-basins-IC-from-point-sigma=0.02}(a)), green dots correspond to the basin of attraction of the attractor in the traveling wave mode (blue dots, Fig.~\ref{example-of-regimes-for-basins-IC-from-point-sigma=0.02}(b)), and red dots denote the basin of attraction of the attractor for the multichimera state (Fig.~\ref{example-of-regimes-for-basins-IC-from-point-sigma=0.02}(c)). Trajectories from the white region diverge to inifinity. Other parameters: $\epsilon=1.17931$, $\lambda=0.12$, $R=320$, $T_{obs}=100,000$, steps of division of the $(x,y)$ phase plane are $\Delta x=0.005$ and $\Delta y=0.005$}
	\label{basins-IC-from-point}
\end{figure}

\begin{figure}[ht]
	\centering
\begin{tabular}{c}
\includegraphics[width=.95\columnwidth]{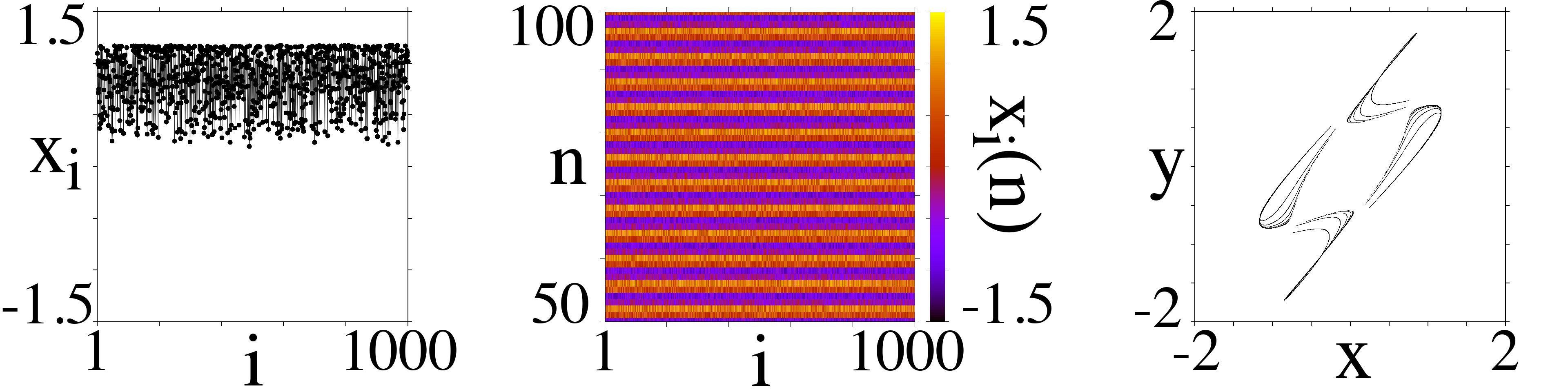} \\
\hspace{8pt} (a)\\
\includegraphics[width=.95\columnwidth]{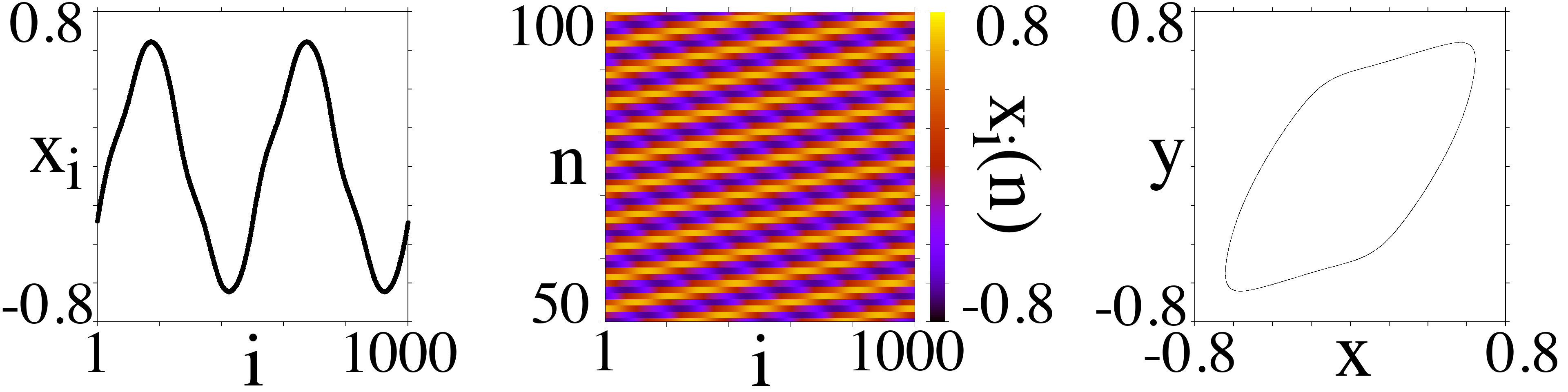}  \\
\hspace{8pt} (b)\\
\includegraphics[width=.95\columnwidth]{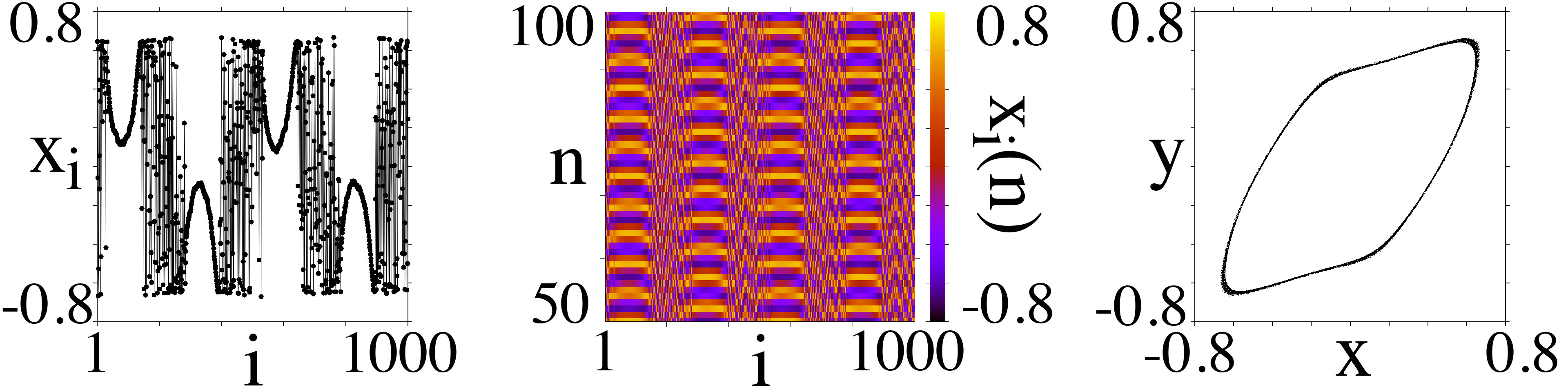}  \\
\hspace{8pt} (c)\\
\end{tabular}
\caption{Snapshots of  variable $x_i$ (left column), space-time plots (middle column), and projections of multidimensional attractors of the system in the $(x,y)$ phase plane (right column) for the incoherent regime (a),  the traveling wave mode (b), and the multichimera state (c). Other parameters: $\epsilon=1.17931$, $\lambda=0.12$, $\sigma=0.02$,   $R=320$, $T_{\rm obs}=100,000$.}
	\label{example-of-regimes-for-basins-IC-from-point-sigma=0.02}
\end{figure}

\begin{figure}[ht]
	\centering
\begin{tabular}{c}
\includegraphics[width=.95\columnwidth]{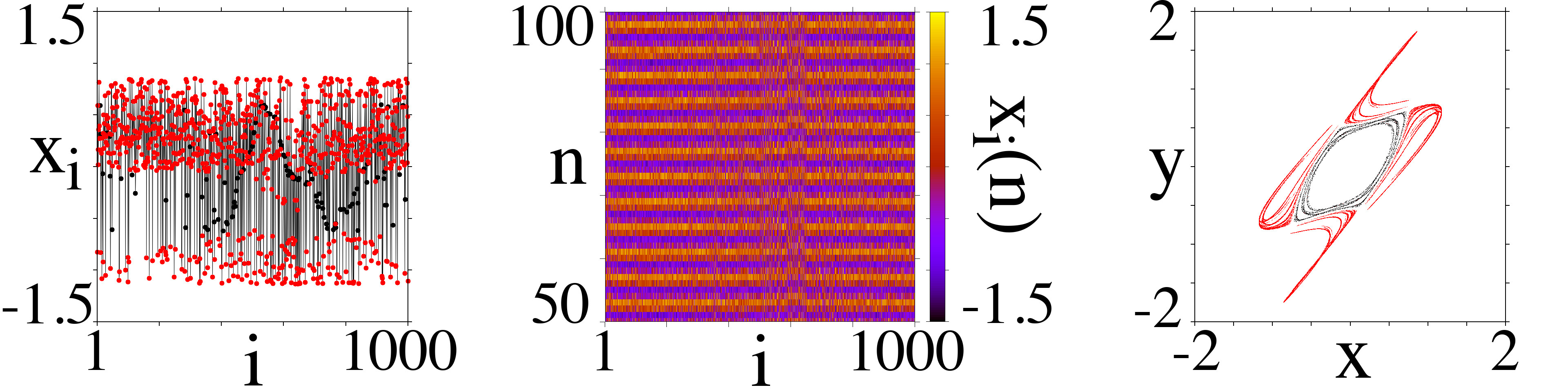} \\
\hspace{8pt} (a)\\
\includegraphics[width=.95\columnwidth]{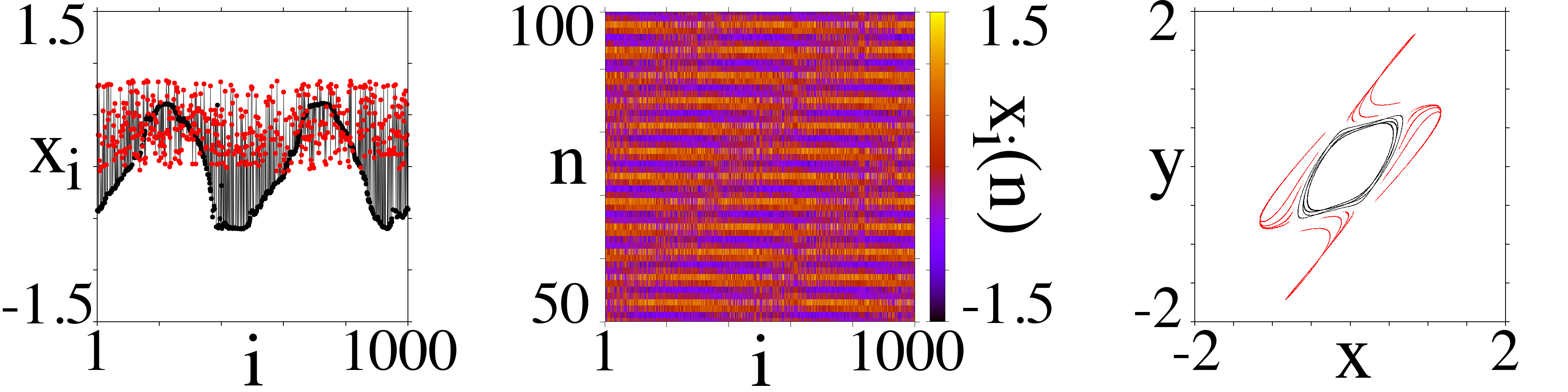}  \\
\hspace{8pt} (b)\\
\includegraphics[width=.95\columnwidth]{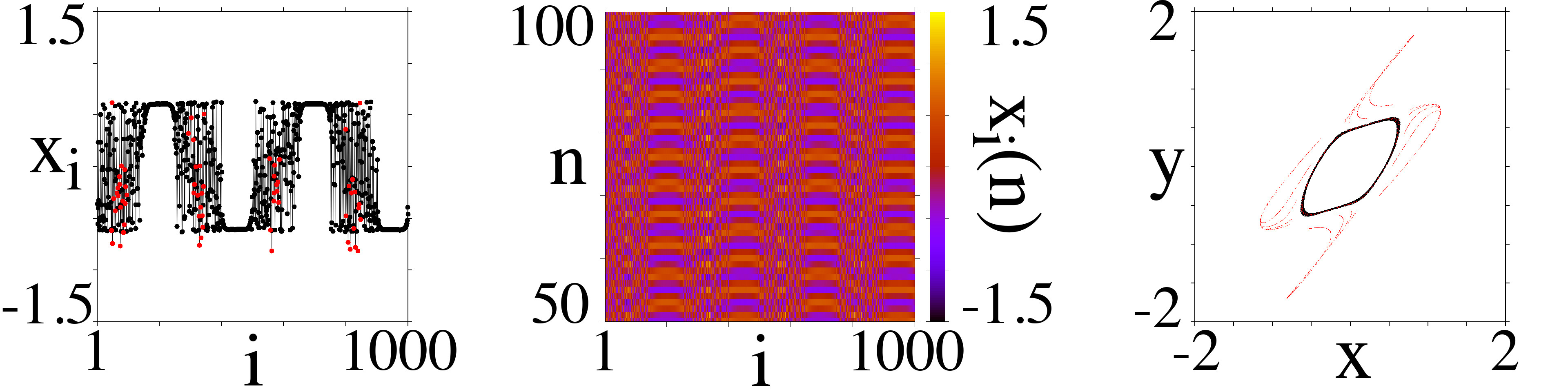}  \\
\hspace{8pt} (c)\\
\end{tabular}
\caption{Snapshots of variable $x_i$ (left column), space-time plots (middle column), and projections of multidimensional attractors of the system in the $(x,y)$ phase plane (right column) for (a),(b) two incoherent regimes, and (c) the coexistence of the multichimera state (black dots) and nodes switched between typical state and solitary state (red dots). Other parameters: $\epsilon=1.17931$, $\lambda=0.12$, $\sigma=0.03$, $R=320$, $T_{\rm obs}=100,000$.}
	\label{example-of-regimes-for-basins-IC-from-point-sigma=0.03}
\end{figure}

Figure~\ref{basins-IC-from-point} shows sections of basins of attraction and projections of multidimensional attractors of the system for three modes which are established in the network under study for a different choice of initial conditions from a small neighborhood (the size of $10^{-8}$) of the chosen phase point into the divided grid (see the description above) and for three different values of the coupling strength $\sigma$. The dynamical regimes which coexist in the network at $\sigma=0.02$ are exemplified in 
Fig.~\ref{example-of-regimes-for-basins-IC-from-point-sigma=0.02} and include the incoherent mode (Fig.~\ref{example-of-regimes-for-basins-IC-from-point-sigma=0.02}(a)), the traveling wave mode (Fig.~\ref{example-of-regimes-for-basins-IC-from-point-sigma=0.02}(b)), and the multichimera state (Fig.~\ref{example-of-regimes-for-basins-IC-from-point-sigma=0.02}(c)). 
As can bee seen in Fig.~\ref{basins-IC-from-point}(a), the traveling wave regime has the largest basin of attraction (green dots), and thus, plays a dominant role in defining the network dynamics when randomly distributed initial conditions are chosen. 
However, as the coupling strength slightly increases, the basins of attractions of all the coexisting modes can visibly change. This evolution can be seen in Fig.~\ref{basins-IC-from-point}(b),(c) for two slightly increasing values of $\sigma$. As follows from the figures, now the dominant role belongs to the multichimera state (red dots). 
At $\sigma=0.03$, the network dynamics can become more complicated, as depicted in Fig.~\ref{example-of-regimes-for-basins-IC-from-point-sigma=0.03}. Three more modes are added to the regimes presented above.  
They include two incoherent regimes (Fig.~\ref{example-of-regimes-for-basins-IC-from-point-sigma=0.03}(a),(b)) and the multichimera state with several switched nodes which jump between the attractor for the typical states and the one for the solitary nodes (Fig.~\ref{example-of-regimes-for-basins-IC-from-point-sigma=0.03}(c)). 
In our numerical simulation, these regimes are referred to as the multichimera state. 
Note that the shape of the basins of attraction qualitatively coincides with the basins of attraction of the single discrete-time oscillator for the chosen parameters (see Fig.~\ref{regimes_single_map}(f)).

\begin{figure}[ht]
	\centering
\begin{tabular}{ccc}
\includegraphics[width=.33\columnwidth]{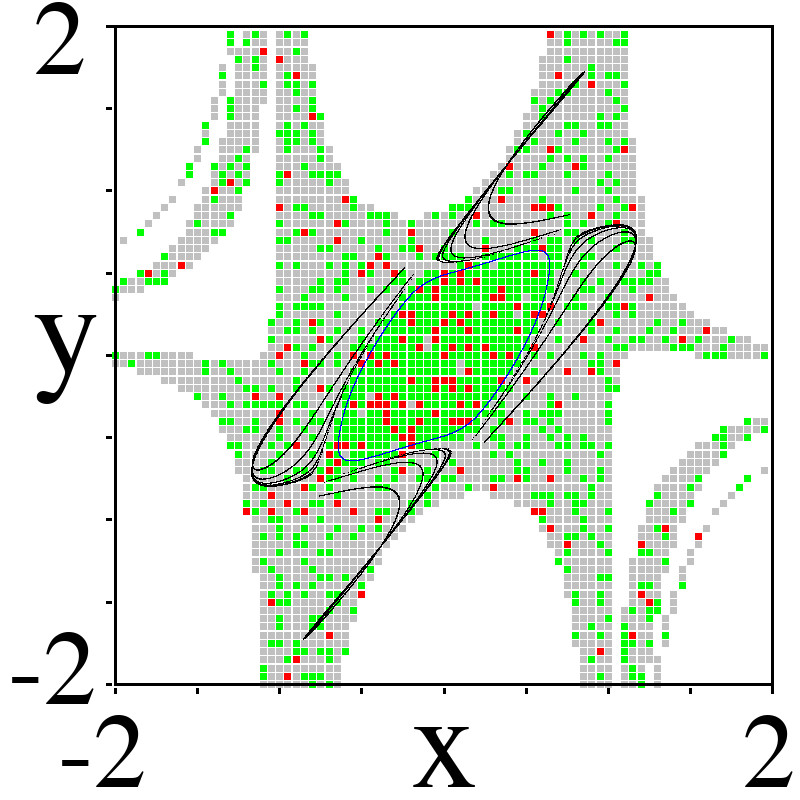} &
\includegraphics[width=.33\columnwidth]{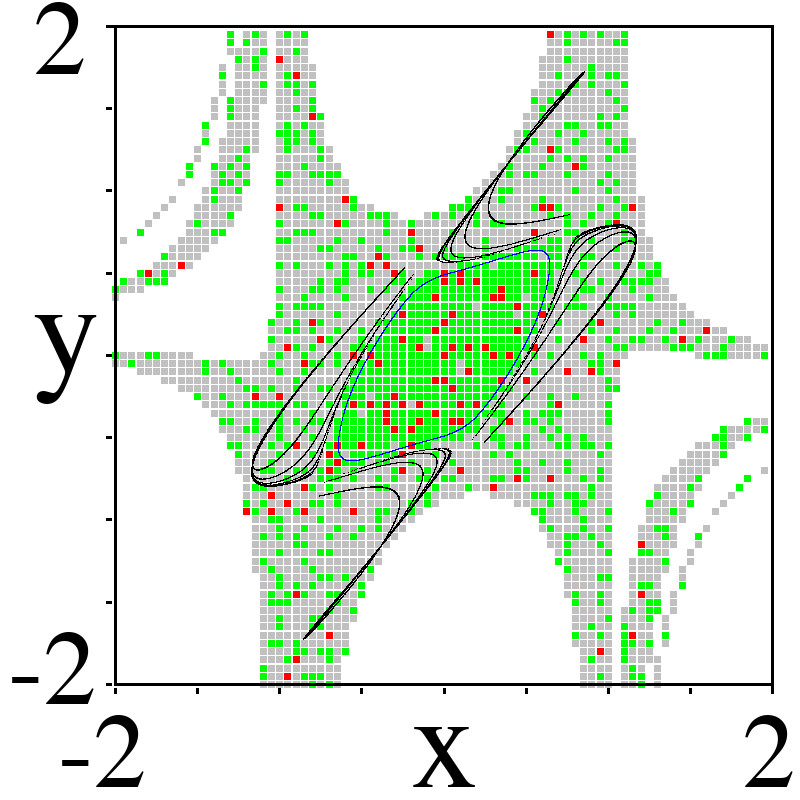}  &
\includegraphics[width=.33\columnwidth]{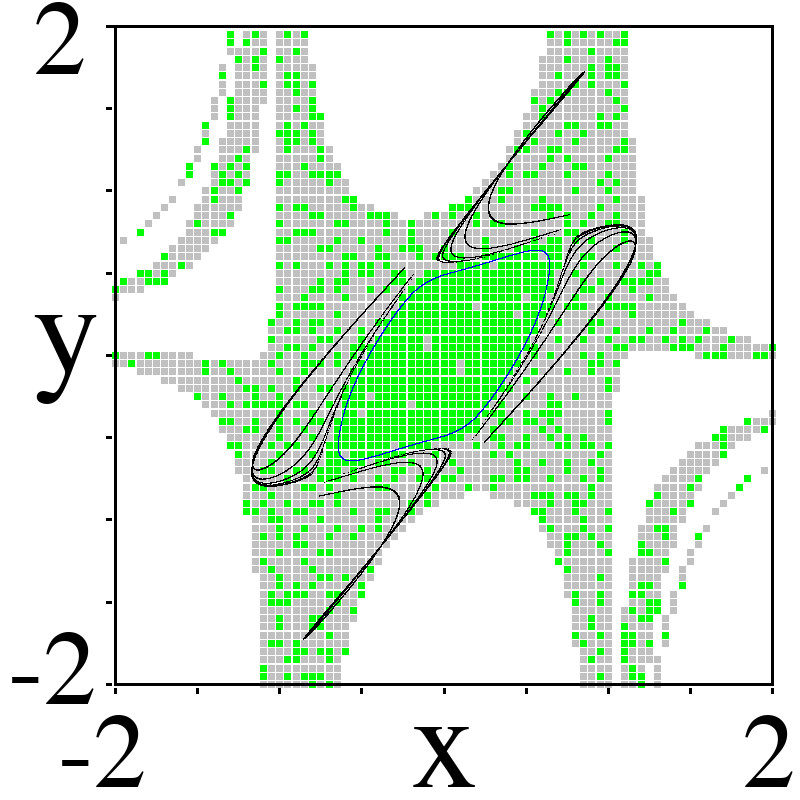}\\
\hspace{8pt} (a) & \hspace{8pt} (b) & \hspace{8pt} (c)\\
\end{tabular}
\begin{tabular}{ccc}
\includegraphics[width=.33\columnwidth]{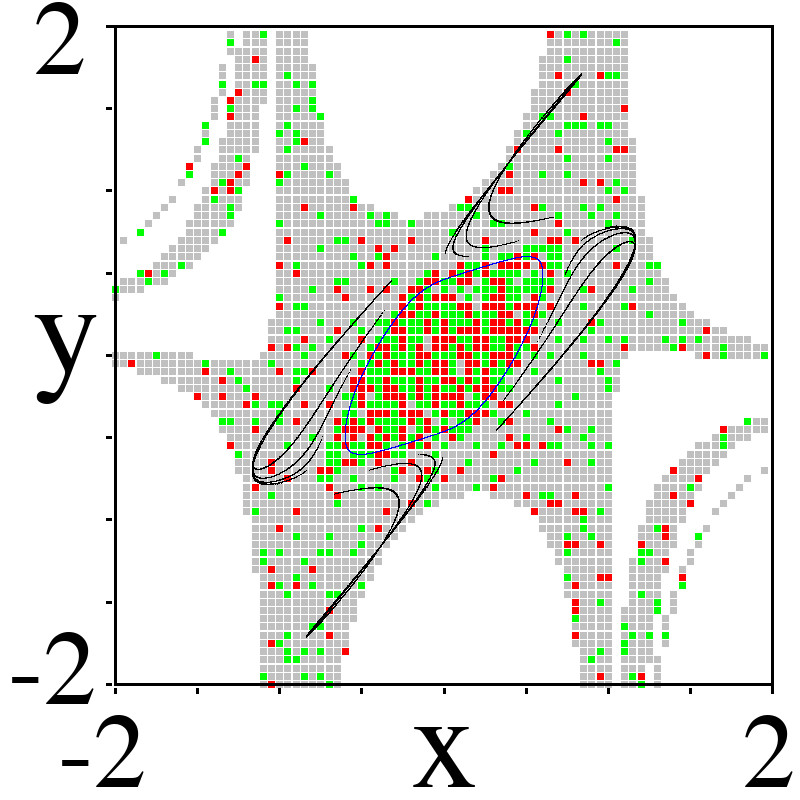} &
\includegraphics[width=.33\columnwidth]{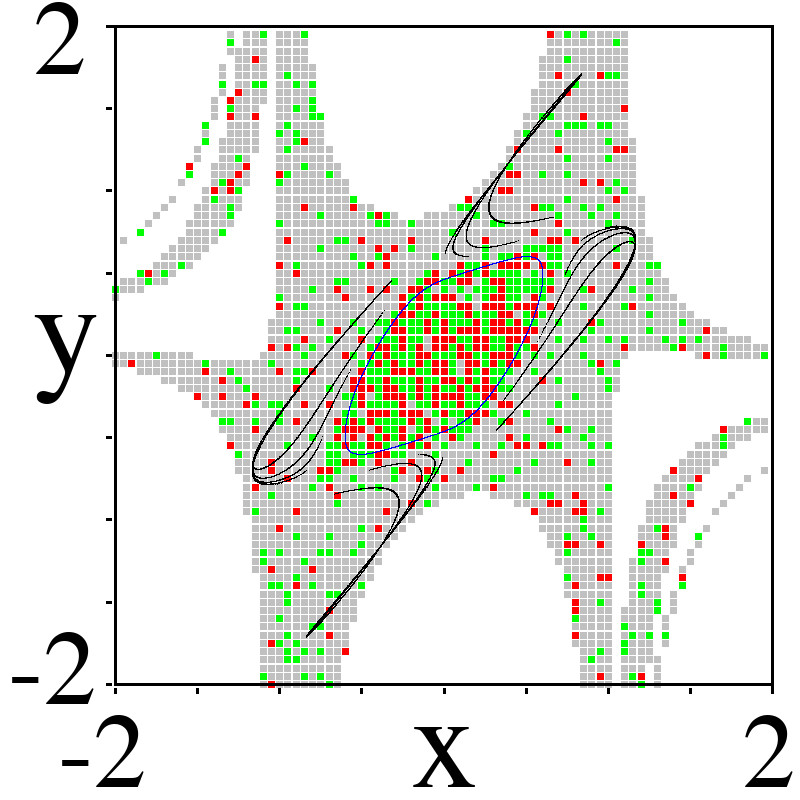}  &
\includegraphics[width=.33\columnwidth]{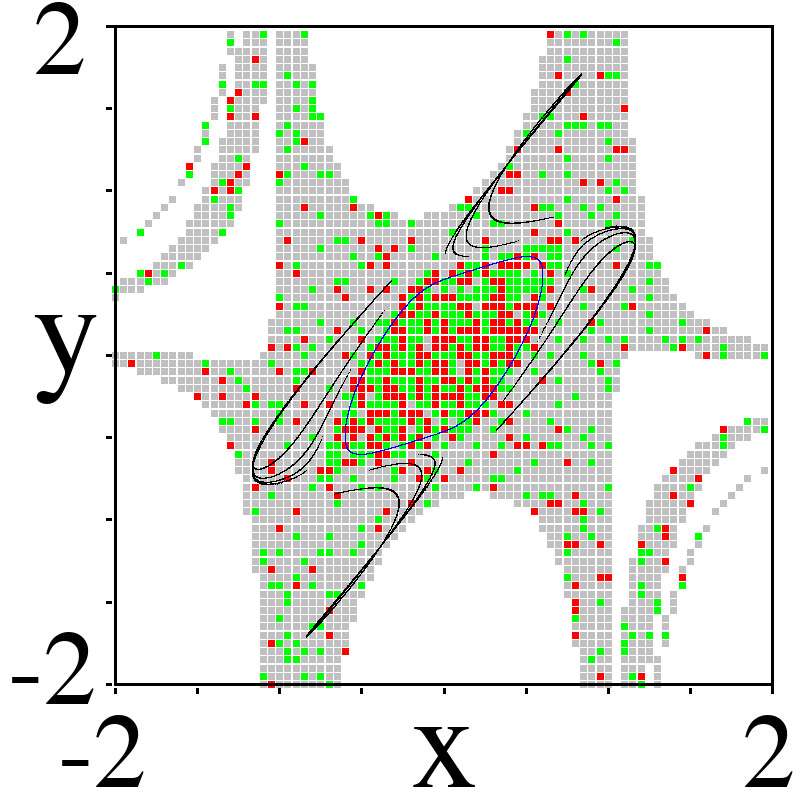}\\
\hspace{8pt} (d) & \hspace{8pt} (e) & \hspace{8pt} (f)\\
\end{tabular}
\caption{Section of basins of attraction for two values of the coupling strength: (a)-(c) $\sigma=0.02$  and (d)-(f) $\sigma=0.03$, and for three different values of the observation time: (a),(d) $T_{\rm obs}=100,000$, (b),(e) $T_{\rm obs}=500,000$, and (c),(f) $T_{\rm obs}=1,000,000$. Grey dots show the basin of attraction for the attractor in the incoherent regime (black dots, Fig.~\ref{example-of-regimes-for-basins-IC-from-point-sigma=0.02}(a)), green dots correspond to the basin of attraction of the attractor in the traveling wave mode (blue dots, Fig.~\ref{example-of-regimes-for-basins-IC-from-point-sigma=0.02}(b)), and red dots denote the basin of attraction of the attractor for the multichimera state (Fig.~\ref{example-of-regimes-for-basins-IC-from-point-sigma=0.02}(c)). Trajectories from the white region diverge to infinity. Other parameters:  $\epsilon=1.17931$, $\lambda=0.12$, $R=320$, steps of division of the $(x,y)$ phase plane are $\Delta x=0.05$ and $\Delta y=0.05$}
	\label{basins-at-diff-time}
\end{figure}

As mentioned above, the multichimera state is a transient mode, and the network dynamics eventually converts to the traveling wave mode. However, the transition time from multichimera states to traveling waves depends not only on the initial conditions but also on the parameters of the network. We analyze this dependence with respect to the coupling strength $\sigma$ by increasing the observation time. The numerical results are presented in Fig.~\ref{basins-at-diff-time}.
It is seen that for $\sigma=0.02$ and with increasing $T_{\rm obs}$, the basin of attraction of the attractor for  the multichimera state (red dots) becomes less: some red points turn green ones which correspond to the basin of attraction for the traveling wave mode (Fig.~\ref{basins-at-diff-time}(a),b)). At 
$T_{\rm obs}=1,000,000$, there are no red points at all (Fig.\ref{basins-at-diff-time}(c)) which means that  the observed multichimera states switch to the traveling wave mode.
On the other hand, at $\sigma=0.03$, the multichimera state is more stable and only several red dots turn green (Compare Fig.~\ref{basins-at-diff-time}(d)-(f)). Moreover, at this coupling strength, some grey dots become green or red, that means that at certain initial conditions, the incoherent regimes switch to either the multichimera state or the traveling wave mode.
Note that these sections of basins of attraction (Fig.~\ref{basins-at-diff-time}) were constructed with a larger step size on the variables ($x,y$) than those shown in Fig.~\ref{basins-IC-from-point} since we only need to qualitatively demonstrate the transition 
to the traveling wave regime and calculations for larger observation times require more computational power or time. 

\subsection{\label{sec:chimera-noise}Noise influence on multichimera states}

During numerical computations, the following question may arise: Is it possible to increase the observation time of multichimera states and to delay the transition to traveling waves? In this regard, we study the effect of external noise on the network \eqref{system} for the parameters and the initial conditions corresponding to the existence of multichimera state. In the presence of additive noise, the network equations are written as follows:

     \begin{eqnarray}\label{system_noise}
x_{i}(n+1)&=&f(x_{i}(n),y_{i}(n))+\\\nonumber 
&+&\frac{\sigma}{2R}\sum\limits_{j=i-R}^{i+R}[f(x_{j}(n),y_{j}(n))-f(x_{i}(n),y_{i}(n))]+D\psi_{i}(n)\\\nonumber 
y_{i}(n+1)&=&g(x_{i}(n),y_{i}(n)),~~~~~~x_{i \pm N}(n) \equiv x_{i}(n),\\\nonumber
     \end{eqnarray}

\noindent  where  $\psi$ is a  discrete noise source whose values are uniformly distributed in the interval $[-0.5,0.5]$, and $D$ is the noise intensity. The noise source influences the network elements  over a certain period of time, which is denoted as $n_{\rm exp}$, after that it is switched off. The moment of turning on the noise generator is selected in two ways: before the dissapearance of multichimera state and after that.

\begin{figure}[ht]
	\centering
\begin{tabular}{cc}
\includegraphics[width=.48\columnwidth]{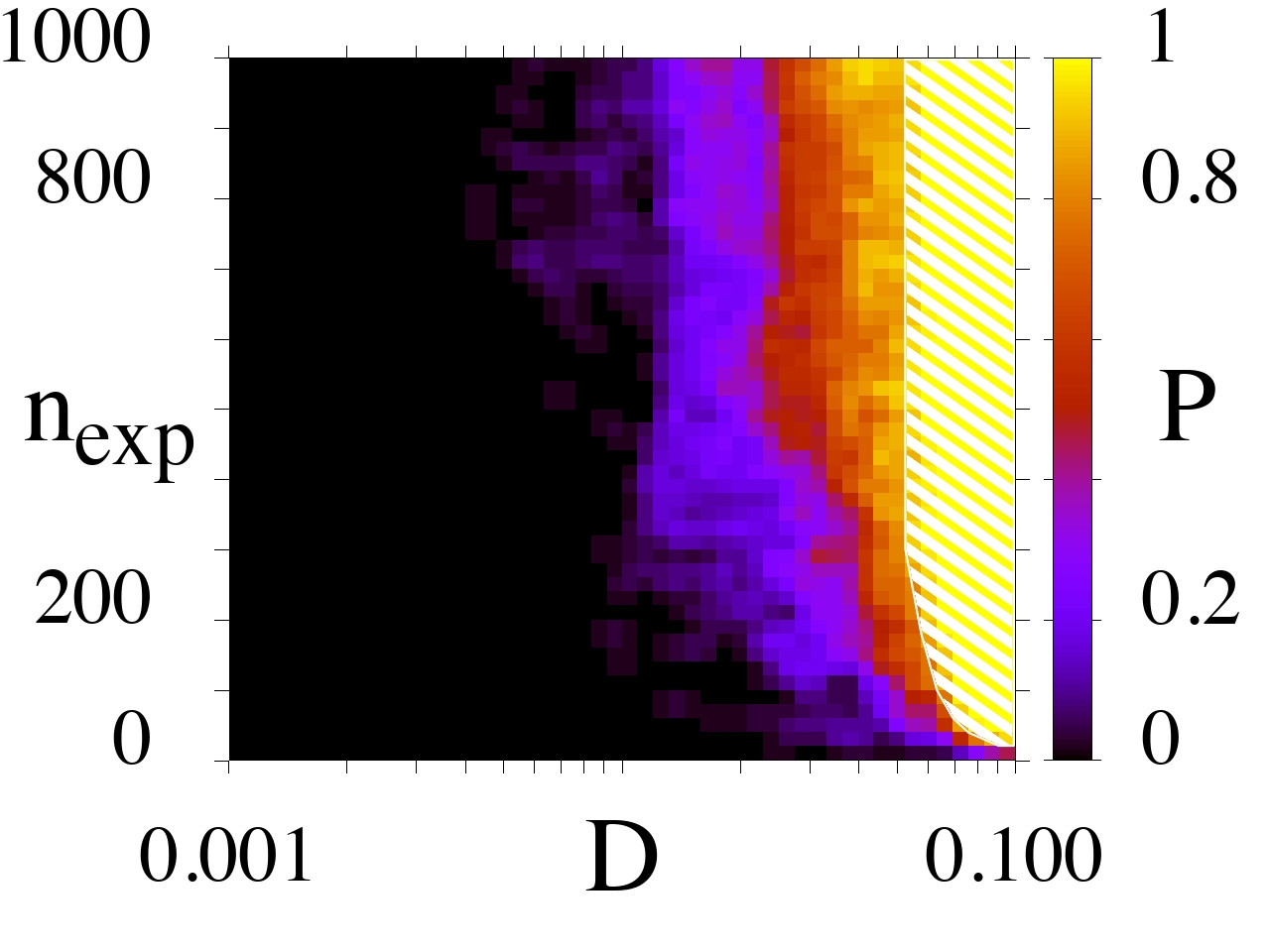} &
\includegraphics[width=.48\columnwidth]{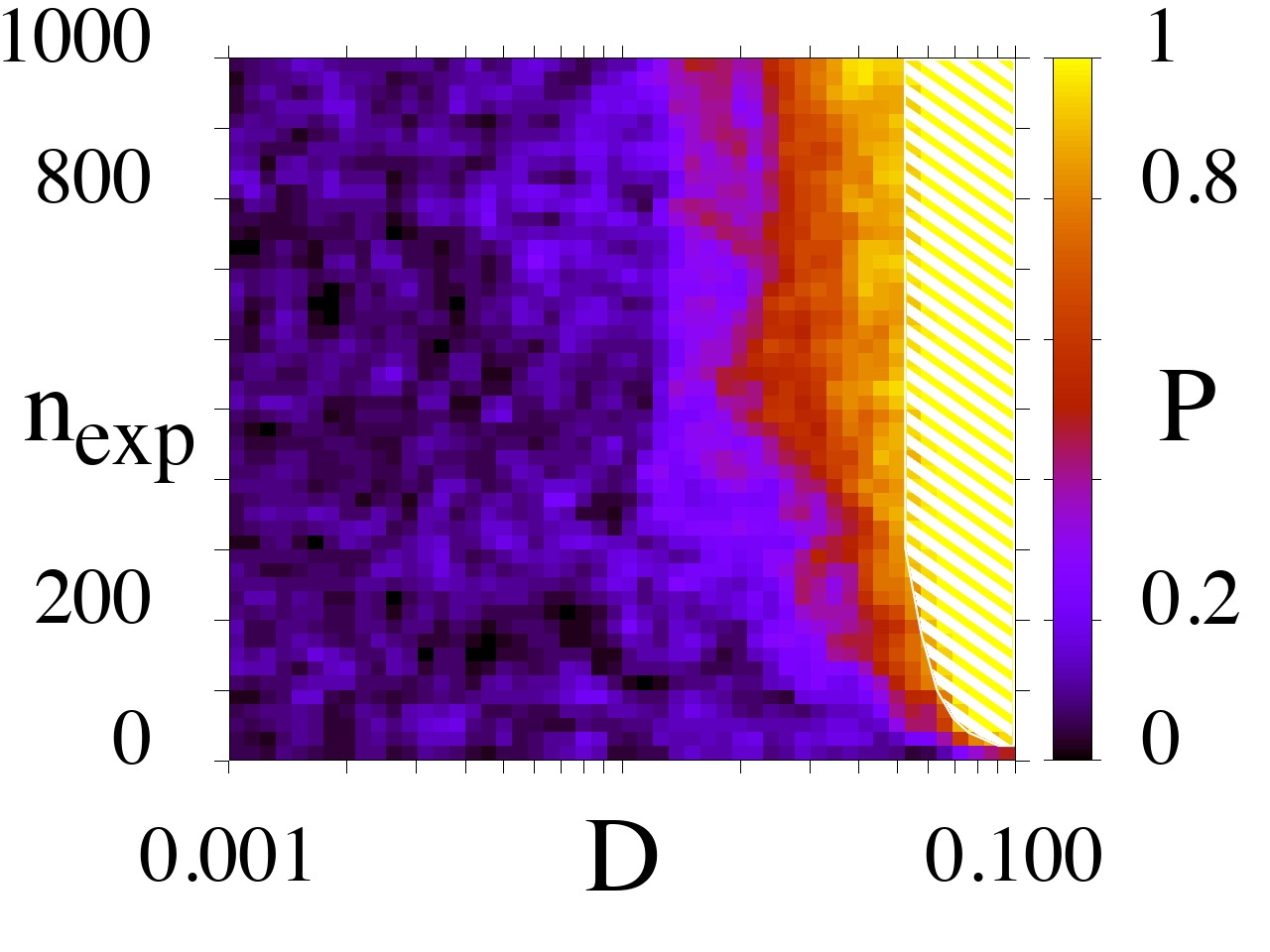}  \\
\hspace{8pt} (a) & \hspace{8pt} (b)\\
\end{tabular}
\begin{tabular}{cc}
\includegraphics[width=.48\columnwidth]{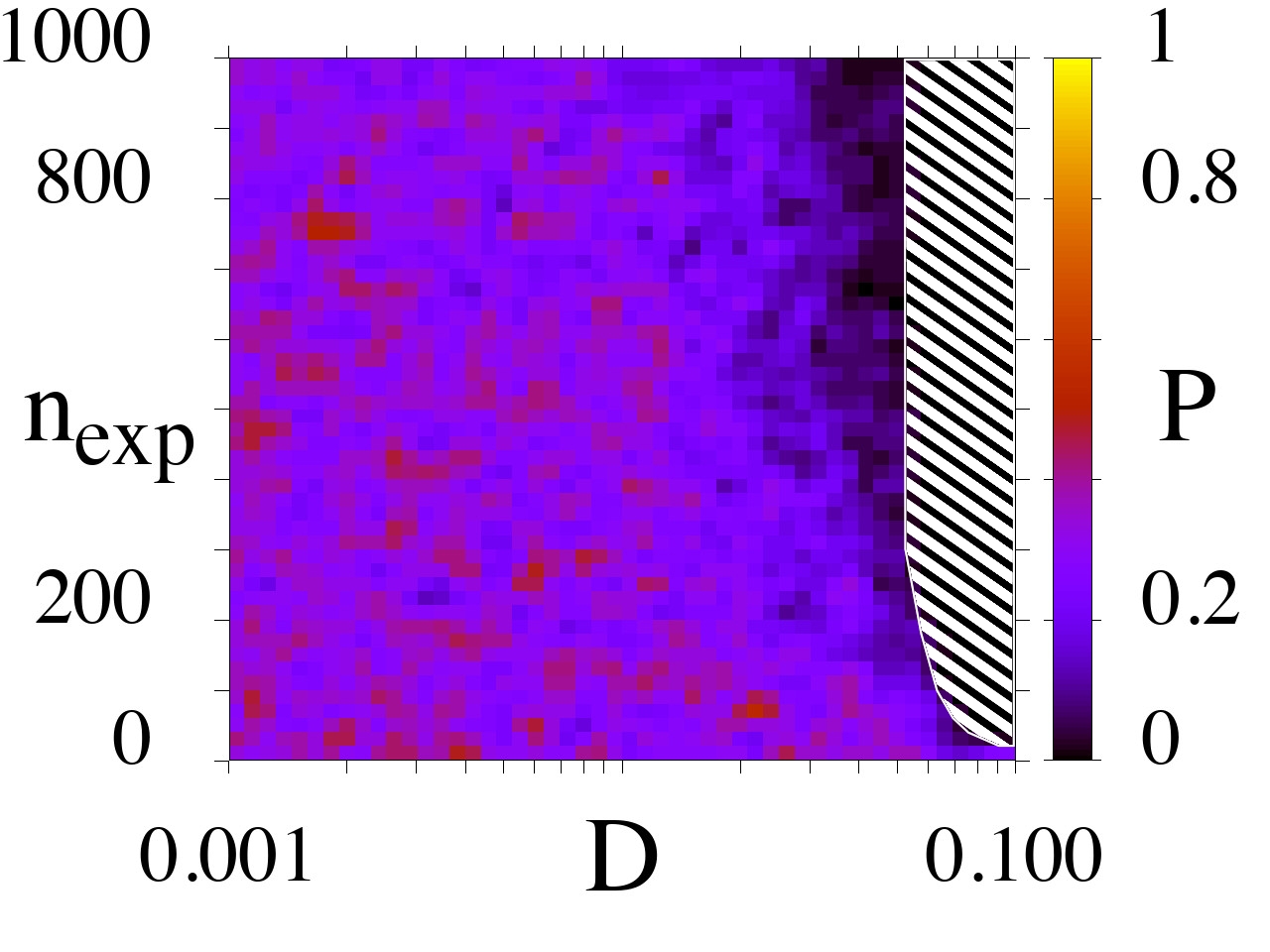} &
\includegraphics[width=.48\columnwidth]{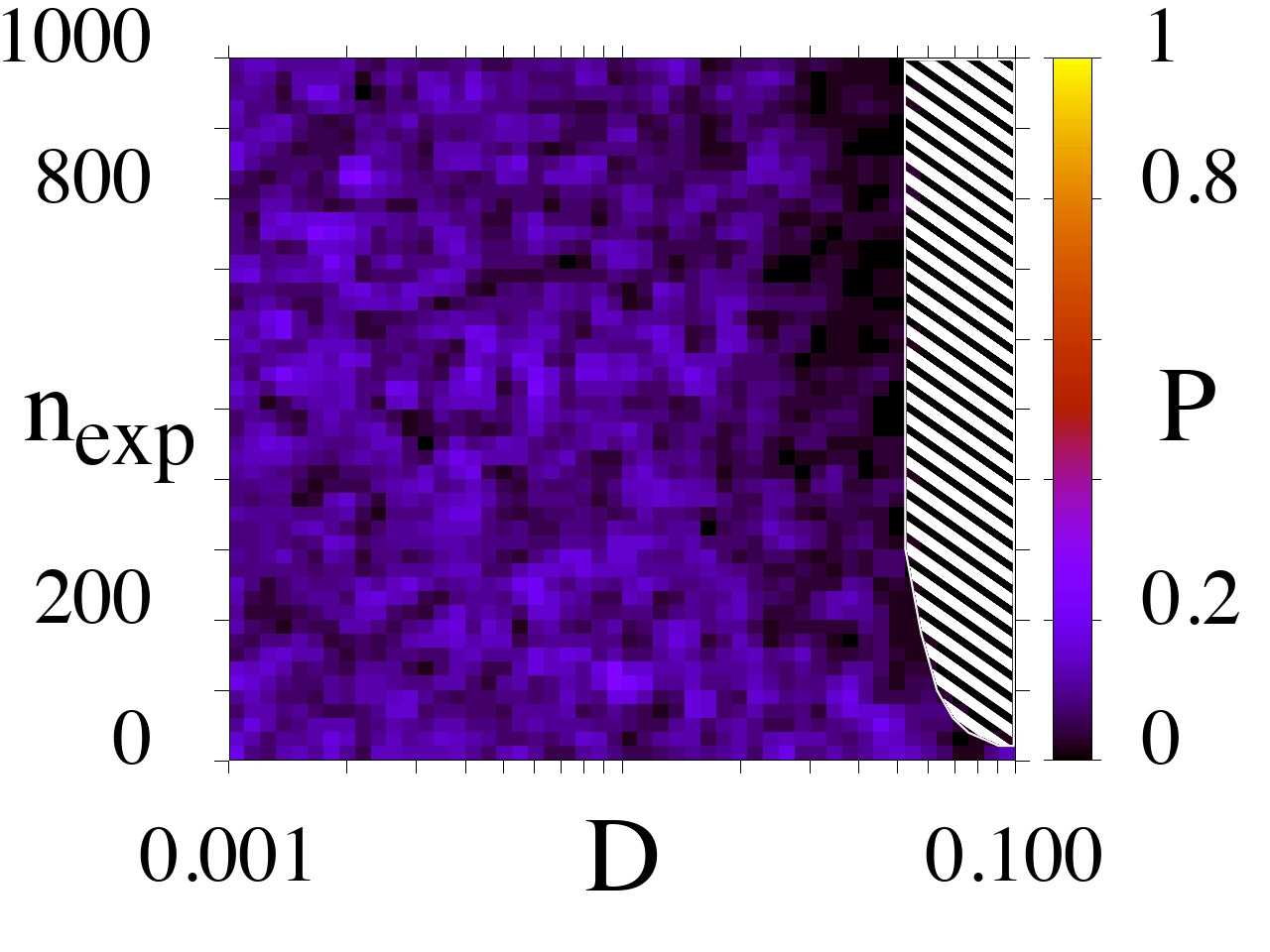}  \\
\hspace{8pt} (c) & \hspace{8pt} (d)\\
\end{tabular}
\caption{Impact of the additive noise with intensity $D$ and exposure time $n_{\rm exp}$ on the network \eqref{system_noise} before the transition from multichimera states to traveling waves. Probability distributions $P$ of observing the multichimera state with short and long lifetime: (a) $T_{\rm life}<10,000$, (b) $T_{\rm life}<100,000$, (c) $T_{\rm life}>1,000,000$, (d) $ T_{\rm life}>2,000,000$. To construct the distributions, fifteen different noise realizations were used. Trajectories from a white-shaded regions diverge to infinity.   Other parameters: $\epsilon=1.17931$, $\lambda=0.12$, $\sigma=0.02$, $R=320$}
	\label{noise-before-death}
\end{figure}

The multichimera state illustrated in Fig.~\ref{chimera-to-traveling-wave} has been chosen as the initial state. The noise source is introduced into all elements of the network at the moment $n=100,000$. The total lifetime (starting with $n=0$) of this chimera without external influence is $T_{\rm life}\approx 920,000$. Thus, we can explore the possibility of suppression of multichimera structure and a faster transition to traveling waves, as well as obtaining a more stable chimera with a longer lifetime.

Figure~\ref{noise-before-death} shows probability distributions of short and long lifetimes of the multichimera state (or the transient time to the traveling wave mode) depending on the noise intensity $D$ and the noise exposure duration $n_{\rm exp}$. The distributions are constructed using 15 different noise realizations. As can be seen from the diagrams, some 
noise realizations of large intensity ($0.08<D<0.1$) can essentially reduce the multichimera lifetime (for example, $T_{\rm life}<10,000$, Fig.~\ref{noise-before-death}(a)) and thus, speed up the transition to the traveling wave.  This can happen for any duration of the noise influence $n_{\rm exp}$. Note that when the noise intensity $D$ is varied within the range $0.08<D<0.1$, trajectories may diverge to infinity at some sets of the noise source (a white-shaded region in Fig.~\ref{noise-before-death}). 
Our calculations indicate that  there is a small but non-zero probability to increase the multichimera lifetime even more than its ``nominal'' value ($T_{\rm life}\approx 820,000$) by applying additive noise. This is demonstrated by the probability distributions shown in  Fig.~\ref{noise-before-death}(c),(d). Despite the noise duration and its intensity, long-living multichimera states can be observed and thus, more calculation (observation) time is needed to switch to the traveling wave mode.   

\begin{figure}[ht]
	\centering
\begin{tabular}{cc}
\includegraphics[width=.48\columnwidth]{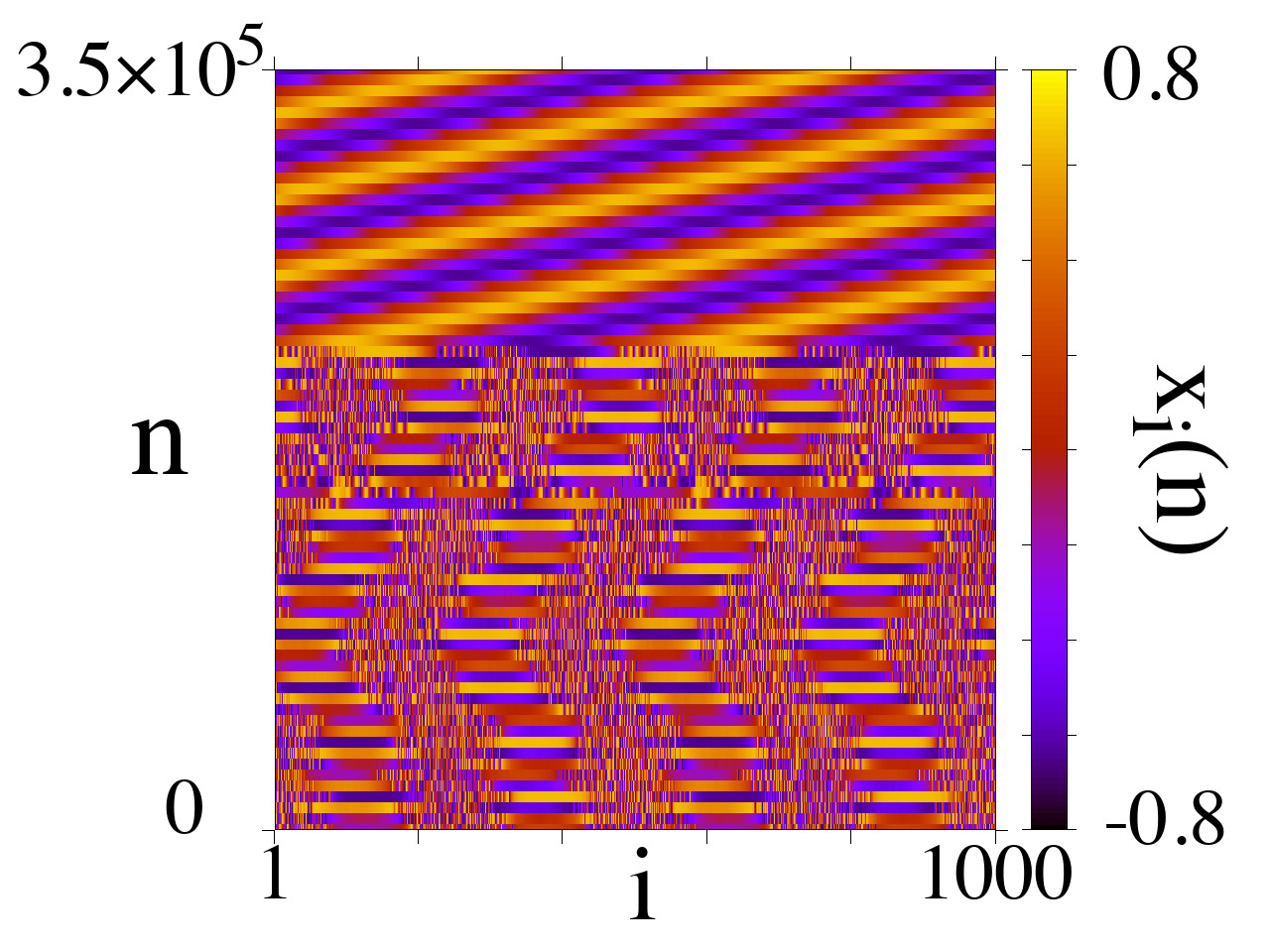} &
\includegraphics[width=.48\columnwidth]{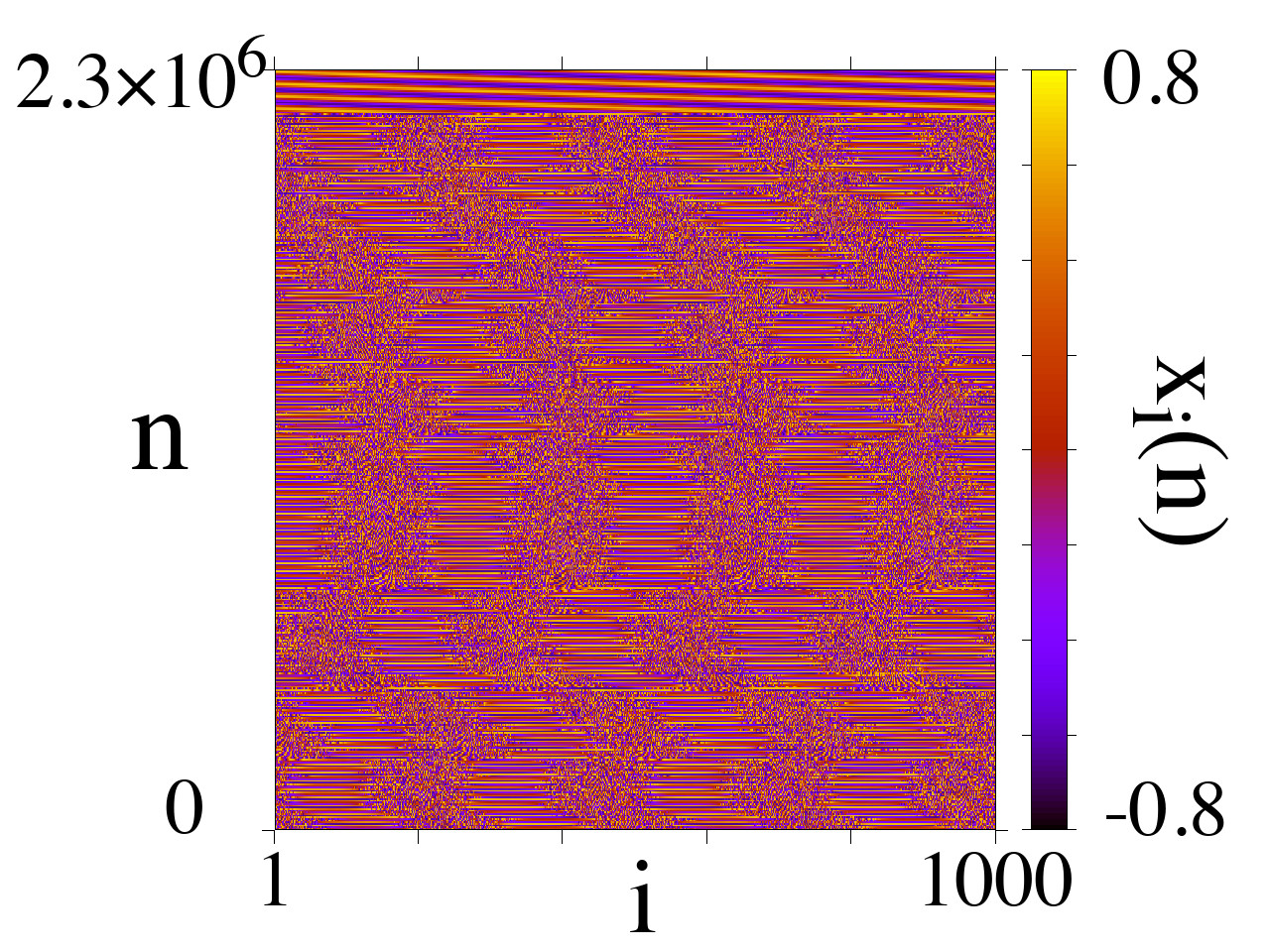}  \\
\hspace{8pt} (a) & \hspace{8pt} (b)\\
\end{tabular}
\begin{tabular}{cc}
\includegraphics[width=.48\columnwidth]{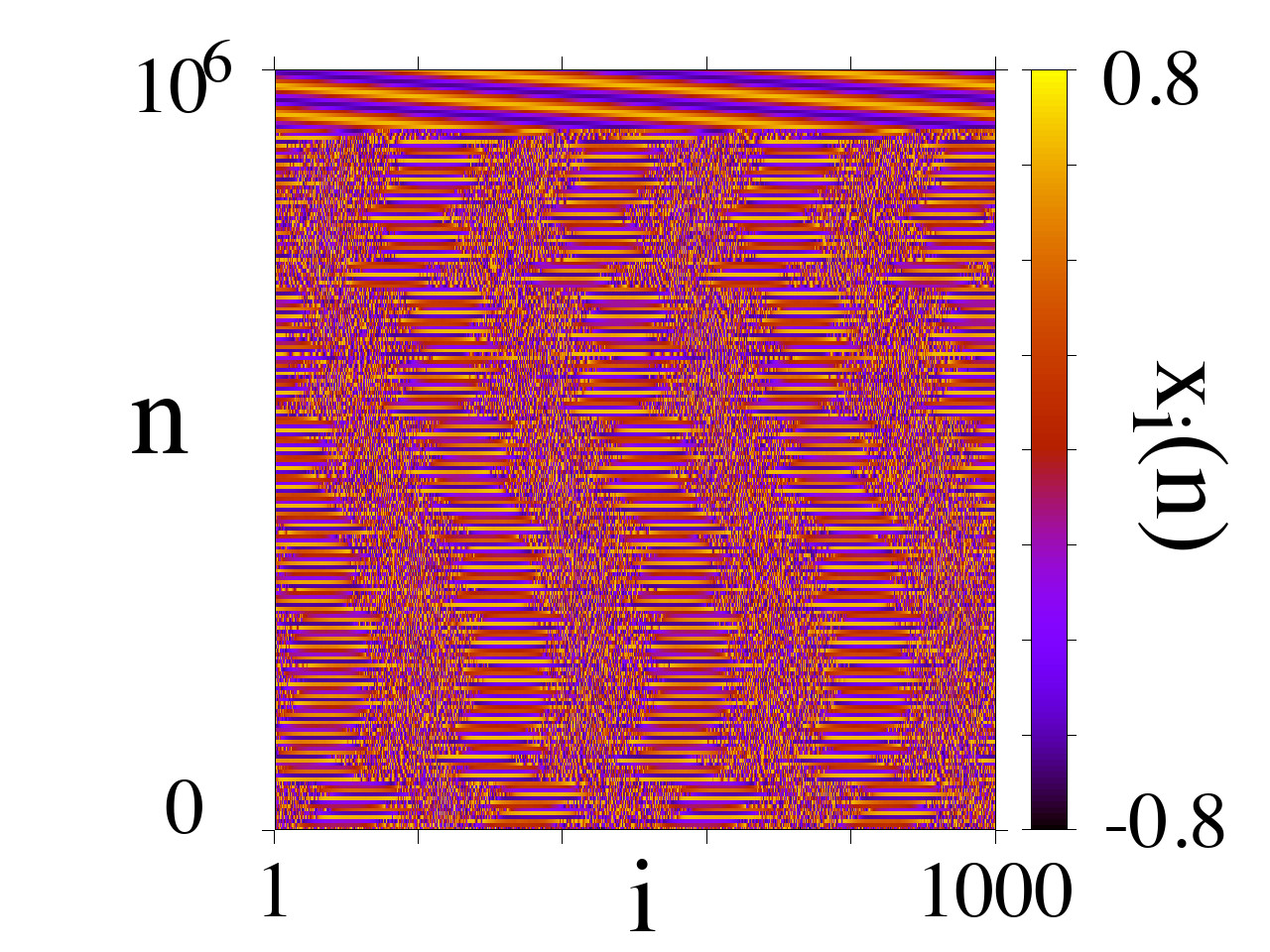} &
\includegraphics[width=.48\columnwidth]{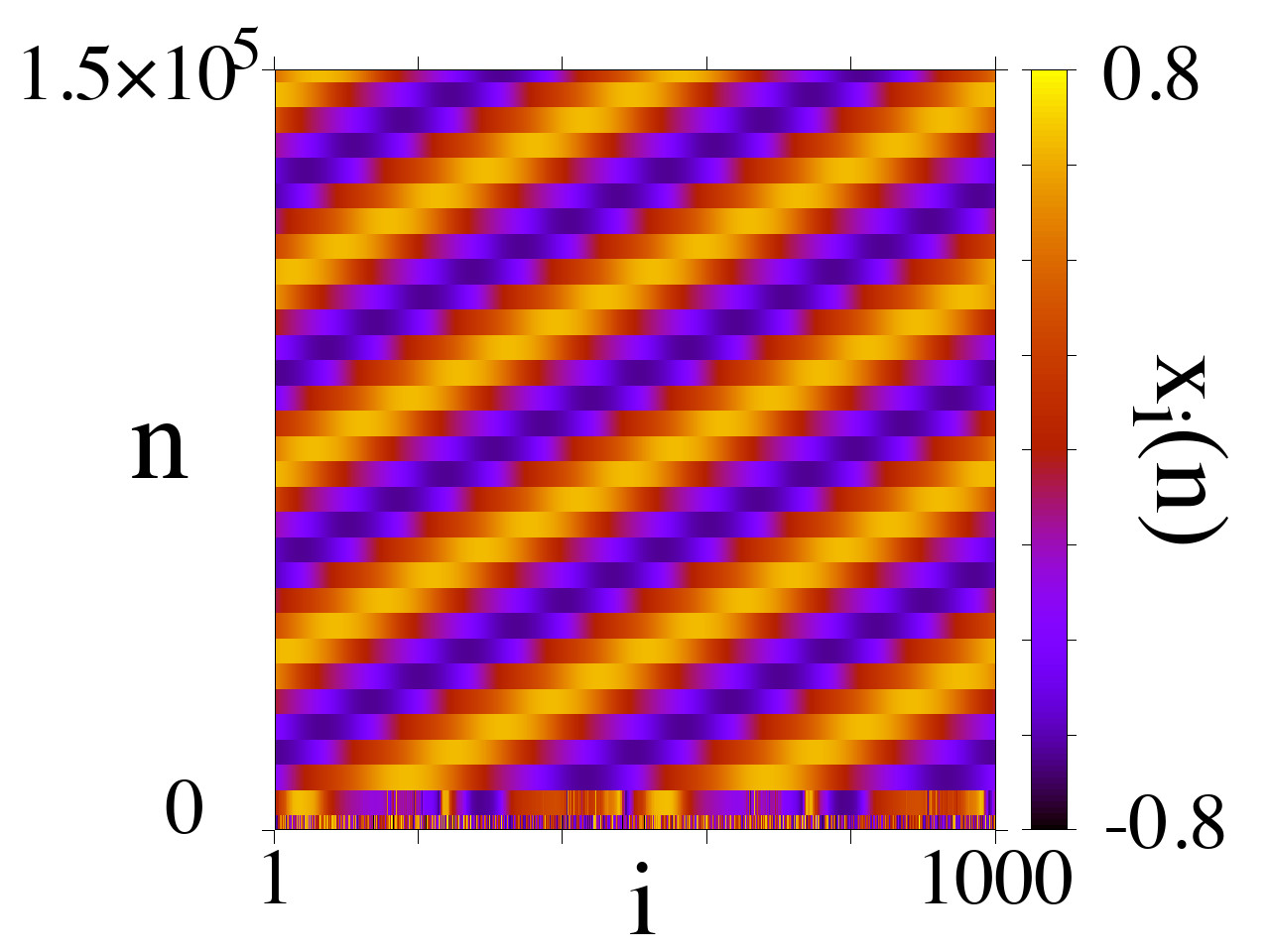}  \\
\hspace{8pt} (c) & \hspace{8pt} (d)\\
\end{tabular}
\caption{Space-time plots illustrating the transition from the multichimera to the traveling wave in the presence of additive noise: (a) $D=10^{-5}$, $n_{\rm exp}=51$, (b) $D=10^{-5}$, $n_{\rm exp}=101$, (c) $D=0.04$, $n_{\rm exp}=301$ -- the first realization of  noise, and (d) $D=0.04$, $n_{\rm exp}=301$ -- the second realization of noise. Every 5000th iteration is displayed in the space-time plots.  Other parameters: $\epsilon=1.17931$, $\lambda=0.12$, $\sigma=0.02$, $R=320$. The time $n=0$ corresponds to the inclusion of additive noise.}
	\label{dynamics-noise-before-death}
\end{figure}

The transition from the multichimera state to the traveling wave in the presence of additive noise is illustrated  in Fig.~\ref{dynamics-noise-before-death} by several space-time plots. It is seen that short-term weak noise ($D=10^{-5}$, $n_{\rm exp}=51$) can suppress the multichimera state and induce a fast transition to the traveling wave (Fig.~\ref{dynamics-noise-before-death}(a)). If weak noise ($D=10^{-5}$) influences the multichimera state a bit longer ($n_{\rm exp}=101$), the chimera lifetime can be  increased (Fig.~\ref{dynamics-noise-before-death}(b)).   The space-time diagrams plotted  in Fig.~\ref{dynamics-noise-before-death}(c),(d) show how two different realizations of noise with the same parameters can effect the process of  changing  the network dynamics. In the first case (Fig.~\ref{dynamics-noise-before-death}(c)), the noise induces a more stable and long-living multichimera state, while the second realization of noise provides a faster switching to the traveling wave mode (Fig.~\ref{dynamics-noise-before-death}(d)).  

\begin{figure}[ht]
	\centering
\begin{tabular}{cc}
\includegraphics[width=.48\columnwidth]{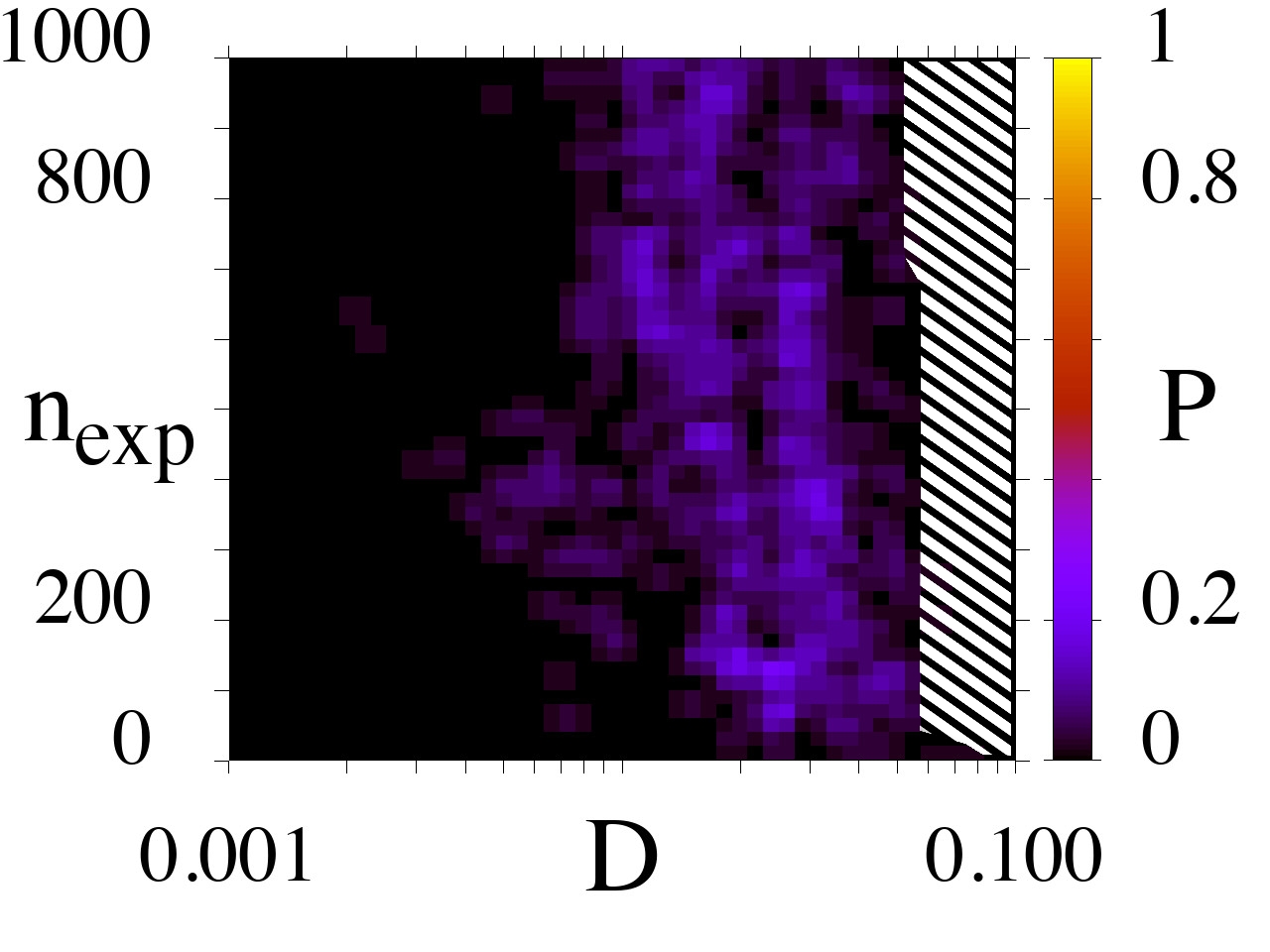} &
\includegraphics[width=.48\columnwidth]{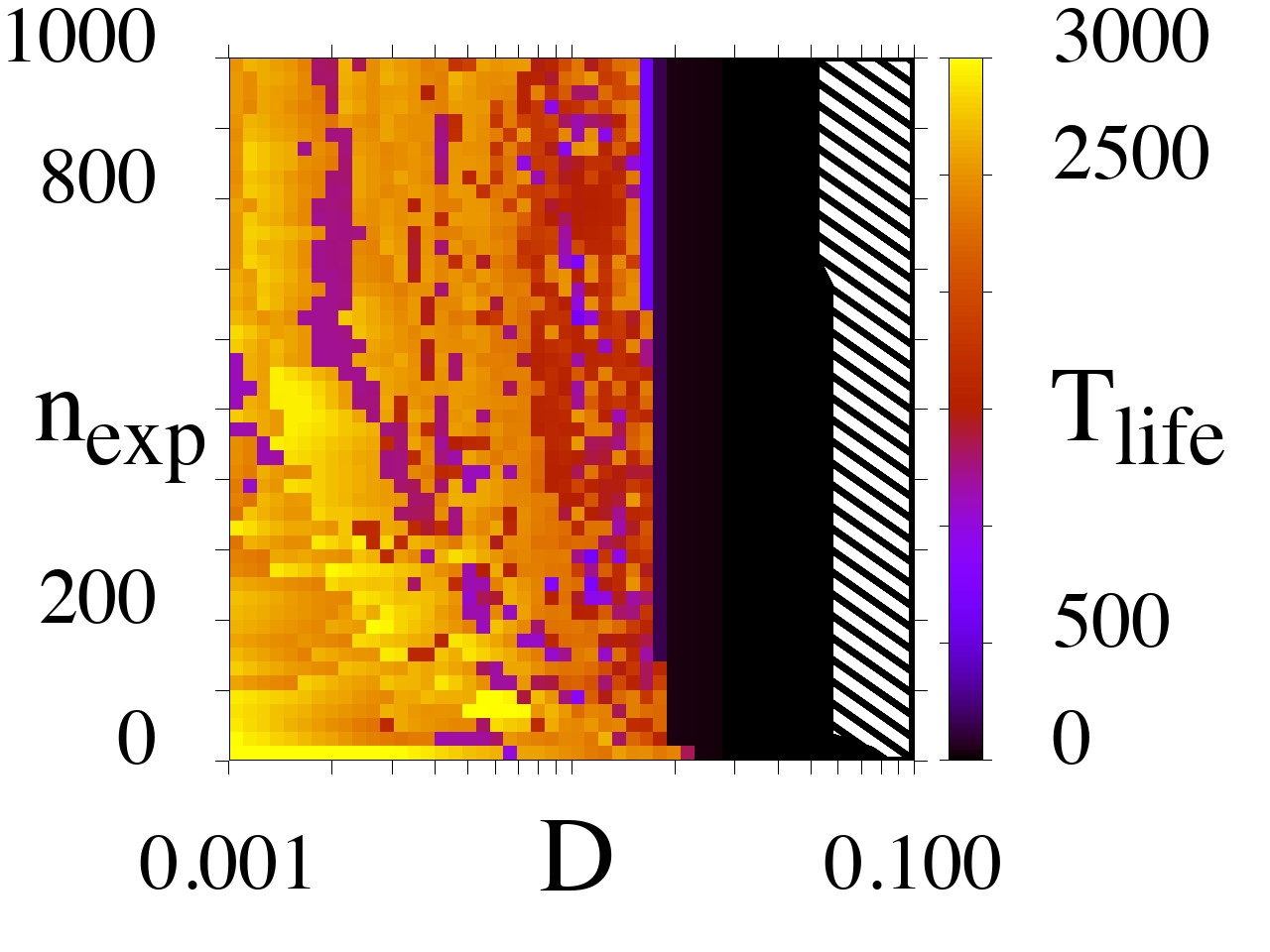}  \\
\hspace{8pt} (a) & \hspace{8pt} (b)\\
\end{tabular}
\begin{tabular}{cc}
\includegraphics[width=.48\columnwidth]{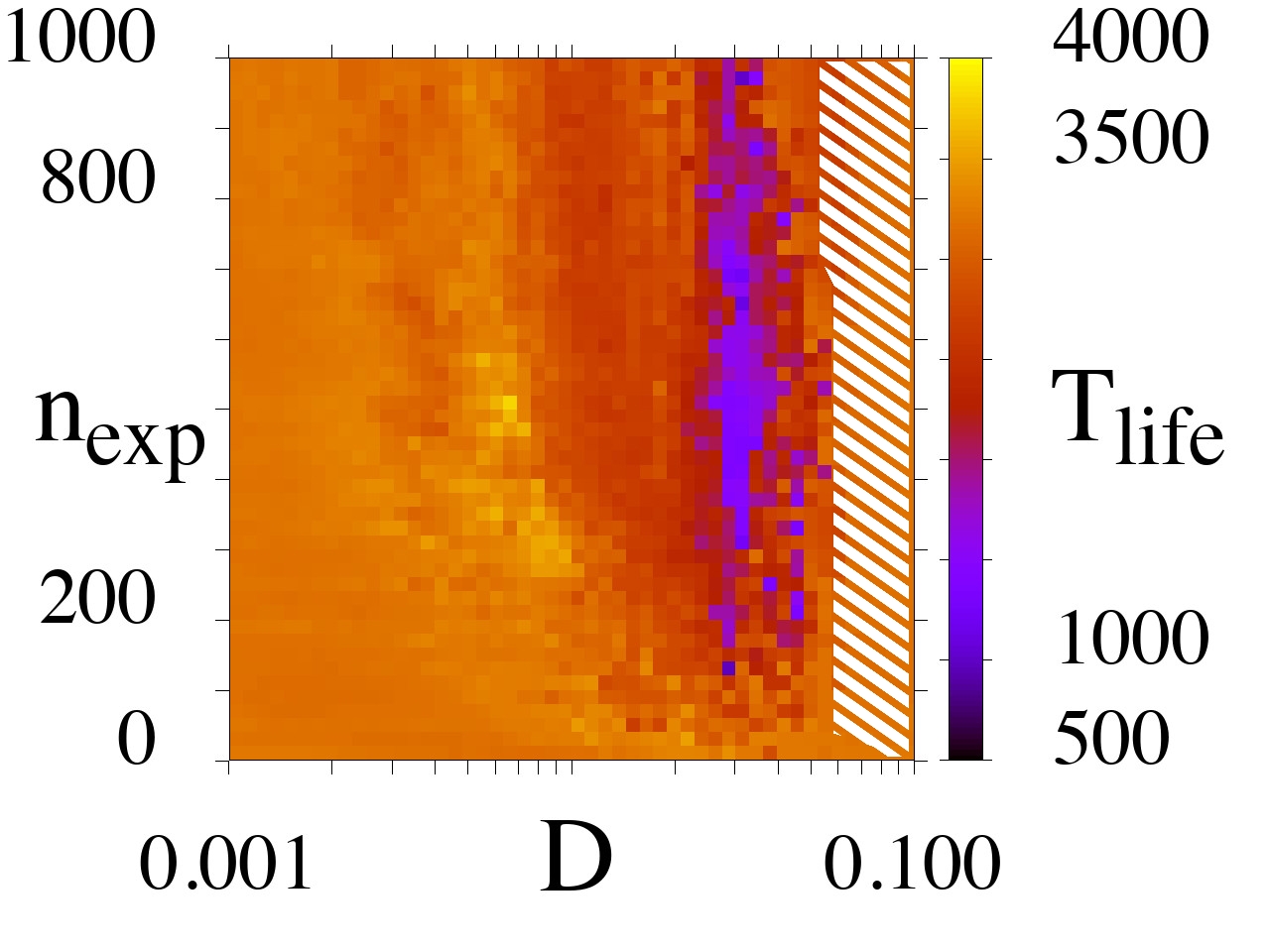} &
\includegraphics[width=.48\columnwidth]{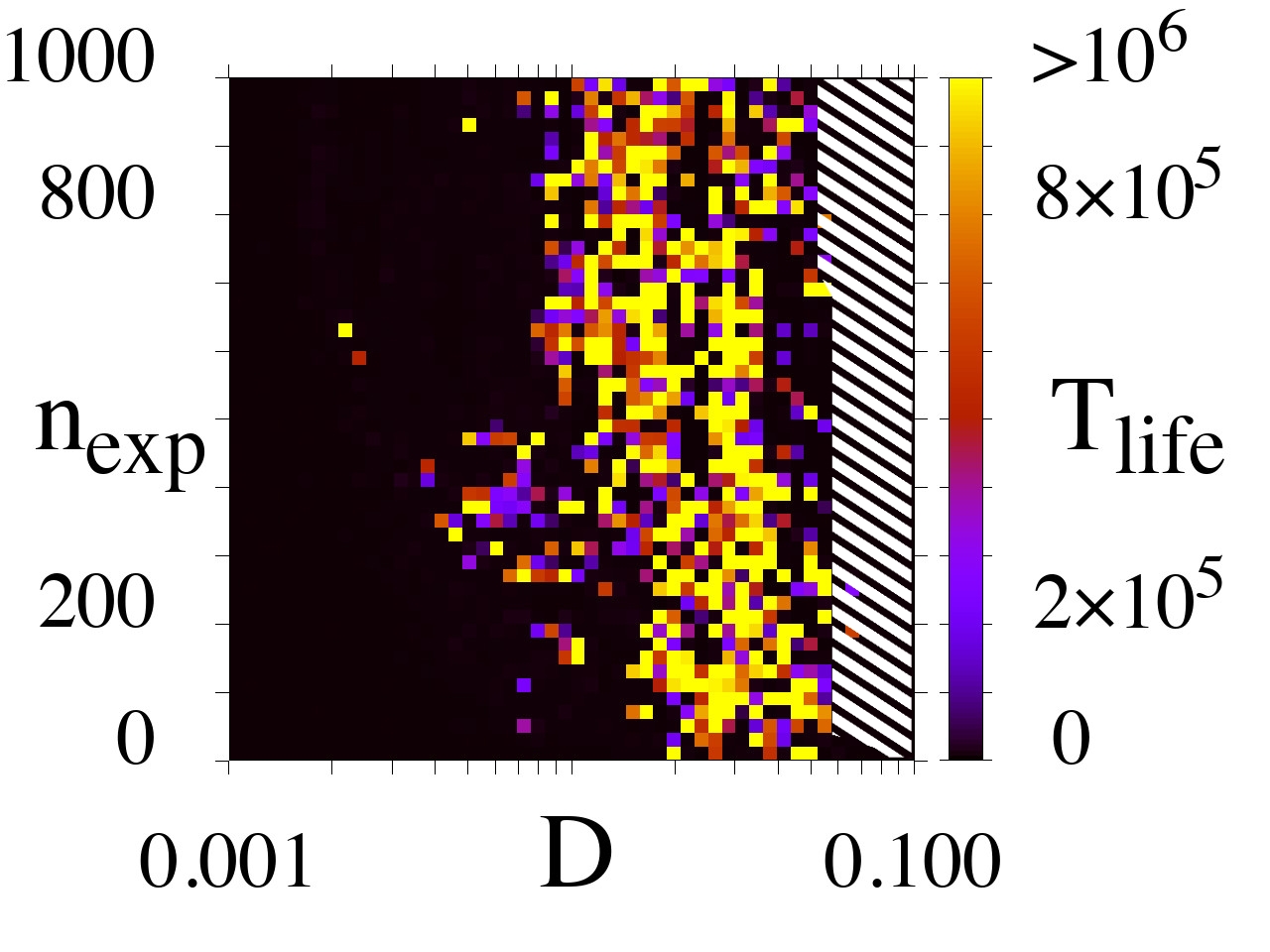}  \\
\hspace{8pt} (c) & \hspace{8pt} (d)\\
\end{tabular}
\caption{Impact of additive noise with intensity $D$ and exposure time $n_{\rm exp}$ on the network \eqref{system_noise} after transition from multichimera states to traveling waves. (a) Probability $P$ of the appearance of multichimera state with the lifetime  $T_{\rm life}>10,000$. Distributions of minimum (b), median (c), and maximum (d) lifetimes of the noise-induced multichimera state.  Trajectories from a white-shaded regions diverge to infinity. To construct the distributions, fifteen noise realizations were used. Other parameters: $\epsilon=1.17931$, $\lambda=0.12$, $\sigma=0.02$, $R=320$.}
	\label{noise-after-death}
\end{figure}

Let us consider another case of introducing noise in the network of discrete van der Pol oscillators. The noise source is switched on after the transition from multichimera state to the traveling wave mode. In other words, in this case the noise effects the traveling wave. As a result, a multichimera state can be induced, but it is characterized by a short lifetime that follows from the probability distribution shown in Fig.~\ref{noise-after-death}(a).  
To analyze in more detail the evolution of the noise-induced multichimera,  we plot distributions of minimum (Fig.~\ref{noise-after-death}(b)), median (Fig.~\ref{noise-after-death}(c)) and maximum (Fig.~\ref{noise-after-death}(d)) lifetimes $T_{\rm life}$ of the multichimera in the ($D, n_{\rm exp}$) parameter plane. 
As before, calculations are performed for 15 different realizations of the noise source. 
It is seen that for all values of the noise parameters, the minimum lifetime of the revived multichimera state does not exceed $T_{\rm life}=3,000$ (Fig.~\ref{noise-after-death}(b)), and for $D>0.02$ the minimum lifetime is zero, that is, the multichimera is not induced. 
More informative is the distribution of the median value of the lifetime (Fig.~\ref{noise-after-death}(c)). This plot shows that  for most values of the noise parameters, the lifetime is at the level of $T_{\rm life}\approx 3,000$ (orange color), and for $D\in[0.025,0.05]$ and $n_{\rm exp}>100$ it generally drops to $T_{\rm life}\approx 1,000$. 
However, there is a non-zero probability of inducing of a long-living multichimera state, which is reflected in the pattern of the maximum distribution of $T_{\rm life}$ (Fig.~\ref{noise-after-death}(d)). It can be seen that the multichimera state with $T_{\rm life}>1\times 10^6$ (yellow color) appears mainly at $D>0.01$.

\section{\label{sec:chimera_solitary}Coexistence of the traveling wave and solitary state}

\begin{figure}[ht]
	\centering
\begin{tabular}{c}
\includegraphics[width=.95\columnwidth]{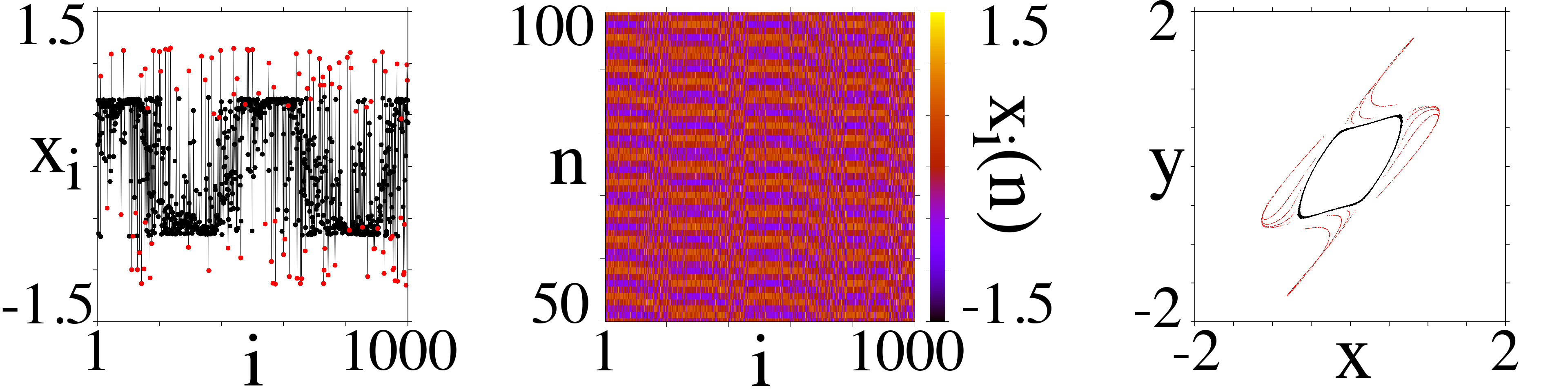} \\
\hspace{8pt} (a)\\
\includegraphics[width=.95\columnwidth]{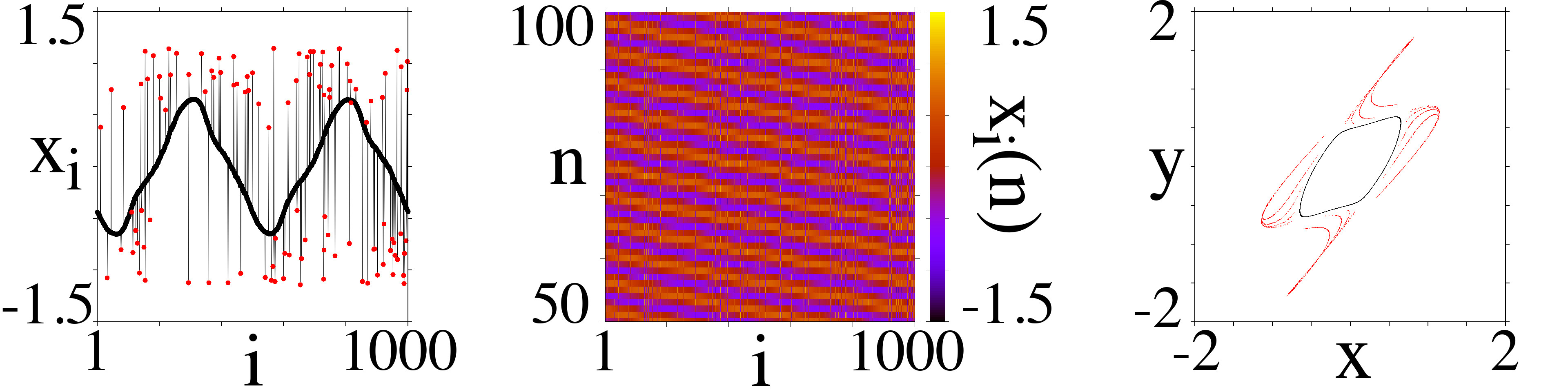}  \\
\hspace{8pt} (b)\\
\includegraphics[width=.95\columnwidth]{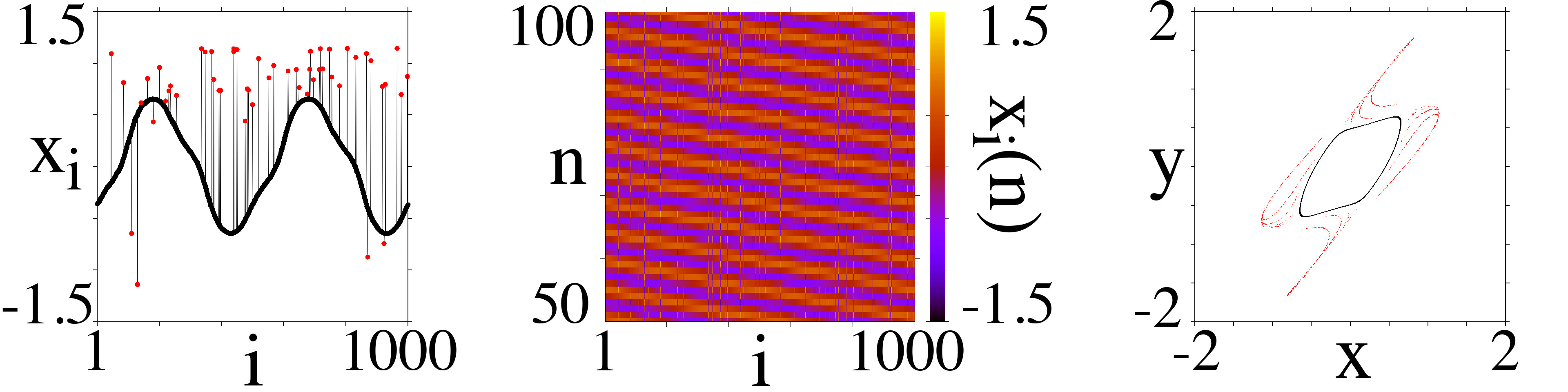} \\
\hspace{8pt} (c)\\
\includegraphics[width=.95\columnwidth]{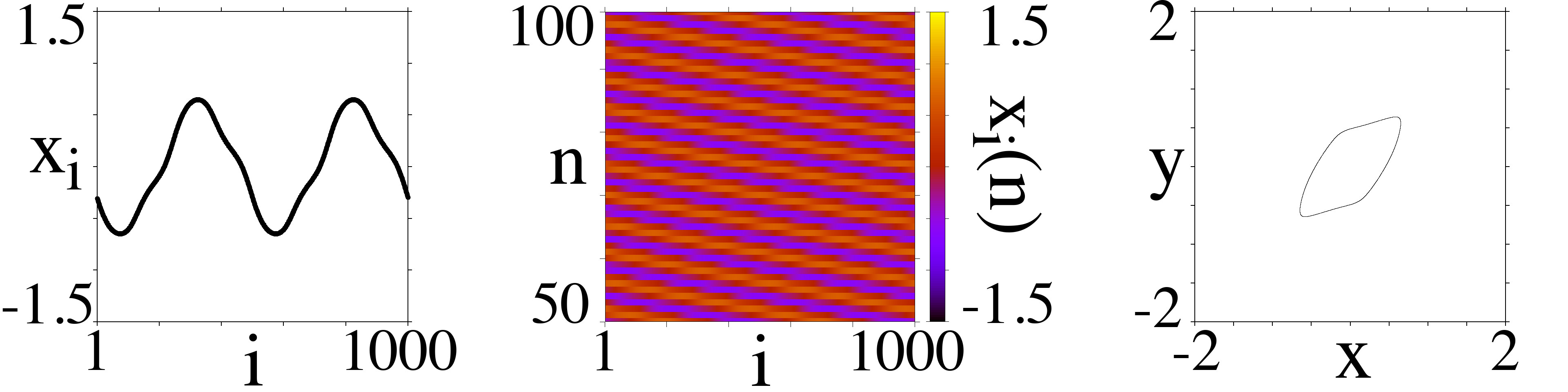}  \\
\hspace{8pt} (d)\\
\end{tabular}\caption{Snapshots of variable $x_i$ (left column), space-time plots (middle column), and projections of multidimensional attractors of the system in the $(x,y)$ phase plane (right column) for different values of the long calculation time: (a) $n=1,200$, (b) $n=3,600$, (c) $n=200,000$, and (d) $n=2,300,000$. Other parameters: $\epsilon=1.19$, $\lambda=0.12$, $\sigma=0.02$, $R=320$.}
	\label{dynamics-wave-solitary}
\end{figure}

During our numerical simulations of the discrete van der Pol oscillators network, we found one more nontrivial regime that deserves more detailed consideration. This dynamical mode is novel and represents the coexistence of traveling wave and solitary nodes and has been exemplified  in Fig.~\ref{map_regimes_ring}, TW+SS and Fig.~\ref{regimes_ring_other},VII. This regime is also transient towards the traveling wave mode, and the transition time  depends on both the network parameters and the initial conditions $(x_{i}(0),y_{i}(0))$,~$i=1,2,\ldots,N=1000$. 
Snapshots, space-time plots and projections of multidimensional attractors of the system are shown in Fig.~\ref{dynamics-wave-solitary} and they illustrate how  dynamics of network \eqref{system} evolves in time as the running computation time increases. It is seen that the regime being similar to the coexistence of the multichimera and solitary nodes (Fig.\ref{dynamics-wave-solitary}(a)) converts to the coexistence of the traveling wave and  solitary nodes (Fig.\ref{dynamics-wave-solitary}(b)). The number of solitary nodes gradually decreases with increasing in the observation time (Fig.\ref{dynamics-wave-solitary}(c)), and eventually only the traveling wave remains (Fig.\ref{dynamics-wave-solitary}(d)).

\begin{figure}[ht]
	\centering
\begin{tabular}{cc}
\includegraphics[width=.48\columnwidth]{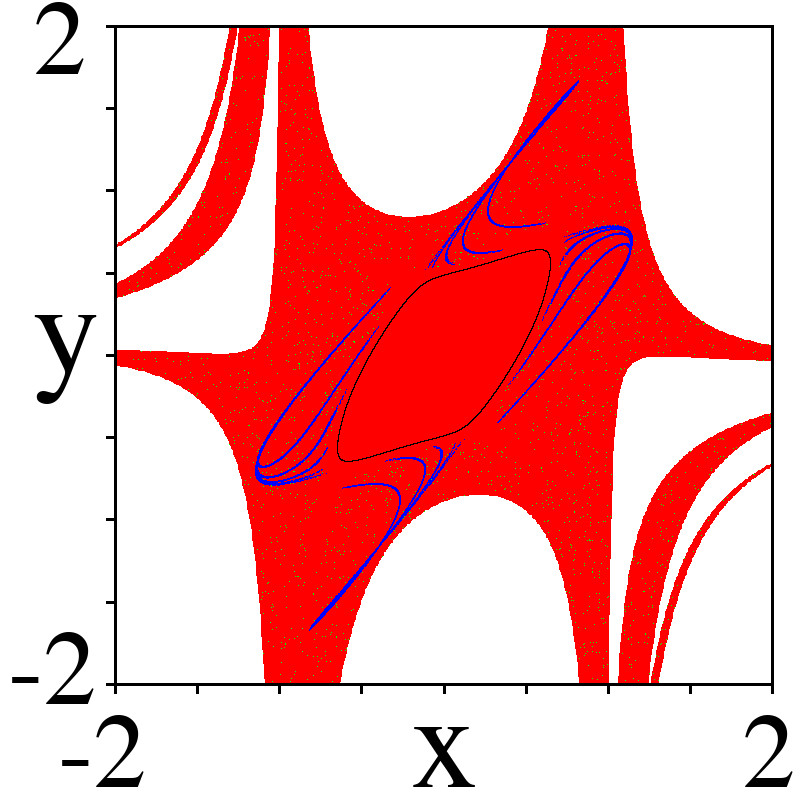} &
\includegraphics[width=.48\columnwidth]{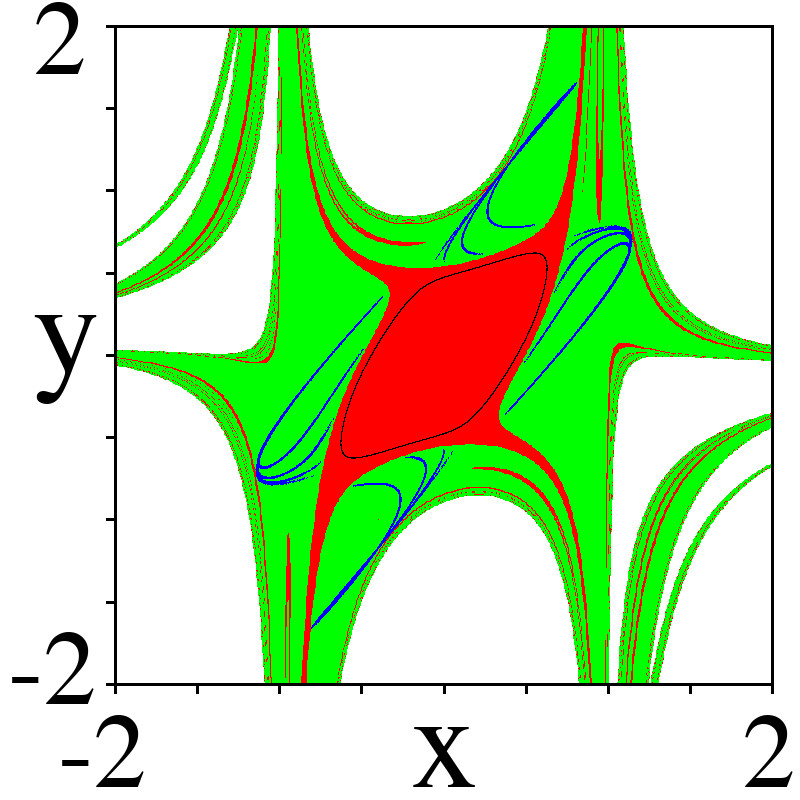}  \\
\hspace{8pt} (a) & \hspace{8pt} (b)\\
\end{tabular}
\begin{tabular}{cc}
\includegraphics[width=.48\columnwidth]{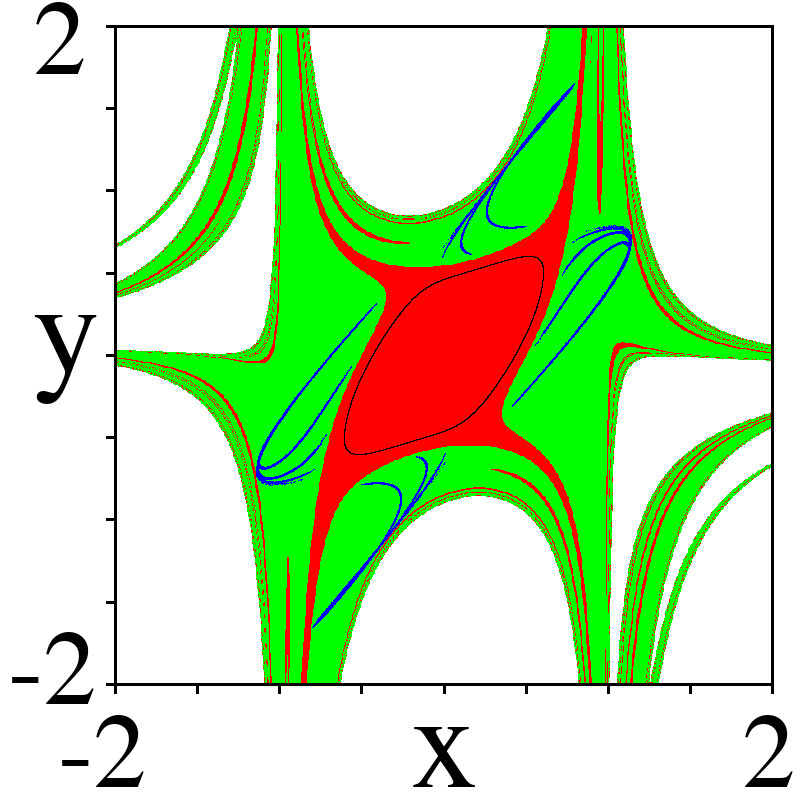} &
\includegraphics[width=.48\columnwidth]{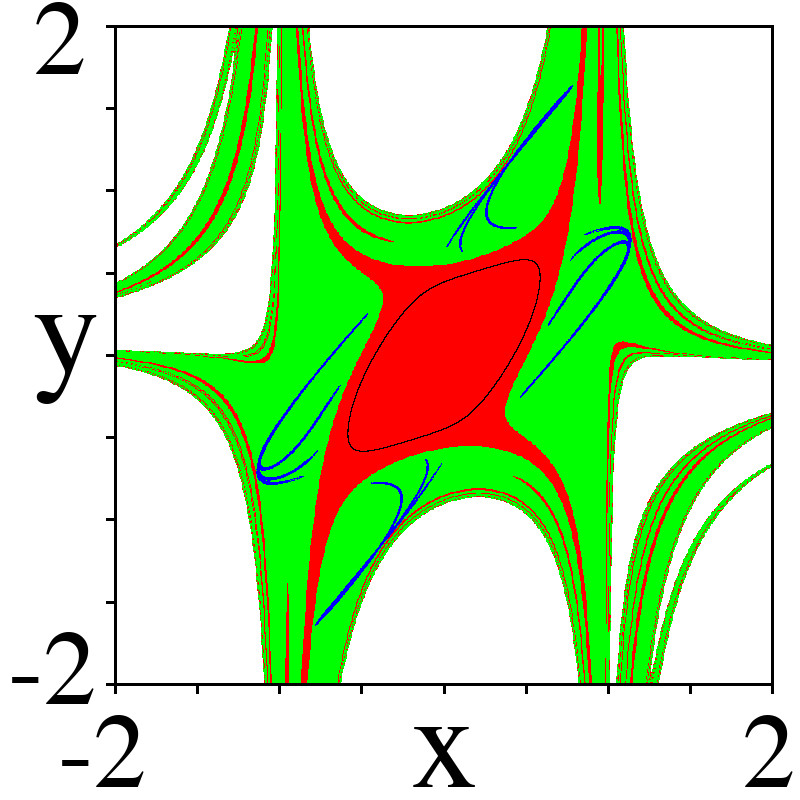}  \\
\hspace{8pt} (c) & \hspace{8pt} (d)\\
\end{tabular}
\caption{Section of basins of attraction for the typical state (traveling wave) (red points) and the solitary state (green points) in the network of discrete van der Pol oscillators at four different values of the coupling strength (in increasing order):  (a) $\sigma=0.02$, (b) $\sigma=0.025$, (c) $\sigma=0.03$, and (d) $\sigma=0.035$. The black cycle is the attractor for the typical state and in blue is the chaotic attractor for the solitary state. Other parameters: $\epsilon=1.19$, $\lambda=0.12$, $R=320$, steps of division of the $(x_{1},y_{1})$ phase plane are $\Delta x=0.005$ and $\Delta y=0.005$, initial conditions of the other elements correspond to the purely traveling wave (Fig.~\ref{dynamics-wave-solitary}(d)).}
	\label{basins-solitary-node}
\end{figure}

As in Section IV(A), we now investigate how the coupling strength $\sigma$ effects the observation of the coexisting regimes. We construct a section of basins of attraction for a typical state, i.e., when an element belongs to the traveling wave, and for the solitary state. 
Figure~\ref{basins-solitary-node} displays the changes in the section of basins of attractions of the states as mentioned above for four increasing values of the coupling strength. The initial spatiotemporal structure corresponds to the purely traveling wave (Fig.~\ref{dynamics-wave-solitary}(d)), and we change the amplitude of  the first element only and trace the behavior of this element. At $\sigma=0.02$, the basin of attraction of the typical state (the black projection of the multidimensional attractor) is riddled (Fig.~\ref{basins-solitary-node}(a), red region), since there are finite inclusions of the basin of attraction of the solitary node (green points and the blue projection of the multidimensional attractor). As $\sigma$ increases, the basin of attraction of the typical state is  significantly reduced and the solitary state begins to dominate in the network dynamics (Fig.~\ref{basins-solitary-node}(b)-(d)). In these cases, the structure of the section of the basins of attraction is homogeneous, and their form remains almost unchanged when the parameter $\epsilon$ and the initial profile of the network dynamics change.

\begin{figure}[ht]
	\centering
\begin{tabular}{c}
\includegraphics[width=.95\columnwidth]{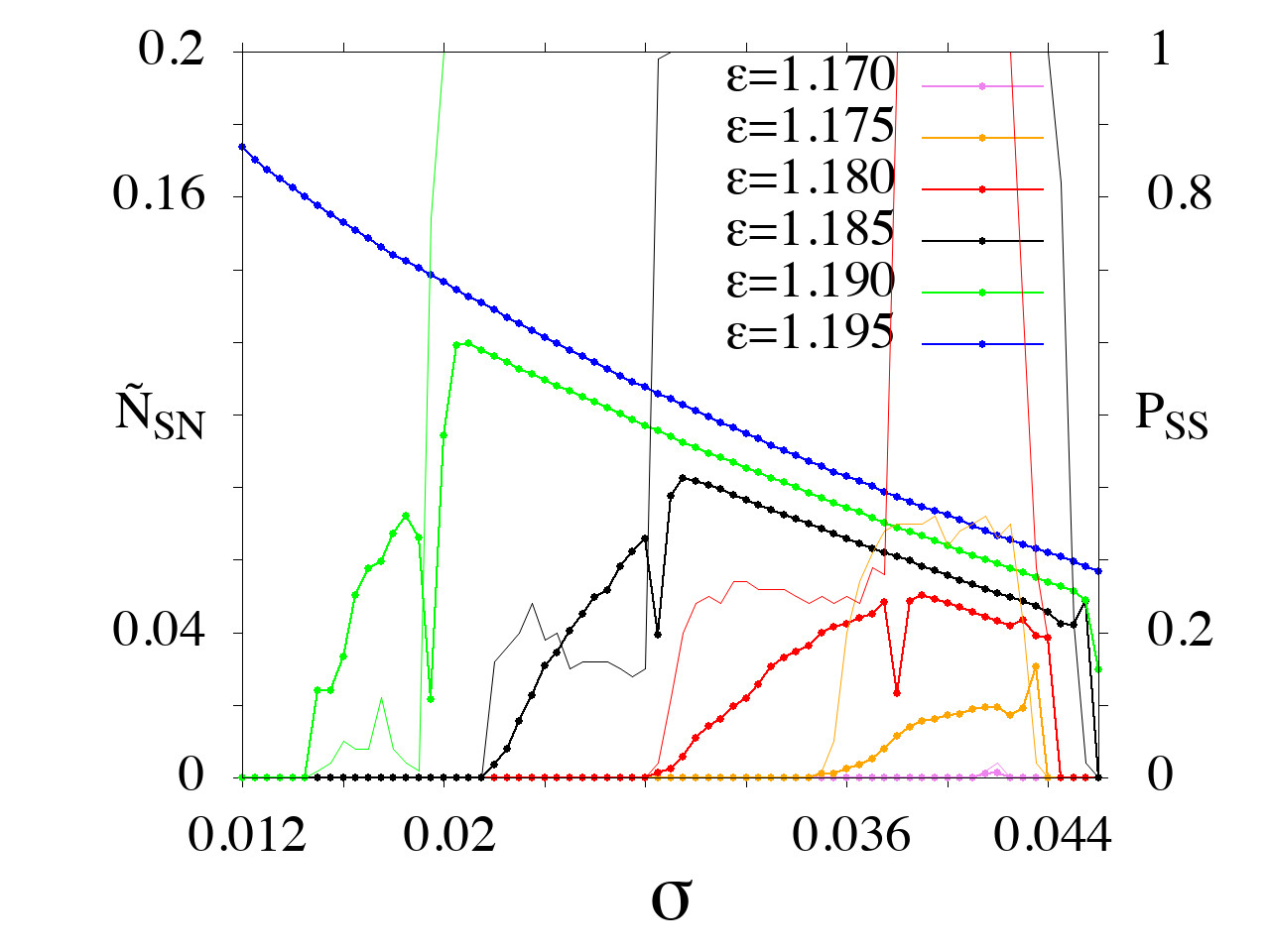} \\
\hspace{8pt} (a)\\
\includegraphics[width=.95\columnwidth]{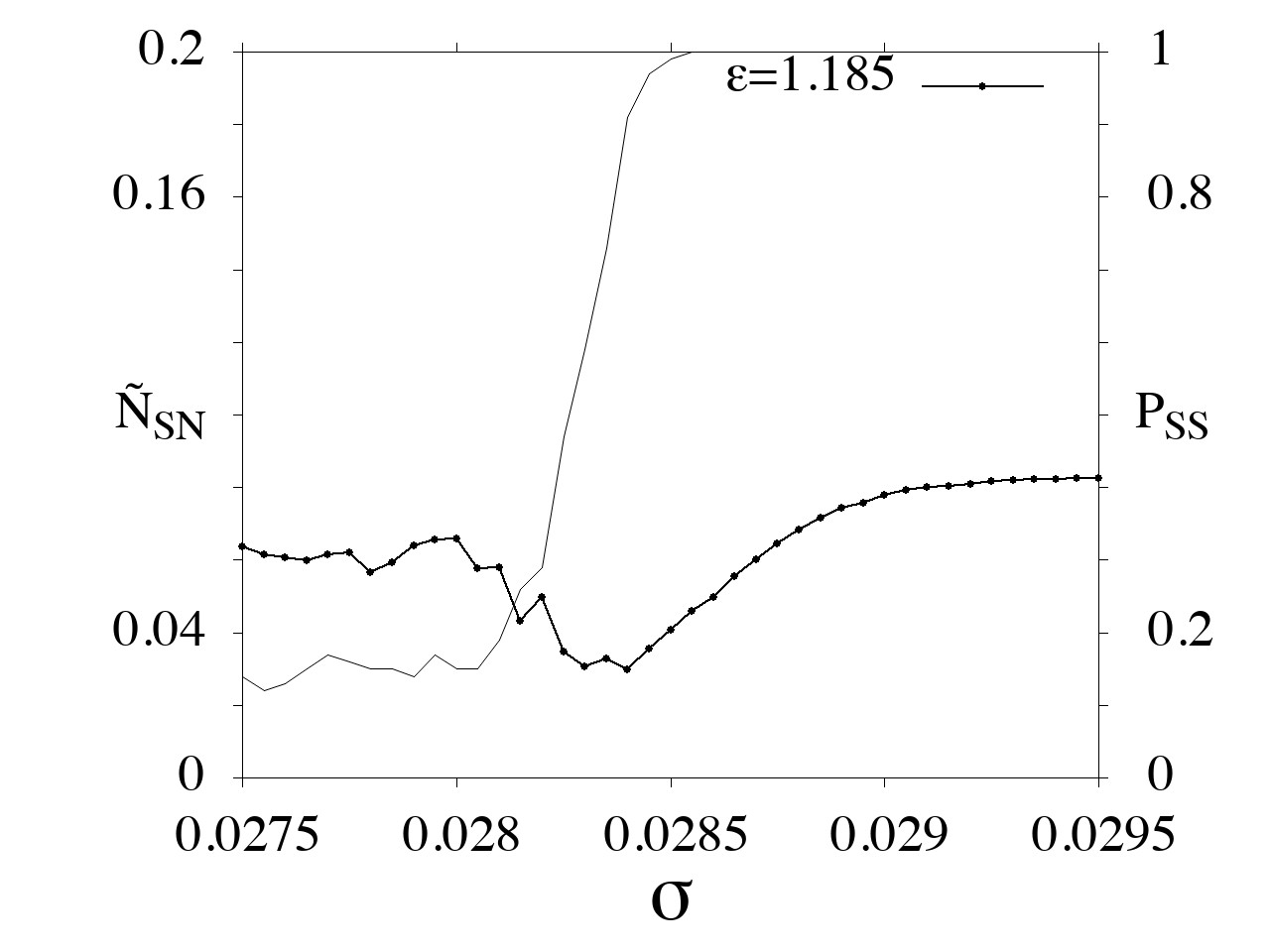}  \\
\hspace{8pt} (b)\\
\end{tabular}
\caption{(a) Dependences of the number of solitary nodes $\tilde{N}_{\rm SN}$ \eqref{N_sn} (dotted curves, the left $y$-axis scale) and of the probability of observing the solitary state $P_{\rm SS}$ (solid curves, the right $y$-axis scale) on the coupling strength $\sigma$ at different values of the parameter $\epsilon$. (b) Enlarged fragment of $\tilde{N}_{\rm SN}$ and $P_{\rm SS}(\sigma)$ dependences for $\epsilon=1.185$. Hundred initial conditions are chosen to be uniformly distributed in the interval $(x_{i}(0),y_{i}(0))\in[-0.5,0.5]$, $i=1,2,\ldots,N=1000$.
Other parameters: $\lambda=0.12$, $R=320$,  $T_{\rm obs}=100,000$. }
	\label{number-sn}
\end{figure}

Our studies show that the number of solitary nodes can depend on the parameter $\epsilon$ and the coupling strength $\sigma$. This dependence is analyzed by calculating a normalized number of solitary nodes averaged over the initial conditions $\tilde{N}_{\rm SN}$:

      \begin{eqnarray}\label{N_sn}
\tilde{N}_{\rm SN}=\frac{ \sum\limits_{IC_{\rm SS}}N_{\rm SN}/N}{IC_{\rm SS}},
     \end{eqnarray}

\noindent where $N_{\rm SN}$ is a number of solitary nodes, $N=1000$ is the total number of elements in the network \eqref{system}, $IC_{\rm SS}$ is a number of the initial conditions which lead to the solitary state. Numerical results for the dependences $\tilde{N}_{\rm SN}$ on $\sigma$ are shown in Fig.~\ref{number-sn}(a) (dotted curves, the left $y$-axis scale) for different values of $\epsilon$. The coupling $\sigma$ is varied within the range $[0.012, 0.044]$ since when $\sigma<0.012$ the network exhibits the incoherent dynamics  or multicluster chimera states and at $\sigma>0.044$ we can observe multicluster chimera states.
As it follows from the picture, in the regime of coexistence of the traveling wave and the solitary state ($\epsilon=1.195 $), the number of solitary nodes gradually decreases when the coupling strength increases (Fig.~\ref{number-sn}(a), blue dotted curve). 
When $\epsilon$ decreases and thus the probability of the appearance of the multicluster chimera or the coexistence of the multicluster chimera and the solitary state increases, the number of solitary nodes grows as $\sigma$ increases. As it follows in Fig.~\ref{number-sn}(a), the less the parameter $\epsilon$, the stronger coupling is needed to induce the solitary state. For each value of $\epsilon$ there is a certain threshold or critical value of $\sigma$, exceeding which the number $N_{\rm SN}$ gradually decreases. It can also be emphasized that with decrease in the parameter $\epsilon$, both the maximum number of solitary nodes and width of the coupling strength range within which the solitary state exists in the network also decrease. 

Calculated dependences of the probability of observing the solitary state $P_{\rm SS}$ on the coupling strength $\sigma$ are plotted in Fig.~\ref{number-sn}(a) (solid curves, the right $y$-axis scale).
It should be noted that at $\epsilon=1.195$, $P_{\rm SS}=1$ over the whole considered range of $\sigma$ (Fig.~\ref{number-sn}(a)).
For smaller values of $\epsilon$, e.g., $\epsilon=1.190,~1.185,~1.180$, two regions with respect to $\sigma$ can be distinguished where the probability $P_{\rm SS}$ is essentially different. When the coupling strength is rather weak, the probability of observing the solitary state is much less than $1$, and for large values of $\sigma$, the probability $P_{\rm SS}$ rises sharply to $1$. At $\epsilon<1.180$, such a separation of the $\sigma$ range is no longer observed, and for all values of  $\sigma$, the probability of observing the solitary state is $P_{\rm SS}<1$.

It can be seen in Fig.~\ref{number-sn}(a) that there is a dip in the dependences  $\tilde{N}_{\rm SN}$ versus $\sigma$ (green, black, and red dotted curves) at the moment when the probability of observing the solitary state $P_{\rm SS}$ rises sharply to $1$ ($\sigma_{\rm cr}\approx0.019$ at $\epsilon=1.190$, $\sigma_{\rm cr}\approx0.028$ at $\epsilon=1.185$, $\sigma_{\rm cr}\approx0.0375$ at $\epsilon=1.180$). 
Our studies show that this can be explained by the fact that at $\sigma<\sigma_{\rm cr}$, the lifetime of solitary nodes is rather short and may not match $T_{\rm obs}=100,000$ iterations. This circumstance leads to a low probability of observing the solitary state regime. At $\sigma>\sigma_{\rm cr}$, as noted earlier, the probability of observing the solitary state rises sharply to $P_{\rm SS}=1$, but the lifetime of separate solitary nodes is still less than $T_{\rm obs}=100,000$. Thus,  some of the solitary state regimes (within a certain range of $\sigma$) contain the solitary nodes $N_{\rm SN}\approx80$, while the others have much less solitary nodes.  After averaging the number of solitary nodes over the network elements, we get a dip in the dependences $\tilde{N}_{\rm SN}(\sigma)$.

Figure~\ref{number-sn}(b) shows an enlarged fragment of the dependences $\tilde{N}_{\rm SN}$ and $P_{\rm SS}$ versus $\sigma$ for $\epsilon=1.185$. These plots give more clear evidence of the decrease in the average number of solitary nodes which is accompanied by a sharp increase in the probability $P_{\rm SS}$ at $\sigma_{\rm cr}\approx0.028$.

\section{\label{sec:conclusion}Conclusion}

We have studied numerically the spatiotemporal dynamics of a ring network of nonlocally coupled discrete-time van der Pol oscillators when the control parameters of the individual elements, the coupling strength,  and initial conditions are varied. The results obtained and described for the network of nonlocally coupled discrete van der Pol oscillators give evidence of the possibility of observing a diverse variety of spatiotemporal patterns. Particularly, we have found chimera states of different types, solitary states, and the regimes when the multichimera or the traveling wave coexists with solitary states. It should be noted that the last two modes  have not been observed in other networks before. 

In our work we have explored in detail the transition from the multichimera state to the traveling wave  depending on the observation time and the coupling strength. It has been found that a sufficiently strong coupling increases the multichimera lifetime and thereby prolongs the transient process to the traveling wave regime. On the contrary, the multichimera appears to be a  short-living pattern in the case of weakly coupled van der Pol oscillators. We have constructed basins of attraction for different regimes in the network and found that they have riddle structures. The evolution of the basins of attraction has been investigated depending on the coupling strength and the observation time.

It has also been demonstrated that the multichimera lifetime and thus the transient process duration can be controlled by applying additive noise to the network. Some noise realizations with both large and small intensities can essentially reduce the multichimera lifetime and thus, speed up the transition to the traveling wave. However, there is also a small but non-zero probability that despite the noise intensity and duration, long-living multichimera states can be observed and thus the transient process increases. 

We have also investigated in detail the coexistence of the traveling wave and the solitary state, which has been found in the network of discrete van der Pol oscillators. This mode is transient and results in the traveling wave pattern. It has been established that as the parameter $\epsilon$ increases, the number of solitary nodes grows even when nonlocal coupling is weak. The smaller the parameter $\epsilon$, the stronger coupling is needed to observe the solitary state. 

As an outlook, firstly, it would be interesting to explore the effect of external noise on the regimes of purely solitary state and of the coexistence of the solitary state and the traveling wave. Secondly, it should be investigated how the dynamics of the ring of nonlocally coupled discrete van der Pol oscillators depend on the coupling range.  Moreover, recently a generalized mechanism behind infinite coexistence of stable periodic solutions in two-dimensional discrete maps were formulated~\cite{Muni:2020wy}. Researchers in Ref.~\cite{VaMu22}, found a similar phenomenon of coexistence of single, double-well chimera states, and coherent clusters in a ring-star network of Chua's oscillators but a mechanism behind such phenomenon in the network of oscillators is still unknown to our best knowledge. It would be interesting to account for a conjecture/mechanism behind this phenomenon in networks of oscillators.    

\acknowledgments 
E.R. and G.S. acknowledge financial support from the Deutsche Forschungsgemeinschaft (DFG, German Research Foundation) - Projekt nummer - 163436311 - SFB 910.

\section*{AUTHOR DECLARATIONS}
\subsection*{Conflict of Interest}
The authors have no conflicts to disclose.

\section*{Data Availability Statement}

The data that supports the findings of this study are available within the article.

\bibliography{thebibliography}

\begin{thebibliography}{100}%
\makeatletter
\providecommand \@ifxundefined [1]{%
 \@ifx{#1\undefined}
}%
\providecommand \@ifnum [1]{%
 \ifnum #1\expandafter \@firstoftwo
 \else \expandafter \@secondoftwo
 \fi
}%
\providecommand \@ifx [1]{%
 \ifx #1\expandafter \@firstoftwo
 \else \expandafter \@secondoftwo
 \fi
}%
\providecommand \natexlab [1]{#1}%
\providecommand \enquote  [1]{``#1''}%
\providecommand \bibnamefont  [1]{#1}%
\providecommand \bibfnamefont [1]{#1}%
\providecommand \citenamefont [1]{#1}%
\providecommand \href@noop [0]{\@secondoftwo}%
\providecommand \href [0]{\begingroup \@sanitize@url \@href}%
\providecommand \@href[1]{\@@startlink{#1}\@@href}%
\providecommand \@@href[1]{\endgroup#1\@@endlink}%
\providecommand \@sanitize@url [0]{\catcode `\\12\catcode `\$12\catcode
  `\&12\catcode `\#12\catcode `\^12\catcode `\_12\catcode `\%12\relax}%
\providecommand \@@startlink[1]{}%
\providecommand \@@endlink[0]{}%
\providecommand \url  [0]{\begingroup\@sanitize@url \@url }%
\providecommand \@url [1]{\endgroup\@href {#1}{\urlprefix }}%
\providecommand \urlprefix  [0]{URL }%
\providecommand \Eprint [0]{\href }%
\providecommand \doibase [0]{http://dx.doi.org/}%
\providecommand \selectlanguage [0]{\@gobble}%
\providecommand \bibinfo  [0]{\@secondoftwo}%
\providecommand \bibfield  [0]{\@secondoftwo}%
\providecommand \translation [1]{[#1]}%
\providecommand \BibitemOpen [0]{}%
\providecommand \bibitemStop [0]{}%
\providecommand \bibitemNoStop [0]{.\EOS\space}%
\providecommand \EOS [0]{\spacefactor3000\relax}%
\providecommand \BibitemShut  [1]{\csname bibitem#1\endcsname}%
\let\auto@bib@innerbib\@empty
\bibitem [{\citenamefont {Pikovsky}, \citenamefont {Rosenblum},\ and\
  \citenamefont {Kurths}(2001)}]{Pikovsky:2001uj}%
  \BibitemOpen
  \bibfield  {author} {\bibinfo {author} {\bibfnamefont {A.}~\bibnamefont
  {Pikovsky}}, \bibinfo {author} {\bibfnamefont {M.}~\bibnamefont {Rosenblum}},
  \ and\ \bibinfo {author} {\bibfnamefont {J.}~\bibnamefont {Kurths}},\
  }\href@noop {} {\emph {\bibinfo {title} {Synchronization: {A} universal
  concept in nonlinear science}}}\ (\bibinfo  {publisher} {Cambridge University
  Press},\ \bibinfo {year} {2001})\BibitemShut {NoStop}%
\bibitem [{\citenamefont {Strogatz}(2001)}]{Strogatz:2001aa}%
  \BibitemOpen
  \bibfield  {author} {\bibinfo {author} {\bibfnamefont {S.~H.}\ \bibnamefont
  {Strogatz}},\ }\bibfield  {title} {\enquote {\bibinfo {title} {Exploring
  complex networks},}\ }\href {\doibase 10.1038/35065725} {\bibfield  {journal}
  {\bibinfo  {journal} {Nature}\ }\textbf {\bibinfo {volume} {410}},\ \bibinfo
  {pages} {268--276} (\bibinfo {year} {2001})}\BibitemShut {NoStop}%
\bibitem [{\citenamefont {Albert}\ and\ \citenamefont
  {Barab{\'a}si}(2002)}]{Albert:2002wu}%
  \BibitemOpen
  \bibfield  {author} {\bibinfo {author} {\bibfnamefont {R.}~\bibnamefont
  {Albert}}\ and\ \bibinfo {author} {\bibfnamefont {A.-L.}\ \bibnamefont
  {Barab{\'a}si}},\ }\bibfield  {title} {\enquote {\bibinfo {title}
  {Statistical mechanics of complex networks},}\ }\href@noop {} {\bibfield
  {journal} {\bibinfo  {journal} {Reviews of modern physics}\ }\textbf
  {\bibinfo {volume} {74}},\ \bibinfo {pages} {47} (\bibinfo {year}
  {2002})}\BibitemShut {NoStop}%
\bibitem [{\citenamefont {Newman}(2003)}]{Newman:2003wd}%
  \BibitemOpen
  \bibfield  {author} {\bibinfo {author} {\bibfnamefont {M.~E.}\ \bibnamefont
  {Newman}},\ }\bibfield  {title} {\enquote {\bibinfo {title} {The structure
  and function of complex networks},}\ }\href@noop {} {\bibfield  {journal}
  {\bibinfo  {journal} {SIAM review}\ }\textbf {\bibinfo {volume} {45}},\
  \bibinfo {pages} {167--256} (\bibinfo {year} {2003})}\BibitemShut {NoStop}%
\bibitem [{\citenamefont {Balanov}\ \emph {et~al.}(2009)\citenamefont
  {Balanov}, \citenamefont {Janson}, \citenamefont {Postnov},\ and\
  \citenamefont {Sosnovtseva}}]{Balanov:2009tt}%
  \BibitemOpen
  \bibfield  {author} {\bibinfo {author} {\bibfnamefont {A.}~\bibnamefont
  {Balanov}}, \bibinfo {author} {\bibfnamefont {N.}~\bibnamefont {Janson}},
  \bibinfo {author} {\bibfnamefont {D.}~\bibnamefont {Postnov}}, \ and\
  \bibinfo {author} {\bibfnamefont {O.}~\bibnamefont {Sosnovtseva}},\
  }\href@noop {} {\emph {\bibinfo {title} {Synchronization: $F$rom Simple to
  Complex}}}\ (\bibinfo  {publisher} {Springer},\ \bibinfo {year}
  {2009})\BibitemShut {NoStop}%
\bibitem [{\citenamefont {Boccaletti}\ \emph {et~al.}(2018)\citenamefont
  {Boccaletti}, \citenamefont {Pisarchik}, \citenamefont {Del~Genio},\ and\
  \citenamefont {Amann}}]{Boccaletti:2018up}%
  \BibitemOpen
  \bibfield  {author} {\bibinfo {author} {\bibfnamefont {S.}~\bibnamefont
  {Boccaletti}}, \bibinfo {author} {\bibfnamefont {A.~N.}\ \bibnamefont
  {Pisarchik}}, \bibinfo {author} {\bibfnamefont {C.~I.}\ \bibnamefont
  {Del~Genio}}, \ and\ \bibinfo {author} {\bibfnamefont {A.}~\bibnamefont
  {Amann}},\ }\href@noop {} {\emph {\bibinfo {title} {Synchronization: {F}rom
  coupled systems to complex networks}}}\ (\bibinfo  {publisher} {Cambridge
  University Press},\ \bibinfo {year} {2018})\BibitemShut {NoStop}%
\bibitem [{\citenamefont {Cardillo}\ \emph {et~al.}(2013)\citenamefont
  {Cardillo}, \citenamefont {Zanin}, \citenamefont {G{\'o}mez-Gardenes},
  \citenamefont {Romance}, \citenamefont {Garc{\'\i}a~del Amo},\ and\
  \citenamefont {Boccaletti}}]{Cardillo:2013wg}%
  \BibitemOpen
  \bibfield  {author} {\bibinfo {author} {\bibfnamefont {A.}~\bibnamefont
  {Cardillo}}, \bibinfo {author} {\bibfnamefont {M.}~\bibnamefont {Zanin}},
  \bibinfo {author} {\bibfnamefont {J.}~\bibnamefont {G{\'o}mez-Gardenes}},
  \bibinfo {author} {\bibfnamefont {M.}~\bibnamefont {Romance}}, \bibinfo
  {author} {\bibfnamefont {A.~J.}\ \bibnamefont {Garc{\'\i}a~del Amo}}, \ and\
  \bibinfo {author} {\bibfnamefont {S.}~\bibnamefont {Boccaletti}},\ }\bibfield
   {title} {\enquote {\bibinfo {title} {Modeling the multi-layer nature of the
  {E}uropean {A}ir {T}ransport {N}etwork: {R}esilience and passengers
  re-scheduling under random failures},}\ }\href@noop {} {\bibfield  {journal}
  {\bibinfo  {journal} {The European Physical Journal Special Topics}\ }\textbf
  {\bibinfo {volume} {215}},\ \bibinfo {pages} {23--33} (\bibinfo {year}
  {2013})}\BibitemShut {NoStop}%
\bibitem [{\citenamefont {Jaros}, \citenamefont {Maistrenko},\ and\
  \citenamefont {Kapitaniak}(2015)}]{Jaros:2015tv}%
  \BibitemOpen
  \bibfield  {author} {\bibinfo {author} {\bibfnamefont {P.}~\bibnamefont
  {Jaros}}, \bibinfo {author} {\bibfnamefont {Y.}~\bibnamefont {Maistrenko}}, \
  and\ \bibinfo {author} {\bibfnamefont {T.}~\bibnamefont {Kapitaniak}},\
  }\bibfield  {title} {\enquote {\bibinfo {title} {Chimera states on the route
  from coherence to rotating waves},}\ }\href@noop {} {\bibfield  {journal}
  {\bibinfo  {journal} {Physical Review E}\ }\textbf {\bibinfo {volume} {91}},\
  \bibinfo {pages} {022907} (\bibinfo {year} {2015})}\BibitemShut {NoStop}%
\bibitem [{\citenamefont {Semenova}\ \emph
  {et~al.}(2017{\natexlab{a}})\citenamefont {Semenova}, \citenamefont
  {Strelkova}, \citenamefont {Anishchenko},\ and\ \citenamefont
  {Zakharova}}]{Semenova:2017tt}%
  \BibitemOpen
  \bibfield  {author} {\bibinfo {author} {\bibfnamefont {N.}~\bibnamefont
  {Semenova}}, \bibinfo {author} {\bibfnamefont {G.}~\bibnamefont {Strelkova}},
  \bibinfo {author} {\bibfnamefont {V.}~\bibnamefont {Anishchenko}}, \ and\
  \bibinfo {author} {\bibfnamefont {A.}~\bibnamefont {Zakharova}},\ }\bibfield
  {title} {\enquote {\bibinfo {title} {Temporal intermittency and the lifetime
  of chimera states in ensembles of nonlocally coupled chaotic oscillators},}\
  }\href@noop {} {\bibfield  {journal} {\bibinfo  {journal} {Chaos: An
  Interdisciplinary Journal of Nonlinear Science}\ }\textbf {\bibinfo {volume}
  {27}},\ \bibinfo {pages} {061102} (\bibinfo {year}
  {2017}{\natexlab{a}})}\BibitemShut {NoStop}%
\bibitem [{\citenamefont {Amari}(1977)}]{Amari:1977wi}%
  \BibitemOpen
  \bibfield  {author} {\bibinfo {author} {\bibfnamefont {S.-i.}\ \bibnamefont
  {Amari}},\ }\bibfield  {title} {\enquote {\bibinfo {title} {Dynamics of
  pattern formation in lateral-inhibition type neural fields},}\ }\href@noop {}
  {\bibfield  {journal} {\bibinfo  {journal} {Biological cybernetics}\ }\textbf
  {\bibinfo {volume} {27}},\ \bibinfo {pages} {77--87} (\bibinfo {year}
  {1977})}\BibitemShut {NoStop}%
\bibitem [{\citenamefont {Compte}\ \emph {et~al.}(2000)\citenamefont {Compte},
  \citenamefont {Brunel}, \citenamefont {Goldman-Rakic},\ and\ \citenamefont
  {Wang}}]{Compte:2000ul}%
  \BibitemOpen
  \bibfield  {author} {\bibinfo {author} {\bibfnamefont {A.}~\bibnamefont
  {Compte}}, \bibinfo {author} {\bibfnamefont {N.}~\bibnamefont {Brunel}},
  \bibinfo {author} {\bibfnamefont {P.~S.}\ \bibnamefont {Goldman-Rakic}}, \
  and\ \bibinfo {author} {\bibfnamefont {X.-J.}\ \bibnamefont {Wang}},\
  }\bibfield  {title} {\enquote {\bibinfo {title} {Synaptic mechanisms and
  network dynamics underlying spatial working memory in a cortical network
  model},}\ }\href@noop {} {\bibfield  {journal} {\bibinfo  {journal} {Cerebral
  cortex}\ }\textbf {\bibinfo {volume} {10}},\ \bibinfo {pages} {910--923}
  (\bibinfo {year} {2000})}\BibitemShut {NoStop}%
\bibitem [{\citenamefont {Panaggio}\ and\ \citenamefont
  {Abrams}(2015)}]{Panaggio:2015uu}%
  \BibitemOpen
  \bibfield  {author} {\bibinfo {author} {\bibfnamefont {M.~J.}\ \bibnamefont
  {Panaggio}}\ and\ \bibinfo {author} {\bibfnamefont {D.~M.}\ \bibnamefont
  {Abrams}},\ }\bibfield  {title} {\enquote {\bibinfo {title} {Chimera states:
  coexistence of coherence and incoherence in networks of coupled
  oscillators},}\ }\href@noop {} {\bibfield  {journal} {\bibinfo  {journal}
  {Nonlinearity}\ }\textbf {\bibinfo {volume} {28}},\ \bibinfo {pages} {R67}
  (\bibinfo {year} {2015})}\BibitemShut {NoStop}%
\bibitem [{\citenamefont {Sch{\"o}ll}(2016)}]{Scholl:2016vm}%
  \BibitemOpen
  \bibfield  {author} {\bibinfo {author} {\bibfnamefont {E.}~\bibnamefont
  {Sch{\"o}ll}},\ }\bibfield  {title} {\enquote {\bibinfo {title}
  {Synchronization patterns and chimera states in complex networks: Interplay
  of topology and dynamics},}\ }\href@noop {} {\bibfield  {journal} {\bibinfo
  {journal} {The European Physical Journal Special Topics}\ }\textbf {\bibinfo
  {volume} {225}},\ \bibinfo {pages} {891--919} (\bibinfo {year}
  {2016})}\BibitemShut {NoStop}%
\bibitem [{\citenamefont {Muni}, \citenamefont {Padhee},\ and\ \citenamefont
  {Pati}(2018)}]{Muni:2018vd}%
  \BibitemOpen
  \bibfield  {author} {\bibinfo {author} {\bibfnamefont {S.~S.}\ \bibnamefont
  {Muni}}, \bibinfo {author} {\bibfnamefont {S.}~\bibnamefont {Padhee}}, \ and\
  \bibinfo {author} {\bibfnamefont {K.~C.}\ \bibnamefont {Pati}},\ }\bibfield
  {title} {\enquote {\bibinfo {title} {A study on the synchronization aspect of
  star connected identical {C}hua's circuits},}\ \ }(\bibinfo  {publisher}
  {IEEE},\ \bibinfo {year} {2018})\ pp.\ \bibinfo {pages} {1--6}\BibitemShut
  {NoStop}%
\bibitem [{\citenamefont {Belyaev}\ and\ \citenamefont
  {Ryashko}(2020)}]{Belyaev:2020we}%
  \BibitemOpen
  \bibfield  {author} {\bibinfo {author} {\bibfnamefont {A.}~\bibnamefont
  {Belyaev}}\ and\ \bibinfo {author} {\bibfnamefont {L.}~\bibnamefont
  {Ryashko}},\ }\bibfield  {title} {\enquote {\bibinfo {title} {Regular and
  chaotic regimes in the system of coupled populations},}\ \ }(\bibinfo
  {publisher} {AIP Publishing LLC},\ \bibinfo {year} {2020})\ p.\ \bibinfo
  {pages} {070023}\BibitemShut {NoStop}%
\bibitem [{\citenamefont {Kuramoto}\ and\ \citenamefont
  {Battogtokh}(2002)}]{Kuramoto:2002uu}%
  \BibitemOpen
  \bibfield  {author} {\bibinfo {author} {\bibfnamefont {Y.}~\bibnamefont
  {Kuramoto}}\ and\ \bibinfo {author} {\bibfnamefont {D.}~\bibnamefont
  {Battogtokh}},\ }\bibfield  {title} {\enquote {\bibinfo {title} {Coexistence
  of coherence and incoherence in nonlocally coupled phase oscillators},}\
  }\href@noop {} {\bibfield  {journal} {\bibinfo  {journal} {Nonlinear
  Phenomena in Complex Systems}\ }\textbf {\bibinfo {volume} {5}},\ \bibinfo
  {pages} {380--385} (\bibinfo {year} {2002})}\BibitemShut {NoStop}%
\bibitem [{\citenamefont {Abrams}\ and\ \citenamefont
  {Strogatz}(2004)}]{Abrams:2004vx}%
  \BibitemOpen
  \bibfield  {author} {\bibinfo {author} {\bibfnamefont {D.~M.}\ \bibnamefont
  {Abrams}}\ and\ \bibinfo {author} {\bibfnamefont {S.~H.}\ \bibnamefont
  {Strogatz}},\ }\bibfield  {title} {\enquote {\bibinfo {title} {Chimera states
  for coupled oscillators},}\ }\href@noop {} {\bibfield  {journal} {\bibinfo
  {journal} {Physical review letters}\ }\textbf {\bibinfo {volume} {93}},\
  \bibinfo {pages} {174102} (\bibinfo {year} {2004})}\BibitemShut {NoStop}%
\bibitem [{\citenamefont {Omelchenko}\ \emph {et~al.}(2011)\citenamefont
  {Omelchenko}, \citenamefont {Maistrenko}, \citenamefont {H{\"o}vel},\ and\
  \citenamefont {Sch{\"o}ll}}]{Omelchenko:2011uc}%
  \BibitemOpen
  \bibfield  {author} {\bibinfo {author} {\bibfnamefont {I.}~\bibnamefont
  {Omelchenko}}, \bibinfo {author} {\bibfnamefont {Y.}~\bibnamefont
  {Maistrenko}}, \bibinfo {author} {\bibfnamefont {P.}~\bibnamefont
  {H{\"o}vel}}, \ and\ \bibinfo {author} {\bibfnamefont {E.}~\bibnamefont
  {Sch{\"o}ll}},\ }\bibfield  {title} {\enquote {\bibinfo {title} {Loss of
  coherence in dynamical networks: spatial chaos and chimera states},}\
  }\href@noop {} {\bibfield  {journal} {\bibinfo  {journal} {Physical Review
  Letters}\ }\textbf {\bibinfo {volume} {106}},\ \bibinfo {pages} {234102}
  (\bibinfo {year} {2011})}\BibitemShut {NoStop}%
\bibitem [{\citenamefont {Dudkowski}, \citenamefont {Maistrenko},\ and\
  \citenamefont {Kapitaniak}(2014)}]{Dudkowski:2014vm}%
  \BibitemOpen
  \bibfield  {author} {\bibinfo {author} {\bibfnamefont {D.}~\bibnamefont
  {Dudkowski}}, \bibinfo {author} {\bibfnamefont {Y.}~\bibnamefont
  {Maistrenko}}, \ and\ \bibinfo {author} {\bibfnamefont {T.}~\bibnamefont
  {Kapitaniak}},\ }\bibfield  {title} {\enquote {\bibinfo {title} {Different
  types of chimera states: An interplay between spatial and dynamical chaos},}\
  }\href@noop {} {\bibfield  {journal} {\bibinfo  {journal} {Physical Review
  E}\ }\textbf {\bibinfo {volume} {90}},\ \bibinfo {pages} {032920} (\bibinfo
  {year} {2014})}\BibitemShut {NoStop}%
\bibitem [{\citenamefont {Slepnev}, \citenamefont {Bukh},\ and\ \citenamefont
  {Vadivasova}(2017)}]{Slepnev:2017tz}%
  \BibitemOpen
  \bibfield  {author} {\bibinfo {author} {\bibfnamefont {A.~V.}\ \bibnamefont
  {Slepnev}}, \bibinfo {author} {\bibfnamefont {A.~V.}\ \bibnamefont {Bukh}}, \
  and\ \bibinfo {author} {\bibfnamefont {T.~E.}\ \bibnamefont {Vadivasova}},\
  }\bibfield  {title} {\enquote {\bibinfo {title} {Stationary and
  non-stationary chimeras in an ensemble of chaotic self-sustained oscillators
  with inertial nonlinearity},}\ }\href@noop {} {\bibfield  {journal} {\bibinfo
   {journal} {Nonlinear Dynamics}\ }\textbf {\bibinfo {volume} {88}},\ \bibinfo
  {pages} {2983--2992} (\bibinfo {year} {2017})}\BibitemShut {NoStop}%
\bibitem [{\citenamefont {Zakharova}, \citenamefont {Kapeller},\ and\
  \citenamefont {Sch{\"o}ll}(2014)}]{Zakharova:2014td}%
  \BibitemOpen
  \bibfield  {author} {\bibinfo {author} {\bibfnamefont {A.}~\bibnamefont
  {Zakharova}}, \bibinfo {author} {\bibfnamefont {M.}~\bibnamefont {Kapeller}},
  \ and\ \bibinfo {author} {\bibfnamefont {E.}~\bibnamefont {Sch{\"o}ll}},\
  }\bibfield  {title} {\enquote {\bibinfo {title} {Chimera death: Symmetry
  breaking in dynamical networks},}\ }\href@noop {} {\bibfield  {journal}
  {\bibinfo  {journal} {Physical Review Letters}\ }\textbf {\bibinfo {volume}
  {112}},\ \bibinfo {pages} {154101} (\bibinfo {year} {2014})}\BibitemShut
  {NoStop}%
\bibitem [{\citenamefont {Ulonska}\ \emph {et~al.}(2016)\citenamefont
  {Ulonska}, \citenamefont {Omelchenko}, \citenamefont {Zakharova},\ and\
  \citenamefont {Sch{\"o}ll}}]{Ulonska:2016tx}%
  \BibitemOpen
  \bibfield  {author} {\bibinfo {author} {\bibfnamefont {S.}~\bibnamefont
  {Ulonska}}, \bibinfo {author} {\bibfnamefont {I.}~\bibnamefont {Omelchenko}},
  \bibinfo {author} {\bibfnamefont {A.}~\bibnamefont {Zakharova}}, \ and\
  \bibinfo {author} {\bibfnamefont {E.}~\bibnamefont {Sch{\"o}ll}},\ }\bibfield
   {title} {\enquote {\bibinfo {title} {Chimera states in networks of van der
  {P}ol oscillators with hierarchical connectivities},}\ }\href@noop {}
  {\bibfield  {journal} {\bibinfo  {journal} {Chaos: An Interdisciplinary
  Journal of Nonlinear Science}\ }\textbf {\bibinfo {volume} {26}},\ \bibinfo
  {pages} {094825} (\bibinfo {year} {2016})}\BibitemShut {NoStop}%
\bibitem [{\citenamefont {Semenova}\ \emph {et~al.}(2016)\citenamefont
  {Semenova}, \citenamefont {Zakharova}, \citenamefont {Anishchenko},\ and\
  \citenamefont {Sch{\"o}ll}}]{Semenova:2016aa}%
  \BibitemOpen
  \bibfield  {author} {\bibinfo {author} {\bibfnamefont {N.}~\bibnamefont
  {Semenova}}, \bibinfo {author} {\bibfnamefont {A.}~\bibnamefont {Zakharova}},
  \bibinfo {author} {\bibfnamefont {V.}~\bibnamefont {Anishchenko}}, \ and\
  \bibinfo {author} {\bibfnamefont {E.}~\bibnamefont {Sch{\"o}ll}},\ }\bibfield
   {title} {\enquote {\bibinfo {title} {Coherence-resonance chimeras in a
  network of excitable elements},}\ }\href@noop {} {\bibfield  {journal}
  {\bibinfo  {journal} {Physical Review Letters}\ }\textbf {\bibinfo {volume}
  {117}},\ \bibinfo {pages} {014102} (\bibinfo {year} {2016})}\BibitemShut
  {NoStop}%
\bibitem [{\citenamefont {Tsigkri-DeSmedt}\ \emph {et~al.}(2016)\citenamefont
  {Tsigkri-DeSmedt}, \citenamefont {Hizanidis}, \citenamefont {H{\"o}vel},\
  and\ \citenamefont {Provata}}]{Tsigkri-DeSmedt:2016wm}%
  \BibitemOpen
  \bibfield  {author} {\bibinfo {author} {\bibfnamefont {N.}~\bibnamefont
  {Tsigkri-DeSmedt}}, \bibinfo {author} {\bibfnamefont {J.}~\bibnamefont
  {Hizanidis}}, \bibinfo {author} {\bibfnamefont {P.}~\bibnamefont
  {H{\"o}vel}}, \ and\ \bibinfo {author} {\bibfnamefont {A.}~\bibnamefont
  {Provata}},\ }\bibfield  {title} {\enquote {\bibinfo {title} {Multi-chimera
  states and transitions in the leaky integrate-and-fire model with nonlocal
  and hierarchical connectivity},}\ }\href@noop {} {\bibfield  {journal}
  {\bibinfo  {journal} {The European Physical Journal Special Topics}\ }\textbf
  {\bibinfo {volume} {225}},\ \bibinfo {pages} {1149--1164} (\bibinfo {year}
  {2016})}\BibitemShut {NoStop}%
\bibitem [{\citenamefont {Buscarino}\ \emph {et~al.}(2015)\citenamefont
  {Buscarino}, \citenamefont {Frasca}, \citenamefont {Gambuzza},\ and\
  \citenamefont {H{\"o}vel}}]{Buscarino:2015vr}%
  \BibitemOpen
  \bibfield  {author} {\bibinfo {author} {\bibfnamefont {A.}~\bibnamefont
  {Buscarino}}, \bibinfo {author} {\bibfnamefont {M.}~\bibnamefont {Frasca}},
  \bibinfo {author} {\bibfnamefont {L.~V.}\ \bibnamefont {Gambuzza}}, \ and\
  \bibinfo {author} {\bibfnamefont {P.}~\bibnamefont {H{\"o}vel}},\ }\bibfield
  {title} {\enquote {\bibinfo {title} {Chimera states in time-varying complex
  networks},}\ }\href@noop {} {\bibfield  {journal} {\bibinfo  {journal}
  {Physical Review E}\ }\textbf {\bibinfo {volume} {91}},\ \bibinfo {pages}
  {022817} (\bibinfo {year} {2015})}\BibitemShut {NoStop}%
\bibitem [{\citenamefont {Omelchenko}\ \emph {et~al.}(2015)\citenamefont
  {Omelchenko}, \citenamefont {Provata}, \citenamefont {Hizanidis},
  \citenamefont {Sch{\"o}ll},\ and\ \citenamefont
  {H{\"o}vel}}]{Omelchenko:2015uu}%
  \BibitemOpen
  \bibfield  {author} {\bibinfo {author} {\bibfnamefont {I.}~\bibnamefont
  {Omelchenko}}, \bibinfo {author} {\bibfnamefont {A.}~\bibnamefont {Provata}},
  \bibinfo {author} {\bibfnamefont {J.}~\bibnamefont {Hizanidis}}, \bibinfo
  {author} {\bibfnamefont {E.}~\bibnamefont {Sch{\"o}ll}}, \ and\ \bibinfo
  {author} {\bibfnamefont {P.}~\bibnamefont {H{\"o}vel}},\ }\bibfield  {title}
  {\enquote {\bibinfo {title} {Robustness of chimera states for coupled
  {F}itz{H}ugh-{N}agumo oscillators},}\ }\href@noop {} {\bibfield  {journal}
  {\bibinfo  {journal} {Physical Review E}\ }\textbf {\bibinfo {volume} {91}},\
  \bibinfo {pages} {022917} (\bibinfo {year} {2015})}\BibitemShut {NoStop}%
\bibitem [{\citenamefont {Banerjee}\ \emph {et~al.}(2016)\citenamefont
  {Banerjee}, \citenamefont {Dutta}, \citenamefont {Zakharova},\ and\
  \citenamefont {Sch{\"o}ll}}]{Banerjee:2016tw}%
  \BibitemOpen
  \bibfield  {author} {\bibinfo {author} {\bibfnamefont {T.}~\bibnamefont
  {Banerjee}}, \bibinfo {author} {\bibfnamefont {P.~S.}\ \bibnamefont {Dutta}},
  \bibinfo {author} {\bibfnamefont {A.}~\bibnamefont {Zakharova}}, \ and\
  \bibinfo {author} {\bibfnamefont {E.}~\bibnamefont {Sch{\"o}ll}},\ }\bibfield
   {title} {\enquote {\bibinfo {title} {Chimera patterns induced by
  distance-dependent power-law coupling in ecological networks},}\ }\href@noop
  {} {\bibfield  {journal} {\bibinfo  {journal} {Physical Review E}\ }\textbf
  {\bibinfo {volume} {94}},\ \bibinfo {pages} {032206} (\bibinfo {year}
  {2016})}\BibitemShut {NoStop}%
\bibitem [{\citenamefont {Ghosh}\ \emph {et~al.}(2016)\citenamefont {Ghosh},
  \citenamefont {Kumar}, \citenamefont {Zakharova},\ and\ \citenamefont
  {Jalan}}]{Ghosh:2016vc}%
  \BibitemOpen
  \bibfield  {author} {\bibinfo {author} {\bibfnamefont {S.}~\bibnamefont
  {Ghosh}}, \bibinfo {author} {\bibfnamefont {A.}~\bibnamefont {Kumar}},
  \bibinfo {author} {\bibfnamefont {A.}~\bibnamefont {Zakharova}}, \ and\
  \bibinfo {author} {\bibfnamefont {S.}~\bibnamefont {Jalan}},\ }\bibfield
  {title} {\enquote {\bibinfo {title} {Birth and death of chimera: Interplay of
  delay and multiplexing},}\ }\href@noop {} {\bibfield  {journal} {\bibinfo
  {journal} {EPL (Europhysics Letters)}\ }\textbf {\bibinfo {volume} {115}},\
  \bibinfo {pages} {60005} (\bibinfo {year} {2016})}\BibitemShut {NoStop}%
\bibitem [{\citenamefont {Majhi}, \citenamefont {Perc},\ and\ \citenamefont
  {Ghosh}(2017)}]{Majhi:2017to}%
  \BibitemOpen
  \bibfield  {author} {\bibinfo {author} {\bibfnamefont {S.}~\bibnamefont
  {Majhi}}, \bibinfo {author} {\bibfnamefont {M.}~\bibnamefont {Perc}}, \ and\
  \bibinfo {author} {\bibfnamefont {D.}~\bibnamefont {Ghosh}},\ }\bibfield
  {title} {\enquote {\bibinfo {title} {Chimera states in a multilayer network
  of coupled and uncoupled neurons},}\ }\href@noop {} {\bibfield  {journal}
  {\bibinfo  {journal} {Chaos: an interdisciplinary journal of nonlinear
  science}\ }\textbf {\bibinfo {volume} {27}},\ \bibinfo {pages} {073109 
  1054--1500} (\bibinfo {year} {2017})}\BibitemShut {NoStop}%
\bibitem [{\citenamefont {Kasatkin}\ \emph {et~al.}(2017)\citenamefont
  {Kasatkin}, \citenamefont {Yanchuk}, \citenamefont {Sch{\"o}ll},\ and\
  \citenamefont {Nekorkin}}]{Kasatkin:2017wl}%
  \BibitemOpen
  \bibfield  {author} {\bibinfo {author} {\bibfnamefont {D.}~\bibnamefont
  {Kasatkin}}, \bibinfo {author} {\bibfnamefont {S.}~\bibnamefont {Yanchuk}},
  \bibinfo {author} {\bibfnamefont {E.}~\bibnamefont {Sch{\"o}ll}}, \ and\
  \bibinfo {author} {\bibfnamefont {V.}~\bibnamefont {Nekorkin}},\ }\bibfield
  {title} {\enquote {\bibinfo {title} {Self-organized emergence of multilayer
  structure and chimera states in dynamical networks with adaptive
  couplings},}\ }\href@noop {} {\bibfield  {journal} {\bibinfo  {journal}
  {Physical Review E}\ }\textbf {\bibinfo {volume} {96}},\ \bibinfo {pages}
  {062211} (\bibinfo {year} {2017})}\BibitemShut {NoStop}%
\bibitem [{\citenamefont {Bukh}\ \emph {et~al.}(2017)\citenamefont {Bukh},
  \citenamefont {Rybalova}, \citenamefont {Semenova}, \citenamefont
  {Strelkova},\ and\ \citenamefont {Anishchenko}}]{Bukh:2017vp}%
  \BibitemOpen
  \bibfield  {author} {\bibinfo {author} {\bibfnamefont {A.}~\bibnamefont
  {Bukh}}, \bibinfo {author} {\bibfnamefont {E.}~\bibnamefont {Rybalova}},
  \bibinfo {author} {\bibfnamefont {N.}~\bibnamefont {Semenova}}, \bibinfo
  {author} {\bibfnamefont {G.}~\bibnamefont {Strelkova}}, \ and\ \bibinfo
  {author} {\bibfnamefont {V.}~\bibnamefont {Anishchenko}},\ }\bibfield
  {title} {\enquote {\bibinfo {title} {New type of chimera and mutual
  synchronization of spatiotemporal structures in two coupled ensembles of
  nonlocally interacting chaotic maps},}\ }\href@noop {} {\bibfield  {journal}
  {\bibinfo  {journal} {Chaos: An Interdisciplinary Journal of Nonlinear
  Science}\ }\textbf {\bibinfo {volume} {27}},\ \bibinfo {pages} {111102}
  (\bibinfo {year} {2017})}\BibitemShut {NoStop}%
\bibitem [{\citenamefont {Sawicki}\ \emph {et~al.}(2017)\citenamefont
  {Sawicki}, \citenamefont {Omelchenko}, \citenamefont {Zakharova},\ and\
  \citenamefont {Sch{\"o}ll}}]{Sawicki:2017um}%
  \BibitemOpen
  \bibfield  {author} {\bibinfo {author} {\bibfnamefont {J.}~\bibnamefont
  {Sawicki}}, \bibinfo {author} {\bibfnamefont {I.}~\bibnamefont {Omelchenko}},
  \bibinfo {author} {\bibfnamefont {A.}~\bibnamefont {Zakharova}}, \ and\
  \bibinfo {author} {\bibfnamefont {E.}~\bibnamefont {Sch{\"o}ll}},\ }\bibfield
   {title} {\enquote {\bibinfo {title} {Chimera states in complex networks:
  interplay of fractal topology and delay},}\ }\href@noop {} {\bibfield
  {journal} {\bibinfo  {journal} {The European Physical Journal Special
  Topics}\ }\textbf {\bibinfo {volume} {226}},\ \bibinfo {pages} {1883--1892}
  (\bibinfo {year} {2017})}\BibitemShut {NoStop}%
\bibitem [{\citenamefont {Hagerstrom}\ \emph {et~al.}(2012)\citenamefont
  {Hagerstrom}, \citenamefont {Murphy}, \citenamefont {Roy}, \citenamefont
  {H{\"o}vel}, \citenamefont {Omelchenko},\ and\ \citenamefont
  {Sch{\"o}ll}}]{Hagerstrom:2012vd}%
  \BibitemOpen
  \bibfield  {author} {\bibinfo {author} {\bibfnamefont {A.~M.}\ \bibnamefont
  {Hagerstrom}}, \bibinfo {author} {\bibfnamefont {T.~E.}\ \bibnamefont
  {Murphy}}, \bibinfo {author} {\bibfnamefont {R.}~\bibnamefont {Roy}},
  \bibinfo {author} {\bibfnamefont {P.}~\bibnamefont {H{\"o}vel}}, \bibinfo
  {author} {\bibfnamefont {I.}~\bibnamefont {Omelchenko}}, \ and\ \bibinfo
  {author} {\bibfnamefont {E.}~\bibnamefont {Sch{\"o}ll}},\ }\bibfield  {title}
  {\enquote {\bibinfo {title} {Experimental observation of chimeras in
  coupled-map lattices},}\ }\href@noop {} {\bibfield  {journal} {\bibinfo
  {journal} {Nature Physics}\ }\textbf {\bibinfo {volume} {8}},\ \bibinfo
  {pages} {658--661} (\bibinfo {year} {2012})}\BibitemShut {NoStop}%
\bibitem [{\citenamefont {Larger}, \citenamefont {Penkovsky},\ and\
  \citenamefont {Maistrenko}(2013)}]{Larger:2013ub}%
  \BibitemOpen
  \bibfield  {author} {\bibinfo {author} {\bibfnamefont {L.}~\bibnamefont
  {Larger}}, \bibinfo {author} {\bibfnamefont {B.}~\bibnamefont {Penkovsky}}, \
  and\ \bibinfo {author} {\bibfnamefont {Y.}~\bibnamefont {Maistrenko}},\
  }\bibfield  {title} {\enquote {\bibinfo {title} {Virtual chimera states for
  delayed-feedback systems},}\ }\href@noop {} {\bibfield  {journal} {\bibinfo
  {journal} {Physical review letters}\ }\textbf {\bibinfo {volume} {111}},\
  \bibinfo {pages} {054103} (\bibinfo {year} {2013})}\BibitemShut {NoStop}%
\bibitem [{\citenamefont {Martens}\ \emph {et~al.}(2013)\citenamefont
  {Martens}, \citenamefont {Thutupalli}, \citenamefont {Fourriere},\ and\
  \citenamefont {Hallatschek}}]{Martens:2013wq}%
  \BibitemOpen
  \bibfield  {author} {\bibinfo {author} {\bibfnamefont {E.~A.}\ \bibnamefont
  {Martens}}, \bibinfo {author} {\bibfnamefont {S.}~\bibnamefont {Thutupalli}},
  \bibinfo {author} {\bibfnamefont {A.}~\bibnamefont {Fourriere}}, \ and\
  \bibinfo {author} {\bibfnamefont {O.}~\bibnamefont {Hallatschek}},\
  }\bibfield  {title} {\enquote {\bibinfo {title} {Chimera states in mechanical
  oscillator networks},}\ }\href@noop {} {\bibfield  {journal} {\bibinfo
  {journal} {Proceedings of the National Academy of Sciences}\ }\textbf
  {\bibinfo {volume} {110}},\ \bibinfo {pages} {10563--10567} (\bibinfo {year}
  {2013})}\BibitemShut {NoStop}%
\bibitem [{\citenamefont {Kapitaniak}\ \emph {et~al.}(2014)\citenamefont
  {Kapitaniak}, \citenamefont {Kuzma}, \citenamefont {Wojewoda}, \citenamefont
  {Czolczynski},\ and\ \citenamefont {Maistrenko}}]{Kapitaniak:2014vc}%
  \BibitemOpen
  \bibfield  {author} {\bibinfo {author} {\bibfnamefont {T.}~\bibnamefont
  {Kapitaniak}}, \bibinfo {author} {\bibfnamefont {P.}~\bibnamefont {Kuzma}},
  \bibinfo {author} {\bibfnamefont {J.}~\bibnamefont {Wojewoda}}, \bibinfo
  {author} {\bibfnamefont {K.}~\bibnamefont {Czolczynski}}, \ and\ \bibinfo
  {author} {\bibfnamefont {Y.}~\bibnamefont {Maistrenko}},\ }\bibfield  {title}
  {\enquote {\bibinfo {title} {Imperfect chimera states for coupled pendula},}\
  }\href@noop {} {\bibfield  {journal} {\bibinfo  {journal} {Scientific
  reports}\ }\textbf {\bibinfo {volume} {4}},\ \bibinfo {pages} {6379}
  (\bibinfo {year} {2014})}\BibitemShut {NoStop}%
\bibitem [{\citenamefont {Gambuzza}\ \emph {et~al.}(2014)\citenamefont
  {Gambuzza}, \citenamefont {Buscarino}, \citenamefont {Chessari},
  \citenamefont {Fortuna}, \citenamefont {Meucci},\ and\ \citenamefont
  {Frasca}}]{Gambuzza:2014vj}%
  \BibitemOpen
  \bibfield  {author} {\bibinfo {author} {\bibfnamefont {L.~V.}\ \bibnamefont
  {Gambuzza}}, \bibinfo {author} {\bibfnamefont {A.}~\bibnamefont {Buscarino}},
  \bibinfo {author} {\bibfnamefont {S.}~\bibnamefont {Chessari}}, \bibinfo
  {author} {\bibfnamefont {L.}~\bibnamefont {Fortuna}}, \bibinfo {author}
  {\bibfnamefont {R.}~\bibnamefont {Meucci}}, \ and\ \bibinfo {author}
  {\bibfnamefont {M.}~\bibnamefont {Frasca}},\ }\bibfield  {title} {\enquote
  {\bibinfo {title} {Experimental investigation of chimera states with
  quiescent and synchronous domains in coupled electronic oscillators},}\
  }\href@noop {} {\bibfield  {journal} {\bibinfo  {journal} {Physical Review
  E}\ }\textbf {\bibinfo {volume} {90}},\ \bibinfo {pages} {032905} (\bibinfo
  {year} {2014})}\BibitemShut {NoStop}%
\bibitem [{\citenamefont {Rosin}\ \emph {et~al.}(2014)\citenamefont {Rosin},
  \citenamefont {Rontani}, \citenamefont {Haynes}, \citenamefont {Sch{\"o}ll},\
  and\ \citenamefont {Gauthier}}]{Rosin:2014wy}%
  \BibitemOpen
  \bibfield  {author} {\bibinfo {author} {\bibfnamefont {D.~P.}\ \bibnamefont
  {Rosin}}, \bibinfo {author} {\bibfnamefont {D.}~\bibnamefont {Rontani}},
  \bibinfo {author} {\bibfnamefont {N.~D.}\ \bibnamefont {Haynes}}, \bibinfo
  {author} {\bibfnamefont {E.}~\bibnamefont {Sch{\"o}ll}}, \ and\ \bibinfo
  {author} {\bibfnamefont {D.~J.}\ \bibnamefont {Gauthier}},\ }\bibfield
  {title} {\enquote {\bibinfo {title} {Transient scaling and resurgence of
  chimera states in networks of {B}oolean phase oscillators},}\ }\href@noop {}
  {\bibfield  {journal} {\bibinfo  {journal} {Physical Review E}\ }\textbf
  {\bibinfo {volume} {90}},\ \bibinfo {pages} {030902} (\bibinfo {year}
  {2014})}\BibitemShut {NoStop}%
\bibitem [{\citenamefont {Schmidt}\ \emph {et~al.}(2014)\citenamefont
  {Schmidt}, \citenamefont {Sch{\"o}nleber}, \citenamefont {Krischer},\ and\
  \citenamefont {Garc{\'\i}a-Morales}}]{Schmidt:2014ui}%
  \BibitemOpen
  \bibfield  {author} {\bibinfo {author} {\bibfnamefont {L.}~\bibnamefont
  {Schmidt}}, \bibinfo {author} {\bibfnamefont {K.}~\bibnamefont
  {Sch{\"o}nleber}}, \bibinfo {author} {\bibfnamefont {K.}~\bibnamefont
  {Krischer}}, \ and\ \bibinfo {author} {\bibfnamefont {V.}~\bibnamefont
  {Garc{\'\i}a-Morales}},\ }\bibfield  {title} {\enquote {\bibinfo {title}
  {Coexistence of synchrony and incoherence in oscillatory media under
  nonlinear global coupling},}\ }\href@noop {} {\bibfield  {journal} {\bibinfo
  {journal} {Chaos: An Interdisciplinary Journal of Nonlinear Science}\
  }\textbf {\bibinfo {volume} {24}},\ \bibinfo {pages} {013102} (\bibinfo
  {year} {2014})}\BibitemShut {NoStop}%
\bibitem [{\citenamefont {Tinsley}, \citenamefont {Nkomo},\ and\ \citenamefont
  {Showalter}(2012)}]{Tinsley:2012tc}%
  \BibitemOpen
  \bibfield  {author} {\bibinfo {author} {\bibfnamefont {M.~R.}\ \bibnamefont
  {Tinsley}}, \bibinfo {author} {\bibfnamefont {S.}~\bibnamefont {Nkomo}}, \
  and\ \bibinfo {author} {\bibfnamefont {K.}~\bibnamefont {Showalter}},\
  }\bibfield  {title} {\enquote {\bibinfo {title} {Chimera and phase-cluster
  states in populations of coupled chemical oscillators},}\ }\href@noop {}
  {\bibfield  {journal} {\bibinfo  {journal} {Nature Physics}\ }\textbf
  {\bibinfo {volume} {8}},\ \bibinfo {pages} {662--665} (\bibinfo {year}
  {2012})}\BibitemShut {NoStop}%
\bibitem [{\citenamefont {Wickramasinghe}\ and\ \citenamefont
  {Kiss}(2013)}]{Wickramasinghe:2013tp}%
  \BibitemOpen
  \bibfield  {author} {\bibinfo {author} {\bibfnamefont {M.}~\bibnamefont
  {Wickramasinghe}}\ and\ \bibinfo {author} {\bibfnamefont {I.~Z.}\
  \bibnamefont {Kiss}},\ }\bibfield  {title} {\enquote {\bibinfo {title}
  {Spatially organized dynamical states in chemical oscillator networks:
  Synchronization, dynamical differentiation, and chimera patterns},}\
  }\href@noop {} {\bibfield  {journal} {\bibinfo  {journal} {PloS one}\
  }\textbf {\bibinfo {volume} {8}},\ \bibinfo {pages} {e80586} (\bibinfo {year}
  {2013})}\BibitemShut {NoStop}%
\bibitem [{\citenamefont {Shena}\ \emph {et~al.}(2017)\citenamefont {Shena},
  \citenamefont {Hizanidis}, \citenamefont {Kovanis},\ and\ \citenamefont
  {Tsironis}}]{Shena:2017ul}%
  \BibitemOpen
  \bibfield  {author} {\bibinfo {author} {\bibfnamefont {J.}~\bibnamefont
  {Shena}}, \bibinfo {author} {\bibfnamefont {J.}~\bibnamefont {Hizanidis}},
  \bibinfo {author} {\bibfnamefont {V.}~\bibnamefont {Kovanis}}, \ and\
  \bibinfo {author} {\bibfnamefont {G.~P.}\ \bibnamefont {Tsironis}},\
  }\bibfield  {title} {\enquote {\bibinfo {title} {Turbulent chimeras in large
  semiconductor laser arrays},}\ }\href@noop {} {\bibfield  {journal} {\bibinfo
   {journal} {Scientific reports}\ }\textbf {\bibinfo {volume} {7}},\ \bibinfo
  {pages} {1--8} (\bibinfo {year} {2017})}\BibitemShut {NoStop}%
\bibitem [{\citenamefont {Maistrenko}, \citenamefont {Penkovsky},\ and\
  \citenamefont {Rosenblum}(2014)}]{Maistrenko:2014tm}%
  \BibitemOpen
  \bibfield  {author} {\bibinfo {author} {\bibfnamefont {Y.}~\bibnamefont
  {Maistrenko}}, \bibinfo {author} {\bibfnamefont {B.}~\bibnamefont
  {Penkovsky}}, \ and\ \bibinfo {author} {\bibfnamefont {M.}~\bibnamefont
  {Rosenblum}},\ }\bibfield  {title} {\enquote {\bibinfo {title} {Solitary
  state at the edge of synchrony in ensembles with attractive and repulsive
  interactions},}\ }\href@noop {} {\bibfield  {journal} {\bibinfo  {journal}
  {Physical Review E}\ }\textbf {\bibinfo {volume} {89}},\ \bibinfo {pages}
  {060901} (\bibinfo {year} {2014})}\BibitemShut {NoStop}%
\bibitem [{\citenamefont {Berner}\ \emph {et~al.}(2020)\citenamefont {Berner},
  \citenamefont {Polanska}, \citenamefont {Sch{\"o}ll},\ and\ \citenamefont
  {Yanchuk}}]{Berner:2020um}%
  \BibitemOpen
  \bibfield  {author} {\bibinfo {author} {\bibfnamefont {R.}~\bibnamefont
  {Berner}}, \bibinfo {author} {\bibfnamefont {A.}~\bibnamefont {Polanska}},
  \bibinfo {author} {\bibfnamefont {E.}~\bibnamefont {Sch{\"o}ll}}, \ and\
  \bibinfo {author} {\bibfnamefont {S.}~\bibnamefont {Yanchuk}},\ }\bibfield
  {title} {\enquote {\bibinfo {title} {Solitary states in adaptive nonlocal
  oscillator networks},}\ }\href@noop {} {\bibfield  {journal} {\bibinfo
  {journal} {The European Physical Journal Special Topics}\ }\textbf {\bibinfo
  {volume} {229}},\ \bibinfo {pages} {2183--2203} (\bibinfo {year}
  {2020})}\BibitemShut {NoStop}%
\bibitem [{\citenamefont {Jaros}\ \emph {et~al.}(2018)\citenamefont {Jaros},
  \citenamefont {Brezetsky}, \citenamefont {Levchenko}, \citenamefont
  {Dudkowski}, \citenamefont {Kapitaniak},\ and\ \citenamefont
  {Maistrenko}}]{Jaros:2018ve}%
  \BibitemOpen
  \bibfield  {author} {\bibinfo {author} {\bibfnamefont {P.}~\bibnamefont
  {Jaros}}, \bibinfo {author} {\bibfnamefont {S.}~\bibnamefont {Brezetsky}},
  \bibinfo {author} {\bibfnamefont {R.}~\bibnamefont {Levchenko}}, \bibinfo
  {author} {\bibfnamefont {D.}~\bibnamefont {Dudkowski}}, \bibinfo {author}
  {\bibfnamefont {T.}~\bibnamefont {Kapitaniak}}, \ and\ \bibinfo {author}
  {\bibfnamefont {Y.}~\bibnamefont {Maistrenko}},\ }\bibfield  {title}
  {\enquote {\bibinfo {title} {Solitary states for coupled oscillators with
  inertia},}\ }\href@noop {} {\bibfield  {journal} {\bibinfo  {journal} {Chaos:
  An Interdisciplinary Journal of Nonlinear Science}\ }\textbf {\bibinfo
  {volume} {28}},\ \bibinfo {pages} {011103} (\bibinfo {year}
  {2018})}\BibitemShut {NoStop}%
\bibitem [{\citenamefont {Wu}\ and\ \citenamefont {Dhamala}(2018)}]{Wu:2018ws}%
  \BibitemOpen
  \bibfield  {author} {\bibinfo {author} {\bibfnamefont {H.}~\bibnamefont
  {Wu}}\ and\ \bibinfo {author} {\bibfnamefont {M.}~\bibnamefont {Dhamala}},\
  }\bibfield  {title} {\enquote {\bibinfo {title} {Dynamics of kuramoto
  oscillators with time-delayed positive and negative couplings},}\ }\href@noop
  {} {\bibfield  {journal} {\bibinfo  {journal} {Physical Review E}\ }\textbf
  {\bibinfo {volume} {98}},\ \bibinfo {pages} {032221} (\bibinfo {year}
  {2018})}\BibitemShut {NoStop}%
\bibitem [{\citenamefont {Semenova}\ \emph {et~al.}(2015)\citenamefont
  {Semenova}, \citenamefont {Zakharova}, \citenamefont {Sch{\"o}ll},\ and\
  \citenamefont {Anishchenko}}]{Semenova:2015tt}%
  \BibitemOpen
  \bibfield  {author} {\bibinfo {author} {\bibfnamefont {N.}~\bibnamefont
  {Semenova}}, \bibinfo {author} {\bibfnamefont {A.}~\bibnamefont {Zakharova}},
  \bibinfo {author} {\bibfnamefont {E.}~\bibnamefont {Sch{\"o}ll}}, \ and\
  \bibinfo {author} {\bibfnamefont {V.}~\bibnamefont {Anishchenko}},\
  }\bibfield  {title} {\enquote {\bibinfo {title} {Does hyperbolicity impede
  emergence of chimera states in networks of nonlocally coupled chaotic
  oscillators?}}\ }\href@noop {} {\bibfield  {journal} {\bibinfo  {journal}
  {EPL (Europhysics Letters)}\ }\textbf {\bibinfo {volume} {112}},\ \bibinfo
  {pages} {40002} (\bibinfo {year} {2015})}\BibitemShut {NoStop}%
\bibitem [{\citenamefont {Rybalova}\ \emph {et~al.}(2017)\citenamefont
  {Rybalova}, \citenamefont {Semenova}, \citenamefont {Strelkova},\ and\
  \citenamefont {Anishchenko}}]{Rybalova:2017tl}%
  \BibitemOpen
  \bibfield  {author} {\bibinfo {author} {\bibfnamefont {E.}~\bibnamefont
  {Rybalova}}, \bibinfo {author} {\bibfnamefont {N.}~\bibnamefont {Semenova}},
  \bibinfo {author} {\bibfnamefont {G.}~\bibnamefont {Strelkova}}, \ and\
  \bibinfo {author} {\bibfnamefont {V.}~\bibnamefont {Anishchenko}},\
  }\bibfield  {title} {\enquote {\bibinfo {title} {Transition from complete
  synchronization to spatio-temporal chaos in coupled chaotic systems with
  nonhyperbolic and hyperbolic attractors},}\ }\href@noop {} {\bibfield
  {journal} {\bibinfo  {journal} {The European Physical Journal Special
  Topics}\ }\textbf {\bibinfo {volume} {226}},\ \bibinfo {pages} {1857--1866}
  (\bibinfo {year} {2017})}\BibitemShut {NoStop}%
\bibitem [{\citenamefont {Semenova}\ \emph
  {et~al.}(2017{\natexlab{b}})\citenamefont {Semenova}, \citenamefont
  {Rybalova}, \citenamefont {Strelkova},\ and\ \citenamefont
  {Anishchenko}}]{Semenova:2017wn}%
  \BibitemOpen
  \bibfield  {author} {\bibinfo {author} {\bibfnamefont {N.~I.}\ \bibnamefont
  {Semenova}}, \bibinfo {author} {\bibfnamefont {E.~V.}\ \bibnamefont
  {Rybalova}}, \bibinfo {author} {\bibfnamefont {G.~I.}\ \bibnamefont
  {Strelkova}}, \ and\ \bibinfo {author} {\bibfnamefont {V.~S.}\ \bibnamefont
  {Anishchenko}},\ }\bibfield  {title} {\enquote {\bibinfo {title}
  {''{C}oherence--incoherence'' transition in ensembles of nonlocally coupled
  chaotic oscillators with nonhyperbolic and hyperbolic attractors},}\
  }\href@noop {} {\bibfield  {journal} {\bibinfo  {journal} {Regular and
  Chaotic Dynamics}\ }\textbf {\bibinfo {volume} {22}},\ \bibinfo {pages}
  {148--162} (\bibinfo {year} {2017}{\natexlab{b}})}\BibitemShut {NoStop}%
\bibitem [{\citenamefont {Semenova}, \citenamefont {Vadivasova},\ and\
  \citenamefont {Anishchenko}(2018)}]{Semenova:2018th}%
  \BibitemOpen
  \bibfield  {author} {\bibinfo {author} {\bibfnamefont {N.}~\bibnamefont
  {Semenova}}, \bibinfo {author} {\bibfnamefont {T.}~\bibnamefont
  {Vadivasova}}, \ and\ \bibinfo {author} {\bibfnamefont {V.}~\bibnamefont
  {Anishchenko}},\ }\bibfield  {title} {\enquote {\bibinfo {title} {Mechanism
  of solitary state appearance in an ensemble of nonlocally coupled {L}ozi
  maps},}\ }\href@noop {} {\bibfield  {journal} {\bibinfo  {journal} {The
  European Physical Journal Special Topics}\ }\textbf {\bibinfo {volume}
  {227}},\ \bibinfo {pages} {1173--1183} (\bibinfo {year} {2018})}\BibitemShut
  {NoStop}%
\bibitem [{\citenamefont {Mikhaylenko}\ \emph {et~al.}(2019)\citenamefont
  {Mikhaylenko}, \citenamefont {Ramlow}, \citenamefont {Jalan},\ and\
  \citenamefont {Zakharova}}]{Mikhaylenko:2019uq}%
  \BibitemOpen
  \bibfield  {author} {\bibinfo {author} {\bibfnamefont {M.}~\bibnamefont
  {Mikhaylenko}}, \bibinfo {author} {\bibfnamefont {L.}~\bibnamefont {Ramlow}},
  \bibinfo {author} {\bibfnamefont {S.}~\bibnamefont {Jalan}}, \ and\ \bibinfo
  {author} {\bibfnamefont {A.}~\bibnamefont {Zakharova}},\ }\bibfield  {title}
  {\enquote {\bibinfo {title} {Weak multiplexing in neural networks: Switching
  between chimera and solitary states},}\ }\href@noop {} {\bibfield  {journal}
  {\bibinfo  {journal} {Chaos: An Interdisciplinary Journal of Nonlinear
  Science}\ }\textbf {\bibinfo {volume} {29}},\ \bibinfo {pages} {023122}
  (\bibinfo {year} {2019})}\BibitemShut {NoStop}%
\bibitem [{\citenamefont {Rybalova}\ \emph
  {et~al.}(2019{\natexlab{a}})\citenamefont {Rybalova}, \citenamefont
  {Anishchenko}, \citenamefont {Strelkova},\ and\ \citenamefont
  {Zakharova}}]{Rybalova:2019vg}%
  \BibitemOpen
  \bibfield  {author} {\bibinfo {author} {\bibfnamefont {E.}~\bibnamefont
  {Rybalova}}, \bibinfo {author} {\bibfnamefont {V.}~\bibnamefont
  {Anishchenko}}, \bibinfo {author} {\bibfnamefont {G.}~\bibnamefont
  {Strelkova}}, \ and\ \bibinfo {author} {\bibfnamefont {A.}~\bibnamefont
  {Zakharova}},\ }\bibfield  {title} {\enquote {\bibinfo {title} {Solitary
  states and solitary state chimera in neural networks},}\ }\href@noop {}
  {\bibfield  {journal} {\bibinfo  {journal} {Chaos: An Interdisciplinary
  Journal of Nonlinear Science}\ }\textbf {\bibinfo {volume} {29}},\ \bibinfo
  {pages} {071106} (\bibinfo {year} {2019}{\natexlab{a}})}\BibitemShut
  {NoStop}%
\bibitem [{\citenamefont {Sch{\"u}len}\ \emph {et~al.}(2019)\citenamefont
  {Sch{\"u}len}, \citenamefont {Ghosh}, \citenamefont {Kachhvah}, \citenamefont
  {Zakharova},\ and\ \citenamefont {Jalan}}]{Schulen:2019td}%
  \BibitemOpen
  \bibfield  {author} {\bibinfo {author} {\bibfnamefont {L.}~\bibnamefont
  {Sch{\"u}len}}, \bibinfo {author} {\bibfnamefont {S.}~\bibnamefont {Ghosh}},
  \bibinfo {author} {\bibfnamefont {A.~D.}\ \bibnamefont {Kachhvah}}, \bibinfo
  {author} {\bibfnamefont {A.}~\bibnamefont {Zakharova}}, \ and\ \bibinfo
  {author} {\bibfnamefont {S.}~\bibnamefont {Jalan}},\ }\bibfield  {title}
  {\enquote {\bibinfo {title} {Delay engineered solitary states in complex
  networks},}\ }\href@noop {} {\bibfield  {journal} {\bibinfo  {journal}
  {Chaos, Solitons \& Fractals}\ }\textbf {\bibinfo {volume} {128}},\ \bibinfo
  {pages} {290--296} (\bibinfo {year} {2019})}\BibitemShut {NoStop}%
\bibitem [{\citenamefont {Sch{\"u}len}\ \emph {et~al.}(2021)\citenamefont
  {Sch{\"u}len}, \citenamefont {Janzen}, \citenamefont {Medeiros},\ and\
  \citenamefont {Zakharova}}]{Schulen:2021tn}%
  \BibitemOpen
  \bibfield  {author} {\bibinfo {author} {\bibfnamefont {L.}~\bibnamefont
  {Sch{\"u}len}}, \bibinfo {author} {\bibfnamefont {D.~A.}\ \bibnamefont
  {Janzen}}, \bibinfo {author} {\bibfnamefont {E.~S.}\ \bibnamefont
  {Medeiros}}, \ and\ \bibinfo {author} {\bibfnamefont {A.}~\bibnamefont
  {Zakharova}},\ }\bibfield  {title} {\enquote {\bibinfo {title} {Solitary
  states in multiplex neural networks: Onset and vulnerability},}\ }\href@noop
  {} {\bibfield  {journal} {\bibinfo  {journal} {Chaos, Solitons \& Fractals}\
  }\textbf {\bibinfo {volume} {145}},\ \bibinfo {pages} {110670} (\bibinfo
  {year} {2021})}\BibitemShut {NoStop}%
\bibitem [{\citenamefont {Taher}, \citenamefont {Olmi},\ and\ \citenamefont
  {Sch{\"o}ll}(2019)}]{Taher:2019wz}%
  \BibitemOpen
  \bibfield  {author} {\bibinfo {author} {\bibfnamefont {H.}~\bibnamefont
  {Taher}}, \bibinfo {author} {\bibfnamefont {S.}~\bibnamefont {Olmi}}, \ and\
  \bibinfo {author} {\bibfnamefont {E.}~\bibnamefont {Sch{\"o}ll}},\ }\bibfield
   {title} {\enquote {\bibinfo {title} {Enhancing power grid synchronization
  and stability through time-delayed feedback control},}\ }\href@noop {}
  {\bibfield  {journal} {\bibinfo  {journal} {Physical Review E}\ }\textbf
  {\bibinfo {volume} {100}},\ \bibinfo {pages} {062306} (\bibinfo {year}
  {2019})}\BibitemShut {NoStop}%
\bibitem [{\citenamefont {Hellmann}\ \emph {et~al.}(2020)\citenamefont
  {Hellmann}, \citenamefont {Schultz}, \citenamefont {Jaros}, \citenamefont
  {Levchenko}, \citenamefont {Kapitaniak}, \citenamefont {Kurths},\ and\
  \citenamefont {Maistrenko}}]{Hellmann:2020wr}%
  \BibitemOpen
  \bibfield  {author} {\bibinfo {author} {\bibfnamefont {F.}~\bibnamefont
  {Hellmann}}, \bibinfo {author} {\bibfnamefont {P.}~\bibnamefont {Schultz}},
  \bibinfo {author} {\bibfnamefont {P.}~\bibnamefont {Jaros}}, \bibinfo
  {author} {\bibfnamefont {R.}~\bibnamefont {Levchenko}}, \bibinfo {author}
  {\bibfnamefont {T.}~\bibnamefont {Kapitaniak}}, \bibinfo {author}
  {\bibfnamefont {J.}~\bibnamefont {Kurths}}, \ and\ \bibinfo {author}
  {\bibfnamefont {Y.}~\bibnamefont {Maistrenko}},\ }\bibfield  {title}
  {\enquote {\bibinfo {title} {Network-induced multistability through lossy
  coupling and exotic solitary states},}\ }\href@noop {} {\bibfield  {journal}
  {\bibinfo  {journal} {Nature communications}\ }\textbf {\bibinfo {volume}
  {11}},\ \bibinfo {pages} {1--9} (\bibinfo {year} {2020})}\BibitemShut
  {NoStop}%
\bibitem [{\citenamefont {Berner}, \citenamefont {Yanchuk},\ and\ \citenamefont
  {Sch{\"o}ll}(2021)}]{Berner:2021wd}%
  \BibitemOpen
  \bibfield  {author} {\bibinfo {author} {\bibfnamefont {R.}~\bibnamefont
  {Berner}}, \bibinfo {author} {\bibfnamefont {S.}~\bibnamefont {Yanchuk}}, \
  and\ \bibinfo {author} {\bibfnamefont {E.}~\bibnamefont {Sch{\"o}ll}},\
  }\bibfield  {title} {\enquote {\bibinfo {title} {What adaptive neuronal
  networks teach us about power grids},}\ }\href@noop {} {\bibfield  {journal}
  {\bibinfo  {journal} {Physical Review E}\ }\textbf {\bibinfo {volume}
  {103}},\ \bibinfo {pages} {042315} (\bibinfo {year} {2021})}\BibitemShut
  {NoStop}%
\bibitem [{\citenamefont {Kemeth}\ \emph {et~al.}(2016)\citenamefont {Kemeth},
  \citenamefont {Haugland}, \citenamefont {Schmidt}, \citenamefont
  {Kevrekidis},\ and\ \citenamefont {Krischer}}]{Kemeth:2016vc}%
  \BibitemOpen
  \bibfield  {author} {\bibinfo {author} {\bibfnamefont {F.~P.}\ \bibnamefont
  {Kemeth}}, \bibinfo {author} {\bibfnamefont {S.~W.}\ \bibnamefont
  {Haugland}}, \bibinfo {author} {\bibfnamefont {L.}~\bibnamefont {Schmidt}},
  \bibinfo {author} {\bibfnamefont {I.~G.}\ \bibnamefont {Kevrekidis}}, \ and\
  \bibinfo {author} {\bibfnamefont {K.}~\bibnamefont {Krischer}},\ }\bibfield
  {title} {\enquote {\bibinfo {title} {A classification scheme for chimera
  states},}\ }\href@noop {} {\bibfield  {journal} {\bibinfo  {journal} {Chaos:
  An Interdisciplinary Journal of Nonlinear Science}\ }\textbf {\bibinfo
  {volume} {26}},\ \bibinfo {pages} {094815} (\bibinfo {year}
  {2016})}\BibitemShut {NoStop}%
\bibitem [{\citenamefont {Zakharova}, \citenamefont {Kapeller},\ and\
  \citenamefont {Sch{\"o}ll}(2016)}]{Zakharova:2016vm}%
  \BibitemOpen
  \bibfield  {author} {\bibinfo {author} {\bibfnamefont {A.}~\bibnamefont
  {Zakharova}}, \bibinfo {author} {\bibfnamefont {M.}~\bibnamefont {Kapeller}},
  \ and\ \bibinfo {author} {\bibfnamefont {E.}~\bibnamefont {Sch{\"o}ll}},\
  }\bibfield  {title} {\enquote {\bibinfo {title} {Amplitude chimeras and
  chimera death in dynamical networks},}\ \ }(\bibinfo  {publisher} {IOP
  Publishing},\ \bibinfo {year} {2016})\ p.\ \bibinfo {pages}
  {012018}\BibitemShut {NoStop}%
\bibitem [{\citenamefont {Bogomolov}\ \emph {et~al.}(2017)\citenamefont
  {Bogomolov}, \citenamefont {Slepnev}, \citenamefont {Strelkova},
  \citenamefont {Sch{\"o}ll},\ and\ \citenamefont
  {Anishchenko}}]{Bogomolov:2017wq}%
  \BibitemOpen
  \bibfield  {author} {\bibinfo {author} {\bibfnamefont {S.~A.}\ \bibnamefont
  {Bogomolov}}, \bibinfo {author} {\bibfnamefont {A.~V.}\ \bibnamefont
  {Slepnev}}, \bibinfo {author} {\bibfnamefont {G.~I.}\ \bibnamefont
  {Strelkova}}, \bibinfo {author} {\bibfnamefont {E.}~\bibnamefont
  {Sch{\"o}ll}}, \ and\ \bibinfo {author} {\bibfnamefont {V.~S.}\ \bibnamefont
  {Anishchenko}},\ }\bibfield  {title} {\enquote {\bibinfo {title} {Mechanisms
  of appearance of amplitude and phase chimera states in ensembles of
  nonlocally coupled chaotic systems},}\ }\href@noop {} {\bibfield  {journal}
  {\bibinfo  {journal} {Communications in Nonlinear Science and Numerical
  Simulation}\ }\textbf {\bibinfo {volume} {43}},\ \bibinfo {pages} {25--36}
  (\bibinfo {year} {2017})}\BibitemShut {NoStop}%
\bibitem [{\citenamefont {Shepelev}\ \emph {et~al.}(2017)\citenamefont
  {Shepelev}, \citenamefont {Bukh}, \citenamefont {Strelkova}, \citenamefont
  {Vadivasova},\ and\ \citenamefont {Anishchenko}}]{Shepelev:2017uy}%
  \BibitemOpen
  \bibfield  {author} {\bibinfo {author} {\bibfnamefont {I.~A.}\ \bibnamefont
  {Shepelev}}, \bibinfo {author} {\bibfnamefont {A.~V.}\ \bibnamefont {Bukh}},
  \bibinfo {author} {\bibfnamefont {G.~I.}\ \bibnamefont {Strelkova}}, \bibinfo
  {author} {\bibfnamefont {T.~E.}\ \bibnamefont {Vadivasova}}, \ and\ \bibinfo
  {author} {\bibfnamefont {V.~S.}\ \bibnamefont {Anishchenko}},\ }\bibfield
  {title} {\enquote {\bibinfo {title} {Chimera states in ensembles of bistable
  elements with regular and chaotic dynamics},}\ }\href@noop {} {\bibfield
  {journal} {\bibinfo  {journal} {Nonlinear Dynamics}\ }\textbf {\bibinfo
  {volume} {90}},\ \bibinfo {pages} {2317--2330} (\bibinfo {year}
  {2017})}\BibitemShut {NoStop}%
\bibitem [{\citenamefont {Muni}\ \emph {et~al.}(2022)\citenamefont {Muni},
  \citenamefont {Njıtacke}, \citenamefont {Feudjio}, \citenamefont {Fozin},\
  and\ \citenamefont {Awrejcewicz}}]{MuZe22}%
  \BibitemOpen
  \bibfield  {author} {\bibinfo {author} {\bibfnamefont {S.}~\bibnamefont
  {Muni}}, \bibinfo {author} {\bibfnamefont {Z.}~\bibnamefont {Njıtacke}},
  \bibinfo {author} {\bibfnamefont {C.}~\bibnamefont {Feudjio}}, \bibinfo
  {author} {\bibfnamefont {T.}~\bibnamefont {Fozin}}, \ and\ \bibinfo {author}
  {\bibfnamefont {J.}~\bibnamefont {Awrejcewicz}},\ }\bibfield  {title}
  {\enquote {\bibinfo {title} {Route to chaos and chimera states in a network
  of memristive {H}indmarsh-{R}ose neurons model with external excitation},}\
  }\href@noop {} {\bibfield  {journal} {\bibinfo  {journal} {Chaos Theory and
  Applications}\ }\textbf {\bibinfo {volume} {4}},\ \bibinfo {pages} {119 --
  127} (\bibinfo {year} {2022})}\BibitemShut {NoStop}%
\bibitem [{\citenamefont {Muni}\ and\ \citenamefont {Provata}(2020)}]{Muni20}%
  \BibitemOpen
  \bibfield  {author} {\bibinfo {author} {\bibfnamefont {S.}~\bibnamefont
  {Muni}}\ and\ \bibinfo {author} {\bibfnamefont {A.}~\bibnamefont {Provata}},\
  }\bibfield  {title} {\enquote {\bibinfo {title} {Chimera states in ring--star
  network of {C}hua circuits},}\ }\href@noop {} {\bibfield  {journal} {\bibinfo
   {journal} {Nonlinear. Dyn.}\ }\textbf {\bibinfo {volume} {101}},\ \bibinfo
  {pages} {2509--2521} (\bibinfo {year} {2020})}\BibitemShut {NoStop}%
\bibitem [{\citenamefont {Santos}\ \emph {et~al.}(2022)\citenamefont {Santos},
  \citenamefont {Sales}, \citenamefont {Muni}, \citenamefont {Szezech},
  \citenamefont {Batista}, \citenamefont {Yanchuk},\ and\ \citenamefont
  {Kurths}}]{VaMu22}%
  \BibitemOpen
  \bibfield  {author} {\bibinfo {author} {\bibfnamefont {V.}~\bibnamefont
  {Santos}}, \bibinfo {author} {\bibfnamefont {M.}~\bibnamefont {Sales}},
  \bibinfo {author} {\bibfnamefont {S.}~\bibnamefont {Muni}}, \bibinfo {author}
  {\bibfnamefont {J.}~\bibnamefont {Szezech}}, \bibinfo {author} {\bibfnamefont
  {A.~.}\ \bibnamefont {Batista}}, \bibinfo {author} {\bibfnamefont
  {S.}~\bibnamefont {Yanchuk}}, \ and\ \bibinfo {author} {\bibfnamefont
  {J.}~\bibnamefont {Kurths}},\ }\href@noop {} {\enquote {\bibinfo {title}
  {Identification of single- and double-well coherence-incoherence patterns by
  the binary distance matrix},}\ } (\bibinfo {year} {2022})\BibitemShut
  {NoStop}%
\bibitem [{\citenamefont {Xie}, \citenamefont {Knobloch},\ and\ \citenamefont
  {Kao}(2014)}]{Xie:2014uj}%
  \BibitemOpen
  \bibfield  {author} {\bibinfo {author} {\bibfnamefont {J.}~\bibnamefont
  {Xie}}, \bibinfo {author} {\bibfnamefont {E.}~\bibnamefont {Knobloch}}, \
  and\ \bibinfo {author} {\bibfnamefont {H.-C.}\ \bibnamefont {Kao}},\
  }\bibfield  {title} {\enquote {\bibinfo {title} {Multicluster and traveling
  chimera states in nonlocal phase-coupled oscillators},}\ }\href@noop {}
  {\bibfield  {journal} {\bibinfo  {journal} {Physical Review E}\ }\textbf
  {\bibinfo {volume} {90}},\ \bibinfo {pages} {022919} (\bibinfo {year}
  {2014})}\BibitemShut {NoStop}%
\bibitem [{\citenamefont {Omelchenko}\ \emph {et~al.}(2013)\citenamefont
  {Omelchenko}, \citenamefont {Omel'chenko}, \citenamefont {H{\"o}vel},\ and\
  \citenamefont {Sch{\"o}ll}}]{Omelchenko:2013uv}%
  \BibitemOpen
  \bibfield  {author} {\bibinfo {author} {\bibfnamefont {I.}~\bibnamefont
  {Omelchenko}}, \bibinfo {author} {\bibfnamefont {E.}~\bibnamefont
  {Omel'chenko}}, \bibinfo {author} {\bibfnamefont {P.}~\bibnamefont
  {H{\"o}vel}}, \ and\ \bibinfo {author} {\bibfnamefont {E.}~\bibnamefont
  {Sch{\"o}ll}},\ }\bibfield  {title} {\enquote {\bibinfo {title} {When
  nonlocal coupling between oscillators becomes stronger: patched synchrony or
  multichimera states},}\ }\href@noop {} {\bibfield  {journal} {\bibinfo
  {journal} {Physical review letters}\ }\textbf {\bibinfo {volume} {110}},\
  \bibinfo {pages} {224101} (\bibinfo {year} {2013})}\BibitemShut {NoStop}%
\bibitem [{\citenamefont {Olmi}\ \emph {et~al.}(2015)\citenamefont {Olmi},
  \citenamefont {Martens}, \citenamefont {Thutupalli},\ and\ \citenamefont
  {Torcini}}]{Olmi:2015uv}%
  \BibitemOpen
  \bibfield  {author} {\bibinfo {author} {\bibfnamefont {S.}~\bibnamefont
  {Olmi}}, \bibinfo {author} {\bibfnamefont {E.~A.}\ \bibnamefont {Martens}},
  \bibinfo {author} {\bibfnamefont {S.}~\bibnamefont {Thutupalli}}, \ and\
  \bibinfo {author} {\bibfnamefont {A.}~\bibnamefont {Torcini}},\ }\bibfield
  {title} {\enquote {\bibinfo {title} {Intermittent chaotic chimeras for
  coupled rotators},}\ }\href@noop {} {\bibfield  {journal} {\bibinfo
  {journal} {Physical Review E}\ }\textbf {\bibinfo {volume} {92}},\ \bibinfo
  {pages} {030901} (\bibinfo {year} {2015})}\BibitemShut {NoStop}%
\bibitem [{\citenamefont {Bordyugov}, \citenamefont {Pikovsky},\ and\
  \citenamefont {Rosenblum}(2010)}]{Bordyugov:2010wg}%
  \BibitemOpen
  \bibfield  {author} {\bibinfo {author} {\bibfnamefont {G.}~\bibnamefont
  {Bordyugov}}, \bibinfo {author} {\bibfnamefont {A.}~\bibnamefont {Pikovsky}},
  \ and\ \bibinfo {author} {\bibfnamefont {M.}~\bibnamefont {Rosenblum}},\
  }\bibfield  {title} {\enquote {\bibinfo {title} {Self-emerging and turbulent
  chimeras in oscillator chains},}\ }\href@noop {} {\bibfield  {journal}
  {\bibinfo  {journal} {Physical Review E}\ }\textbf {\bibinfo {volume} {82}},\
  \bibinfo {pages} {035205} (\bibinfo {year} {2010})}\BibitemShut {NoStop}%
\bibitem [{\citenamefont {Bolotov}\ \emph {et~al.}(2018)\citenamefont
  {Bolotov}, \citenamefont {Smirnov}, \citenamefont {Osipov},\ and\
  \citenamefont {Pikovsky}}]{Bolotov:2018uz}%
  \BibitemOpen
  \bibfield  {author} {\bibinfo {author} {\bibfnamefont {M.}~\bibnamefont
  {Bolotov}}, \bibinfo {author} {\bibfnamefont {L.}~\bibnamefont {Smirnov}},
  \bibinfo {author} {\bibfnamefont {G.}~\bibnamefont {Osipov}}, \ and\ \bibinfo
  {author} {\bibfnamefont {A.}~\bibnamefont {Pikovsky}},\ }\bibfield  {title}
  {\enquote {\bibinfo {title} {Simple and complex chimera states in a
  nonlinearly coupled oscillatory medium},}\ }\href@noop {} {\bibfield
  {journal} {\bibinfo  {journal} {Chaos: An Interdisciplinary Journal of
  Nonlinear Science}\ }\textbf {\bibinfo {volume} {28}},\ \bibinfo {pages}
  {045101} (\bibinfo {year} {2018})}\BibitemShut {NoStop}%
\bibitem [{\citenamefont {Maistrenko}\ \emph {et~al.}(2015)\citenamefont
  {Maistrenko}, \citenamefont {Sudakov}, \citenamefont {Osiv},\ and\
  \citenamefont {Maistrenko}}]{Maistrenko:2015vq}%
  \BibitemOpen
  \bibfield  {author} {\bibinfo {author} {\bibfnamefont {Y.}~\bibnamefont
  {Maistrenko}}, \bibinfo {author} {\bibfnamefont {O.}~\bibnamefont {Sudakov}},
  \bibinfo {author} {\bibfnamefont {O.}~\bibnamefont {Osiv}}, \ and\ \bibinfo
  {author} {\bibfnamefont {V.}~\bibnamefont {Maistrenko}},\ }\bibfield  {title}
  {\enquote {\bibinfo {title} {Chimera states in three dimensions},}\
  }\href@noop {} {\bibfield  {journal} {\bibinfo  {journal} {New Journal of
  Physics}\ }\textbf {\bibinfo {volume} {17}},\ \bibinfo {pages} {073037 
  1367--2630} (\bibinfo {year} {2015})}\BibitemShut {NoStop}%
\bibitem [{\citenamefont {Kuramoto}\ and\ \citenamefont
  {Shima}(2003)}]{Kuramoto:2003tj}%
  \BibitemOpen
  \bibfield  {author} {\bibinfo {author} {\bibfnamefont {Y.}~\bibnamefont
  {Kuramoto}}\ and\ \bibinfo {author} {\bibfnamefont {S.-i.}\ \bibnamefont
  {Shima}},\ }\bibfield  {title} {\enquote {\bibinfo {title} {Rotating spirals
  without phase singularity in reaction-diffusion systems},}\ }\href@noop {}
  {\bibfield  {journal} {\bibinfo  {journal} {Progress of Theoretical Physics
  Supplement}\ }\textbf {\bibinfo {volume} {150}},\ \bibinfo {pages} {115--125}
  (\bibinfo {year} {2003})}\BibitemShut {NoStop}%
\bibitem [{\citenamefont {Shima}\ and\ \citenamefont
  {Kuramoto}(2004)}]{Shima:2004aa}%
  \BibitemOpen
  \bibfield  {author} {\bibinfo {author} {\bibfnamefont {S.-i.}\ \bibnamefont
  {Shima}}\ and\ \bibinfo {author} {\bibfnamefont {Y.}~\bibnamefont
  {Kuramoto}},\ }\bibfield  {title} {\enquote {\bibinfo {title} {Rotating
  spiral waves with phase-randomized core in nonlocally coupled oscillators},}\
  }\href@noop {} {\bibfield  {journal} {\bibinfo  {journal} {Physical Review
  E}\ }\textbf {\bibinfo {volume} {69}},\ \bibinfo {pages} {036213} (\bibinfo
  {year} {2004})}\BibitemShut {NoStop}%
\bibitem [{\citenamefont {Bukh}\ and\ \citenamefont
  {Anishchenko}(2019)}]{Bukh:2019wo}%
  \BibitemOpen
  \bibfield  {author} {\bibinfo {author} {\bibfnamefont {A.}~\bibnamefont
  {Bukh}}\ and\ \bibinfo {author} {\bibfnamefont {V.}~\bibnamefont
  {Anishchenko}},\ }\bibfield  {title} {\enquote {\bibinfo {title} {Spiral,
  target, and chimera wave structures in a two-dimensional ensemble of
  nonlocally coupled van der {P}ol oscillators},}\ }\href@noop {} {\bibfield
  {journal} {\bibinfo  {journal} {Technical Physics Letters}\ }\textbf
  {\bibinfo {volume} {45}},\ \bibinfo {pages} {675--678} (\bibinfo {year}
  {2019})}\BibitemShut {NoStop}%
\bibitem [{\citenamefont {Rybalova}, \citenamefont {Strelkova},\ and\
  \citenamefont {Anishchenko}(2018)}]{Rybalova:2018we}%
  \BibitemOpen
  \bibfield  {author} {\bibinfo {author} {\bibfnamefont {E.}~\bibnamefont
  {Rybalova}}, \bibinfo {author} {\bibfnamefont {G.}~\bibnamefont {Strelkova}},
  \ and\ \bibinfo {author} {\bibfnamefont {V.}~\bibnamefont {Anishchenko}},\
  }\bibfield  {title} {\enquote {\bibinfo {title} {Mechanism of realizing a
  solitary state chimera in a ring of nonlocally coupled chaotic maps},}\
  }\href@noop {} {\bibfield  {journal} {\bibinfo  {journal} {Chaos, Solitons \&
  Fractals}\ }\textbf {\bibinfo {volume} {115}},\ \bibinfo {pages} {300--305}
  (\bibinfo {year} {2018})}\BibitemShut {NoStop}%
\bibitem [{\citenamefont {Omelchenko}\ \emph {et~al.}(2012)\citenamefont
  {Omelchenko}, \citenamefont {Riemenschneider}, \citenamefont {H{\"o}vel},
  \citenamefont {Maistrenko},\ and\ \citenamefont
  {Sch{\"o}ll}}]{Omelchenko:2012tv}%
  \BibitemOpen
  \bibfield  {author} {\bibinfo {author} {\bibfnamefont {I.}~\bibnamefont
  {Omelchenko}}, \bibinfo {author} {\bibfnamefont {B.}~\bibnamefont
  {Riemenschneider}}, \bibinfo {author} {\bibfnamefont {P.}~\bibnamefont
  {H{\"o}vel}}, \bibinfo {author} {\bibfnamefont {Y.}~\bibnamefont
  {Maistrenko}}, \ and\ \bibinfo {author} {\bibfnamefont {E.}~\bibnamefont
  {Sch{\"o}ll}},\ }\bibfield  {title} {\enquote {\bibinfo {title} {Transition
  from spatial coherence to incoherence in coupled chaotic systems},}\
  }\href@noop {} {\bibfield  {journal} {\bibinfo  {journal} {Physical Review
  E}\ }\textbf {\bibinfo {volume} {85}},\ \bibinfo {pages} {026212} (\bibinfo
  {year} {2012})}\BibitemShut {NoStop}%
\bibitem [{\citenamefont {Majhi}, \citenamefont {Kapitaniak},\ and\
  \citenamefont {Ghosh}(2019)}]{Majhi:2019wh}%
  \BibitemOpen
  \bibfield  {author} {\bibinfo {author} {\bibfnamefont {S.}~\bibnamefont
  {Majhi}}, \bibinfo {author} {\bibfnamefont {T.}~\bibnamefont {Kapitaniak}}, \
  and\ \bibinfo {author} {\bibfnamefont {D.}~\bibnamefont {Ghosh}},\ }\bibfield
   {title} {\enquote {\bibinfo {title} {Solitary states in multiplex networks
  owing to competing interactions},}\ }\href@noop {} {\bibfield  {journal}
  {\bibinfo  {journal} {Chaos: An Interdisciplinary Journal of Nonlinear
  Science}\ }\textbf {\bibinfo {volume} {29}},\ \bibinfo {pages} {013108}
  (\bibinfo {year} {2019})}\BibitemShut {NoStop}%
\bibitem [{\citenamefont {Wolfrum}\ and\ \citenamefont
  {Omel'chenko}(2011)}]{Wolfrum:2011wb}%
  \BibitemOpen
  \bibfield  {author} {\bibinfo {author} {\bibfnamefont {M.}~\bibnamefont
  {Wolfrum}}\ and\ \bibinfo {author} {\bibfnamefont {E.}~\bibnamefont
  {Omel'chenko}},\ }\bibfield  {title} {\enquote {\bibinfo {title} {Chimera
  states are chaotic transients},}\ }\href@noop {} {\bibfield  {journal}
  {\bibinfo  {journal} {Physical Review E}\ }\textbf {\bibinfo {volume} {84}},\
  \bibinfo {pages} {015201} (\bibinfo {year} {2011})}\BibitemShut {NoStop}%
\bibitem [{\citenamefont {Loos}\ \emph {et~al.}(2016)\citenamefont {Loos},
  \citenamefont {Claussen}, \citenamefont {Sch{\"o}ll},\ and\ \citenamefont
  {Zakharova}}]{Loos:2016aa}%
  \BibitemOpen
  \bibfield  {author} {\bibinfo {author} {\bibfnamefont {S.~A.}\ \bibnamefont
  {Loos}}, \bibinfo {author} {\bibfnamefont {J.~C.}\ \bibnamefont {Claussen}},
  \bibinfo {author} {\bibfnamefont {E.}~\bibnamefont {Sch{\"o}ll}}, \ and\
  \bibinfo {author} {\bibfnamefont {A.}~\bibnamefont {Zakharova}},\ }\bibfield
  {title} {\enquote {\bibinfo {title} {Chimera patterns under the impact of
  noise},}\ }\href@noop {} {\bibfield  {journal} {\bibinfo  {journal} {Physical
  Review E}\ }\textbf {\bibinfo {volume} {93}},\ \bibinfo {pages} {012209}
  (\bibinfo {year} {2016})}\BibitemShut {NoStop}%
\bibitem [{\citenamefont {Rybalova}\ \emph
  {et~al.}(2019{\natexlab{b}})\citenamefont {Rybalova}, \citenamefont
  {Klyushina}, \citenamefont {Anishchenko},\ and\ \citenamefont
  {Strelkova}}]{Rybalova:2019un}%
  \BibitemOpen
  \bibfield  {author} {\bibinfo {author} {\bibfnamefont {E.~V.}\ \bibnamefont
  {Rybalova}}, \bibinfo {author} {\bibfnamefont {D.~Y.}\ \bibnamefont
  {Klyushina}}, \bibinfo {author} {\bibfnamefont {V.~S.}\ \bibnamefont
  {Anishchenko}}, \ and\ \bibinfo {author} {\bibfnamefont {G.~I.}\ \bibnamefont
  {Strelkova}},\ }\bibfield  {title} {\enquote {\bibinfo {title} {Impact of
  noise on the amplitude chimera lifetime in an ensemble of nonlocally coupled
  chaotic maps},}\ }\href@noop {} {\bibfield  {journal} {\bibinfo  {journal}
  {Regular and Chaotic Dynamics}\ }\textbf {\bibinfo {volume} {24}},\ \bibinfo
  {pages} {432--445} (\bibinfo {year} {2019}{\natexlab{b}})}\BibitemShut
  {NoStop}%
\bibitem [{\citenamefont {Bukh}\ \emph {et~al.}(2018)\citenamefont {Bukh},
  \citenamefont {Slepnev}, \citenamefont {Anishchenko},\ and\ \citenamefont
  {Vadivasova}}]{Bukh:2018wa}%
  \BibitemOpen
  \bibfield  {author} {\bibinfo {author} {\bibfnamefont {A.~V.}\ \bibnamefont
  {Bukh}}, \bibinfo {author} {\bibfnamefont {A.~V.}\ \bibnamefont {Slepnev}},
  \bibinfo {author} {\bibfnamefont {V.~S.}\ \bibnamefont {Anishchenko}}, \ and\
  \bibinfo {author} {\bibfnamefont {T.~E.}\ \bibnamefont {Vadivasova}},\
  }\bibfield  {title} {\enquote {\bibinfo {title} {Stability and noise-induced
  transitions in an ensemble of nonlocally coupled chaotic maps},}\ }\href@noop
  {} {\bibfield  {journal} {\bibinfo  {journal} {Regular and Chaotic Dynamics}\
  }\textbf {\bibinfo {volume} {23}},\ \bibinfo {pages} {325--338} (\bibinfo
  {year} {2018})}\BibitemShut {NoStop}%
\bibitem [{\citenamefont {Shepelev}\ \emph {et~al.}(2020)\citenamefont
  {Shepelev}, \citenamefont {Bukh}, \citenamefont {Muni},\ and\ \citenamefont
  {Anishchenko}}]{ShMu20a}%
  \BibitemOpen
  \bibfield  {author} {\bibinfo {author} {\bibfnamefont {I.}~\bibnamefont
  {Shepelev}}, \bibinfo {author} {\bibfnamefont {A.}~\bibnamefont {Bukh}},
  \bibinfo {author} {\bibfnamefont {S.}~\bibnamefont {Muni}}, \ and\ \bibinfo
  {author} {\bibfnamefont {V.}~\bibnamefont {Anishchenko}},\ }\bibfield
  {title} {\enquote {\bibinfo {title} {Role of solitary states in forming
  spatiotemporal patterns in a 2{D} lattice of van der {P}ol oscillators},}\
  }\href@noop {} {\bibfield  {journal} {\bibinfo  {journal} {Chaos, Solitons,
  Fractals}\ }\textbf {\bibinfo {volume} {135:109725}} (\bibinfo {year}
  {2020})}\BibitemShut {NoStop}%
\bibitem [{\citenamefont {Shepelev}, \citenamefont {Bukh},\ and\ \citenamefont
  {Muni}(2020)}]{ShMu20b}%
  \BibitemOpen
  \bibfield  {author} {\bibinfo {author} {\bibfnamefont {I.}~\bibnamefont
  {Shepelev}}, \bibinfo {author} {\bibfnamefont {A.}~\bibnamefont {Bukh}}, \
  and\ \bibinfo {author} {\bibfnamefont {S.}~\bibnamefont {Muni}},\ }\bibfield
  {title} {\enquote {\bibinfo {title} {Quantifying the transition from spiral
  waves to spiral wave chimeras in a lattice of self-sustained oscillators},}\
  }\href {\doibase https://doi.org/10.1134/S1560354720060076} {\bibfield
  {journal} {\bibinfo  {journal} {Regul. Chaot. Dyn.}\ }\textbf {\bibinfo
  {volume} {25}},\ \bibinfo {pages} {597--615} (\bibinfo {year}
  {2020})}\BibitemShut {NoStop}%
\bibitem [{\citenamefont {Shepelev}\ \emph {et~al.}(2021)\citenamefont
  {Shepelev}, \citenamefont {Muni}, \citenamefont {Sch{\"o}ll},\ and\
  \citenamefont {Strelkova}}]{ShMu21a}%
  \BibitemOpen
  \bibfield  {author} {\bibinfo {author} {\bibfnamefont {I.}~\bibnamefont
  {Shepelev}}, \bibinfo {author} {\bibfnamefont {S.}~\bibnamefont {Muni}},
  \bibinfo {author} {\bibfnamefont {E.}~\bibnamefont {Sch{\"o}ll}}, \ and\
  \bibinfo {author} {\bibfnamefont {G.}~\bibnamefont {Strelkova}},\ }\bibfield
  {title} {\enquote {\bibinfo {title} {Repulsive inter-layer coupling induces
  anti-phase synchronization},}\ }\href@noop {} {\bibfield  {journal} {\bibinfo
   {journal} {Chaos}\ }\textbf {\bibinfo {volume} {31, 063116}} (\bibinfo
  {year} {2021})}\BibitemShut {NoStop}%
\bibitem [{\citenamefont {Shepelev}, \citenamefont {Muni},\ and\ \citenamefont
  {Vadivasova}(2021{\natexlab{a}})}]{ShMu21b}%
  \BibitemOpen
  \bibfield  {author} {\bibinfo {author} {\bibfnamefont {I.}~\bibnamefont
  {Shepelev}}, \bibinfo {author} {\bibfnamefont {S.}~\bibnamefont {Muni}}, \
  and\ \bibinfo {author} {\bibfnamefont {T.}~\bibnamefont {Vadivasova}},\
  }\bibfield  {title} {\enquote {\bibinfo {title} {Synchronization of wave
  structures in a heterogeneous multiplex network of 2{D} lattices with
  attractive and repulsive intra-layer coupling},}\ }\href@noop {} {\bibfield
  {journal} {\bibinfo  {journal} {Chaos}\ }\textbf {\bibinfo {volume} {31,
  021104}} (\bibinfo {year} {2021}{\natexlab{a}})}\BibitemShut {NoStop}%
\bibitem [{\citenamefont {Shepelev}, \citenamefont {Muni},\ and\ \citenamefont
  {Vadivasova}(2021{\natexlab{b}})}]{ShMu21c}%
  \BibitemOpen
  \bibfield  {author} {\bibinfo {author} {\bibfnamefont {I.}~\bibnamefont
  {Shepelev}}, \bibinfo {author} {\bibfnamefont {S.}~\bibnamefont {Muni}}, \
  and\ \bibinfo {author} {\bibfnamefont {T.~E.}\ \bibnamefont {Vadivasova}},\
  }\bibfield  {title} {\enquote {\bibinfo {title} {Spatiotemporal patterns in a
  2{D} lattice with linear repulsive and nonlinear attractive coupling},}\
  }\href {\doibase https://doi.org/10.1063/5.0048324} {\bibfield  {journal}
  {\bibinfo  {journal} {Chaos}\ }\textbf {\bibinfo {volume} {31: 043136}}
  (\bibinfo {year} {2021}{\natexlab{b}}),\
  https://doi.org/10.1063/5.0048324}\BibitemShut {NoStop}%
\bibitem [{\citenamefont {Cromer}(1981)}]{Cromer:1981vm}%
  \BibitemOpen
  \bibfield  {author} {\bibinfo {author} {\bibfnamefont {A.}~\bibnamefont
  {Cromer}},\ }\bibfield  {title} {\enquote {\bibinfo {title} {Stable solutions
  using the euler approximation},}\ }\href@noop {} {\bibfield  {journal}
  {\bibinfo  {journal} {American Journal of Physics}\ }\textbf {\bibinfo
  {volume} {49}},\ \bibinfo {pages} {455--459} (\bibinfo {year}
  {1981})}\BibitemShut {NoStop}%
\bibitem [{\citenamefont {Popova}, \citenamefont {Stankevich},\ and\
  \citenamefont {Kuznetsov}(2020)}]{Popova-eng:2020tz}%
  \BibitemOpen
  \bibfield  {author} {\bibinfo {author} {\bibfnamefont {E.~S.}\ \bibnamefont
  {Popova}}, \bibinfo {author} {\bibfnamefont {N.~V.}\ \bibnamefont
  {Stankevich}}, \ and\ \bibinfo {author} {\bibfnamefont {A.~P.}\ \bibnamefont
  {Kuznetsov}},\ }\bibfield  {title} {\enquote {\bibinfo {title} {Cascade of
  invariant curve doubling bifurcations and quasi-periodic {H}{\'e}non
  attractor in the discrete {L}orenz-84 model},}\ }\href@noop {} {\bibfield
  {journal} {\bibinfo  {journal} {Izv. Saratov Univ. (N. S.), Ser. Physics}\
  }\textbf {\bibinfo {volume} {20}},\ \bibinfo {pages} {222--232} (\bibinfo
  {year} {2020})}\BibitemShut {NoStop}%
\bibitem [{\citenamefont {USHIO}\ and\ \citenamefont
  {HIRAI}(1986)}]{USHIO:1986vr}%
  \BibitemOpen
  \bibfield  {author} {\bibinfo {author} {\bibfnamefont {T.}~\bibnamefont
  {USHIO}}\ and\ \bibinfo {author} {\bibfnamefont {K.}~\bibnamefont {HIRAI}},\
  }\bibfield  {title} {\enquote {\bibinfo {title} {Chaos induced by the
  generalized euler method},}\ }\href@noop {} {\bibfield  {journal} {\bibinfo
  {journal} {International journal of systems science}\ }\textbf {\bibinfo
  {volume} {17}},\ \bibinfo {pages} {669--678} (\bibinfo {year}
  {1986})}\BibitemShut {NoStop}%
\bibitem [{\citenamefont {Sprott}(2003)}]{Sprott:2003vc}%
  \BibitemOpen
  \bibfield  {author} {\bibinfo {author} {\bibfnamefont {J.~C.}\ \bibnamefont
  {Sprott}},\ }\href@noop {} {\emph {\bibinfo {title} {Chaos and time-series
  analysis}}},\ Vol.~\bibinfo {volume} {69}\ (\bibinfo {year}
  {2003})\BibitemShut {NoStop}%
\bibitem [{\citenamefont {Zaytcev}(2017)}]{Zaytcev:2017vj}%
  \BibitemOpen
  \bibfield  {author} {\bibinfo {author} {\bibfnamefont {V.~V.}\ \bibnamefont
  {Zaytcev}},\ }\bibfield  {title} {\enquote {\bibinfo {title} {The discrete
  van der pol oscillator: Finite differences and slow amplitudes},}\
  }\href@noop {} {\bibfield  {journal} {\bibinfo  {journal} {Izvestiya VUZ.
  Applied Nonlinear Dynamics}\ }\textbf {\bibinfo {volume} {25}},\ \bibinfo
  {pages} {70--78} (\bibinfo {year} {2017})}\BibitemShut {NoStop}%
\bibitem [{\citenamefont {Nguyen‐Van}\ and\ \citenamefont
  {Hori}(2013)}]{NguyenVan:2013vc}%
  \BibitemOpen
  \bibfield  {author} {\bibinfo {author} {\bibfnamefont {T.}~\bibnamefont
  {Nguyen‐Van}}\ and\ \bibinfo {author} {\bibfnamefont {N.}~\bibnamefont
  {Hori}},\ }\bibfield  {title} {\enquote {\bibinfo {title} {New class of
  discrete‐time models for non‐linear systems through discretisation of
  integration gains},}\ }\href@noop {} {\bibfield  {journal} {\bibinfo
  {journal} {IET Control Theory \& Applications}\ }\textbf {\bibinfo {volume}
  {7}},\ \bibinfo {pages} {80--89} (\bibinfo {year} {2013})}\BibitemShut
  {NoStop}%
\bibitem [{\citenamefont {Astakhov}\ \emph {et~al.}(2001)\citenamefont
  {Astakhov}, \citenamefont {Shabunin}, \citenamefont {Uhm},\ and\
  \citenamefont {Kim}}]{Astakhov:2001wq}%
  \BibitemOpen
  \bibfield  {author} {\bibinfo {author} {\bibfnamefont {V.}~\bibnamefont
  {Astakhov}}, \bibinfo {author} {\bibfnamefont {A.}~\bibnamefont {Shabunin}},
  \bibinfo {author} {\bibfnamefont {W.}~\bibnamefont {Uhm}}, \ and\ \bibinfo
  {author} {\bibfnamefont {S.}~\bibnamefont {Kim}},\ }\bibfield  {title}
  {\enquote {\bibinfo {title} {Multistability formation and synchronization
  loss in coupled h{\'e}non maps: Two sides of the single bifurcational
  mechanism},}\ }\href@noop {} {\bibfield  {journal} {\bibinfo  {journal}
  {Physical Review E}\ }\textbf {\bibinfo {volume} {63}},\ \bibinfo {pages}
  {056212} (\bibinfo {year} {2001})}\BibitemShut {NoStop}%
\bibitem [{\citenamefont {Muni}, \citenamefont {McLachlan},\ and\ \citenamefont
  {Simpson}(2020)}]{Muni:2020wy}%
  \BibitemOpen
  \bibfield  {author} {\bibinfo {author} {\bibfnamefont {S.~S.}\ \bibnamefont
  {Muni}}, \bibinfo {author} {\bibfnamefont {R.~I.}\ \bibnamefont {McLachlan}},
  \ and\ \bibinfo {author} {\bibfnamefont {D.~J.}\ \bibnamefont {Simpson}},\
  }\bibfield  {title} {\enquote {\bibinfo {title} {Homoclinic tangencies with
  infinitely many asymptotically stable single-round periodic solutions},}\
  }\href@noop {} {\bibfield  {journal} {\bibinfo  {journal} {arXiv preprint
  arXiv:2006.01405}\ } (\bibinfo {year} {2020})}\BibitemShut {NoStop}%
\bibitem [{\citenamefont {Benettin}\ \emph {et~al.}(1980)\citenamefont
  {Benettin}, \citenamefont {Galgani}, \citenamefont {Giorgilli},\ and\
  \citenamefont {Strelcyn}}]{Benettin:1980tq}%
  \BibitemOpen
  \bibfield  {author} {\bibinfo {author} {\bibfnamefont {G.}~\bibnamefont
  {Benettin}}, \bibinfo {author} {\bibfnamefont {L.}~\bibnamefont {Galgani}},
  \bibinfo {author} {\bibfnamefont {A.}~\bibnamefont {Giorgilli}}, \ and\
  \bibinfo {author} {\bibfnamefont {J.-M.}\ \bibnamefont {Strelcyn}},\
  }\bibfield  {title} {\enquote {\bibinfo {title} {Lyapunov characteristic
  exponents for smooth dynamical systems and for hamiltonian systems; a method
  for computing all of them. part 1: Theory},}\ }\href@noop {} {\bibfield
  {journal} {\bibinfo  {journal} {Meccanica}\ }\textbf {\bibinfo {volume}
  {15}},\ \bibinfo {pages} {9--20} (\bibinfo {year} {1980})}\BibitemShut
  {NoStop}%
\bibitem [{\citenamefont {Afraimovich}\ and\ \citenamefont
  {Shilnikov}(1991)}]{Afraimovich:1991ul}%
  \BibitemOpen
  \bibfield  {author} {\bibinfo {author} {\bibfnamefont {V.}~\bibnamefont
  {Afraimovich}}\ and\ \bibinfo {author} {\bibfnamefont {L.~P.}\ \bibnamefont
  {Shilnikov}},\ }\bibfield  {title} {\enquote {\bibinfo {title} {Invariant
  two-dimensional tori, their breakdown and stochasticity},}\ }\href@noop {}
  {\bibfield  {journal} {\bibinfo  {journal} {Amer. Math. Soc. Transl}\
  }\textbf {\bibinfo {volume} {149}},\ \bibinfo {pages} {201--212} (\bibinfo
  {year} {1991})}\BibitemShut {NoStop}%
\bibitem [{\citenamefont {Endo}\ and\ \citenamefont
  {Mori}(1978)}]{Endo:1978tf}%
  \BibitemOpen
  \bibfield  {author} {\bibinfo {author} {\bibfnamefont {T.}~\bibnamefont
  {Endo}}\ and\ \bibinfo {author} {\bibfnamefont {S.}~\bibnamefont {Mori}},\
  }\bibfield  {title} {\enquote {\bibinfo {title} {Mode analysis of a ring of a
  large number of mutually coupled van der {P}ol oscillators},}\ }\href@noop {}
  {\bibfield  {journal} {\bibinfo  {journal} {IEEE Transactions on Circuits and
  systems}\ }\textbf {\bibinfo {volume} {25}},\ \bibinfo {pages} {7--18}
  (\bibinfo {year} {1978})}\BibitemShut {NoStop}%
\bibitem [{\citenamefont {Ermentrout}(1985)}]{Ermentrout:1985wo}%
  \BibitemOpen
  \bibfield  {author} {\bibinfo {author} {\bibfnamefont {G.}~\bibnamefont
  {Ermentrout}},\ }\bibfield  {title} {\enquote {\bibinfo {title} {The behavior
  of rings of coupled oscillators},}\ }\href@noop {} {\bibfield  {journal}
  {\bibinfo  {journal} {Journal of mathematical biology}\ }\textbf {\bibinfo
  {volume} {23}},\ \bibinfo {pages} {55--74} (\bibinfo {year}
  {1985})}\BibitemShut {NoStop}%
\bibitem [{\citenamefont {Rybalova}\ \emph
  {et~al.}(2019{\natexlab{c}})\citenamefont {Rybalova}, \citenamefont
  {Strelkova}, \citenamefont {Vadivasova},\ and\ \citenamefont
  {Anishchenko}}]{Rybalova:2019wd}%
  \BibitemOpen
  \bibfield  {author} {\bibinfo {author} {\bibfnamefont {E.~V.}\ \bibnamefont
  {Rybalova}}, \bibinfo {author} {\bibfnamefont {G.~I.}\ \bibnamefont
  {Strelkova}}, \bibinfo {author} {\bibfnamefont {T.~E.}\ \bibnamefont
  {Vadivasova}}, \ and\ \bibinfo {author} {\bibfnamefont {V.~S.}\ \bibnamefont
  {Anishchenko}},\ }\bibfield  {title} {\enquote {\bibinfo {title} {Bistability
  promotes solitary states in ensembles of nonlocally coupled maps},}\ \
  }(\bibinfo  {publisher} {SPIE},\ \bibinfo {year} {2019})\ pp.\ \bibinfo
  {pages} {156--161}\BibitemShut {NoStop}%
\bibitem [{\citenamefont {Santos}\ \emph {et~al.}(2018)\citenamefont {Santos},
  \citenamefont {Szezech~Jr}, \citenamefont {Batista}, \citenamefont {Iarosz},
  \citenamefont {Baptista}, \citenamefont {Ren}, \citenamefont {Grebogi},
  \citenamefont {Viana}, \citenamefont {Caldas},\ and\ \citenamefont
  {Maistrenko}}]{Santos:2018vr}%
  \BibitemOpen
  \bibfield  {author} {\bibinfo {author} {\bibfnamefont {V.~d.}\ \bibnamefont
  {Santos}}, \bibinfo {author} {\bibfnamefont {J.}~\bibnamefont {Szezech~Jr}},
  \bibinfo {author} {\bibfnamefont {A.~M.}\ \bibnamefont {Batista}}, \bibinfo
  {author} {\bibfnamefont {K.~C.}\ \bibnamefont {Iarosz}}, \bibinfo {author}
  {\bibfnamefont {M.~d.~S.}\ \bibnamefont {Baptista}}, \bibinfo {author}
  {\bibfnamefont {H.~P.}\ \bibnamefont {Ren}}, \bibinfo {author} {\bibfnamefont
  {C.}~\bibnamefont {Grebogi}}, \bibinfo {author} {\bibfnamefont {R.~L.}\
  \bibnamefont {Viana}}, \bibinfo {author} {\bibfnamefont {I.}~\bibnamefont
  {Caldas}}, \ and\ \bibinfo {author} {\bibfnamefont {Y.~L.}\ \bibnamefont
  {Maistrenko}},\ }\bibfield  {title} {\enquote {\bibinfo {title} {Riddling:
  Chimera's dilemma},}\ }\href@noop {} {\bibfield  {journal} {\bibinfo
  {journal} {Chaos: An Interdisciplinary Journal of Nonlinear Science}\
  }\textbf {\bibinfo {volume} {28}},\ \bibinfo {pages} {081105} (\bibinfo
  {year} {2018})}\BibitemShut {NoStop}%
\bibitem [{\citenamefont {Dos~Santos}\ \emph {et~al.}(2020)\citenamefont
  {Dos~Santos}, \citenamefont {Borges}, \citenamefont {Iarosz}, \citenamefont
  {Caldas}, \citenamefont {Szezech}, \citenamefont {Viana}, \citenamefont
  {Baptista},\ and\ \citenamefont {Batista}}]{Dos-Santos:2020wv}%
  \BibitemOpen
  \bibfield  {author} {\bibinfo {author} {\bibfnamefont {V.}~\bibnamefont
  {Dos~Santos}}, \bibinfo {author} {\bibfnamefont {F.~S.}\ \bibnamefont
  {Borges}}, \bibinfo {author} {\bibfnamefont {K.~C.}\ \bibnamefont {Iarosz}},
  \bibinfo {author} {\bibfnamefont {I.}~\bibnamefont {Caldas}}, \bibinfo
  {author} {\bibfnamefont {J.~D.}\ \bibnamefont {Szezech}}, \bibinfo {author}
  {\bibfnamefont {R.~L.}\ \bibnamefont {Viana}}, \bibinfo {author}
  {\bibfnamefont {M.~S.}\ \bibnamefont {Baptista}}, \ and\ \bibinfo {author}
  {\bibfnamefont {A.~M.}\ \bibnamefont {Batista}},\ }\bibfield  {title}
  {\enquote {\bibinfo {title} {Basin of attraction for chimera states in a
  network of {R}{\"o}ssler oscillators},}\ }\href@noop {} {\bibfield  {journal}
  {\bibinfo  {journal} {Chaos: An Interdisciplinary Journal of Nonlinear
  Science}\ }\textbf {\bibinfo {volume} {30}},\ \bibinfo {pages} {083115}
  (\bibinfo {year} {2020})}\BibitemShut {NoStop}%
\end{thebibliography}%

\end{document}